\documentclass{aa}

\usepackage{graphicx,epsfig,amsmath,amssymb,color,stmaryrd,multirow,makecell,cuted}
\usepackage{txfonts}
\bibpunct{(}{)}{;}{a}{}{,}
\graphicspath{ {paper_figs/} }

\newcommand{\crit}{{\mathrm{c}}}
\newcommand{\sound}{{\mathrm{s}}}
\newcommand{\grav}{{\mathrm{g}}}
\newcommand{\turb}{{\mathrm{t}}}

\newcommand{\ad}{{\mathrm{ad}}}

\newcommand{\Ms}{{\mathrm{M}_\odot}}

\newcommand{\pc}{{\mathrm{pc}}}
\newcommand{\cc}{{\mathrm{cm^{-3}}}}
\newcommand{\gcc}{{\mathrm{g~cm^{-3}}}}
\newcommand{\AU}{{\mathrm{AU}}}

\newcommand{\p}{_\mathrm{p}}
\newcommand{\e}{_\mathrm{e}}
\def\mathbi#1{\textbf{\em #1}}

\begin{document}

   \title{Stellar mass spectrum within massive collapsing clumps \\
II. Thermodynamics and tidal forces of the first Larson core}


   \titlerunning{Stellar mass spectrum within collapsing clumps}

   \author{
          Yueh-Ning Lee \inst{1,2,3}
          \and
          Patrick Hennebelle\inst{1,2,4}
          }
   \institute{IRFU, CEA, Universit\'{e} Paris-Saclay, F-91191 Gif-sur-Yvette, France\\
              \email{yueh-ning.lee@cea.fr}
         \and
                Universit\'{e} Paris Diderot, AIM, Sorbonne Paris Cit\'{e}, CEA, CNRS, F-91191 Gif-sur-Yvette, France
         \and
                Institut de Physique du Globe de Paris, Sorbonne Paris Cit\'e, Universit\'e Paris Diderot, UMR 7154 CNRS, F-75005 Paris, France
         \and
             LERMA (UMR CNRS 8112), Ecole Normale Sup\'{e}rieure, 75231 Paris Cedex, France\\
             \email{patrick.hennebelle@lra.ens.fr }
             }

  \date{Submitted 6 July 2017; accepted for publication}

 
  \abstract
   {Understanding the origin of the initial mass function (IMF) of stars is a major problem 
for the star formation process and beyond.
     }
   {We investigate the dependence of the peak of the IMF on the physics of the so-called first Larson core, which 
corresponds to the point where the dust becomes opaque to its own radiation.}
   {We perform numerical simulations of  collapsing clouds of $1000~\Ms$ for various 
gas equation of state (eos), paying great attention to the numerical resolution and convergence. 
The initial conditions of these numerical experiments are varied in the companion paper. 
We also develop analytical models that we confront to our numerical results. 
      }
   {If an isothermal eos is used, we show that the peak of the IMF shifts to lower masses with improved 
numerical resolution. When an adiabatic eos is employed, numerical convergence 
is obtained. The peak position varies with the eos and, using an analytical model to infer the mass of the first Larson core, we find that the peak position is about ten times its value. By analyzing the stability of non-linear density 
fluctuations in the vicinity of a point mass and then summing over a reasonable density distribution, 
we find that tidal forces exert a strong stabilizing effect and likely lead to a preferential 
mass several times larger than that of the first Larson core. 
}
   { We propose that in a sufficiently massive and cold cloud, the peak of the IMF is determined by the 
      thermodynamics of the high density adiabatic gas as well as the stabilizing influence of tidal forces. 
The resulting characteristic mass is about ten times the mass of the first Larson core, which altogether leads to a few tenths of solar masses. 
 Since these processes are not related to 
the large scale physical conditions and to the environment, our results suggest a possible explanation for the apparent universality of the peak of the  IMF. }


   \maketitle

\section{Introduction}
Star formation is believed to have major consequences on the structure of our Universe. In particular,
in spite of sustained efforts, the origin of the initial mass function (IMF) is still a matter of debate \citep[e.g.][]{offner2014}. 
While the gravo-turbulent fragmentation of cold dense molecular gas is a widely accepted scenario of star formation, 
it is in apparent contradiction with the observed IMF exhibiting no significant variation among different environments
\citep[e.g.][]{kroupa2001,chabrier2003,bastian2010,offner2014} although \citet{dib2017} found more
variations than usually assumed.

In particular, the physical origin of the peak of the IMF, around $0.3~\Ms$, is still uncertain. 
Several ideas have been proposed, which fall in three categories, not exclusive from one another. 
First, it has been proposed that the thermodynamics of the gas could provide a particular density
to which a Jeans mass is associated. For example, \citet{larson1985} computes the gas temperature 
in dense core conditions and concludes that due to molecular cooling the effective adiabatic 
exponent, $\gamma_\ad$, is expected to be on the order of 0.7 for densities below $\simeq 10^5~\cc$,
while at higher densities, $\gamma_\ad \simeq 1$ due to gas-dust coupling. Since the fragmentation 
sensitively depends on $\gamma_\ad$, the question arises as to whether the change of the effective equation 
of state at $\simeq 10^5~\cc$ could lead to a particular Jeans mass, which assuming a temperature of 
10 K turns out to be $\sim 0.3~\Ms$. Similar, though not identical, line of explanations has been put forward 
by \citet{elmegreen2008} that the thermodynamics of the gas in various environments could result in a 
Jeans mass that weakly depends on the gas density. 
The second category of explanations has been proposed by \citet{Bate09b}, \citet{krumholz2011} and \citet{guszejnov2016}. The idea
is that because of the heating from the accretion luminosity, the gas temperature increases
in the vicinity of the protostar and, therefore, the local increase of the Jeans mass leads to a 
possible {\it self-regulation}. Again, they infer that the Jeans mass has a weaker dependence on the 
gas density than in an isothermal gas. Finally, the third category of explanations invokes a compensation 
between the density and the velocity dispersion which vary in opposite directions with clump masses through 
Larson relations \citep{Hennebelle12,lee2016b}. Considering clumps of higher and higher mass, 
the density decreases with the mass making the Jeans mass higher, while the increasing Mach number  
tends to create denser gas through shocks, yielding a slowly varying characteristic mass. 
None of these explanations is well established yet. 
For example, while it has been suggested by various 
simulations \citep[e.g.][]{Jappsen05,Bate09b} that thermodynamics influences the mass spectrum 
of the objects that form, there is no well established set of simulations demonstrating 
the invariance of the peak of the IMF for a broad range of physical conditions.  

In the companion paper (paper I), we conduct a systematic exploration of the initial conditions
by varying the molecular cloud initial density and turbulence.
The physics is greatly simplified at this stage because our goal is to  identify the physical 
processes at play. 
At relatively high densities ($\gtrsim 10^5~\cc$), the shape of the stellar distribution, 
including the peak position and the high-mass end slope, 
becomes no longer dependent on the global density of the parent molecular cloud.
This means that some physical mechanisms are operating at local scales and do not depend 
on the large ones, say the collapsing clump itself.
At first sight, this result is surprising because the mean Jeans mass changes by more than one order of magnitude
in our series of simulations. Moreover, the thermodynamics treated in these simulations assumes 
that the gas remains isothermal up to a density of a few $10^{10}~\cc$, where the gas becomes 
adiabatic. While in the context of a collapsing clump, the isothermal regime does not provide any characteristic mass, 
the adiabatic regime does, because the Jeans mass increases with increasing density. The smallest Jeans mass, 
estimated at the density where the gas becomes adiabatic, is a few $10^{-3}~\Ms$
\citep{rees1976,whitworth2007} and is, therefore, almost two orders of magnitude smaller than the peak 
position of the stellar distribution around $0.1~\Ms$ observed in the simulations. 
Consequently, the origin of a characteristic mass in our simulations is not obvious 
and needs to be elucidated. 
For this purpose, we perform in this paper a series of simulations with initial conditions 
identical to the ones in paper I,
while altering the gas equation of state (eos).
Through the variation of stellar distribution peak position (around $0.1~\Ms$) according to the varying eos, 
we try to shed light on the mechanisms that  determine the IMF peak mass. 

In the second section, we present the numerical setup employed, including the 
initial conditions and the various eos used through the paper. In the third 
section, we present the results of the numerical simulations.
In the fourth section, we infer the mass of the first Larson cores for
various eos and we study its correlation with the 
mass spectrum peak in the simulations. In the fifth section, we develop an analytical model that
accounts for the factor, on the order of ten, between the mass of the first Larson core
and the peak of the stellar distribution obtained from numerical simulations. 
In the sixth section, we provide a discussion on our results. 
The seventh section concludes the paper.

\setlength{\unitlength}{1cm}
\begin{figure}
\begin{picture} (0,19.3)
\put(0,13){\includegraphics[width=8.7cm]{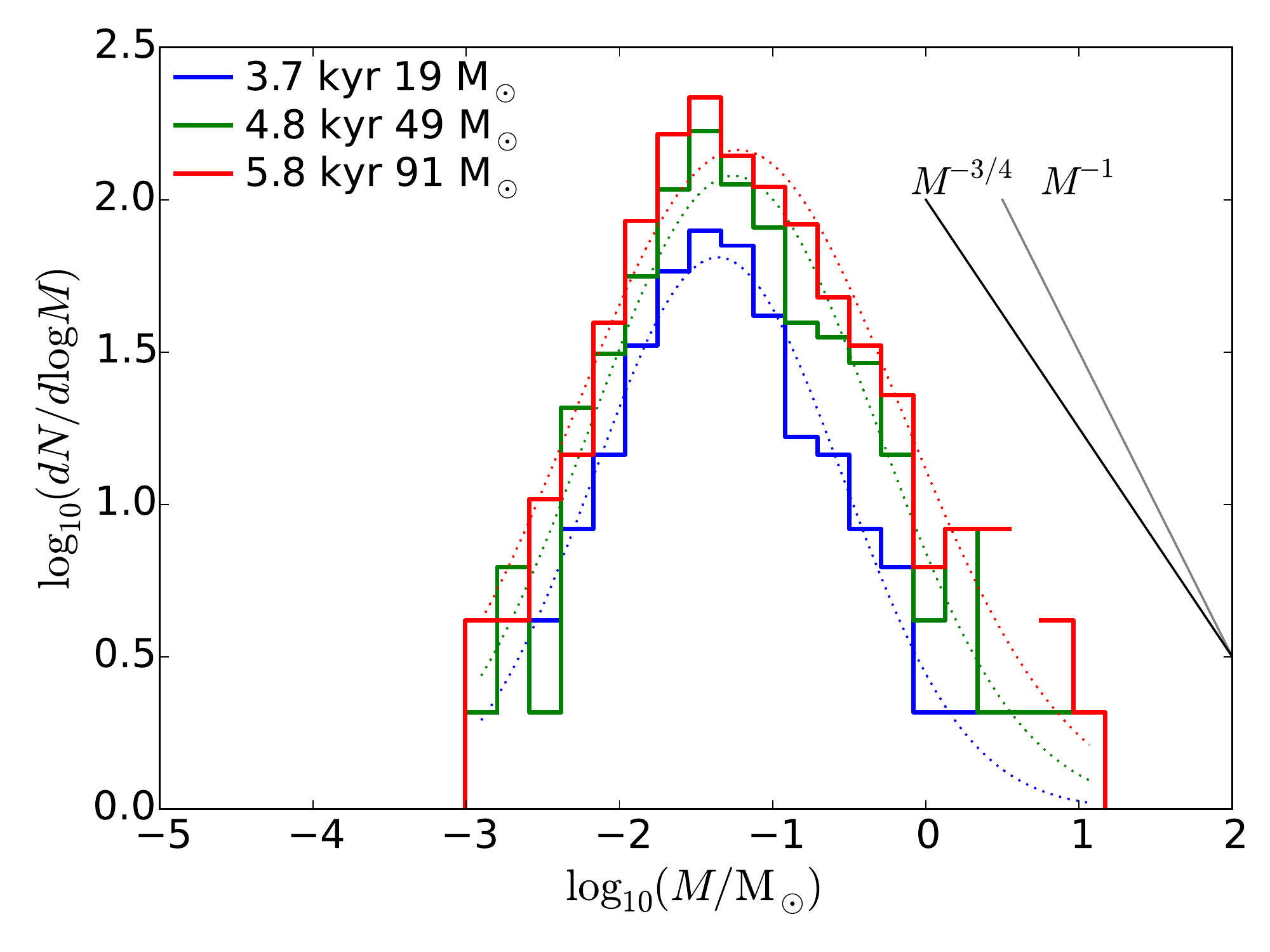}}  
\put(6.5,18.8){I-- --, 17 AU}
\put(0,6.5){\includegraphics[width=8.7cm]{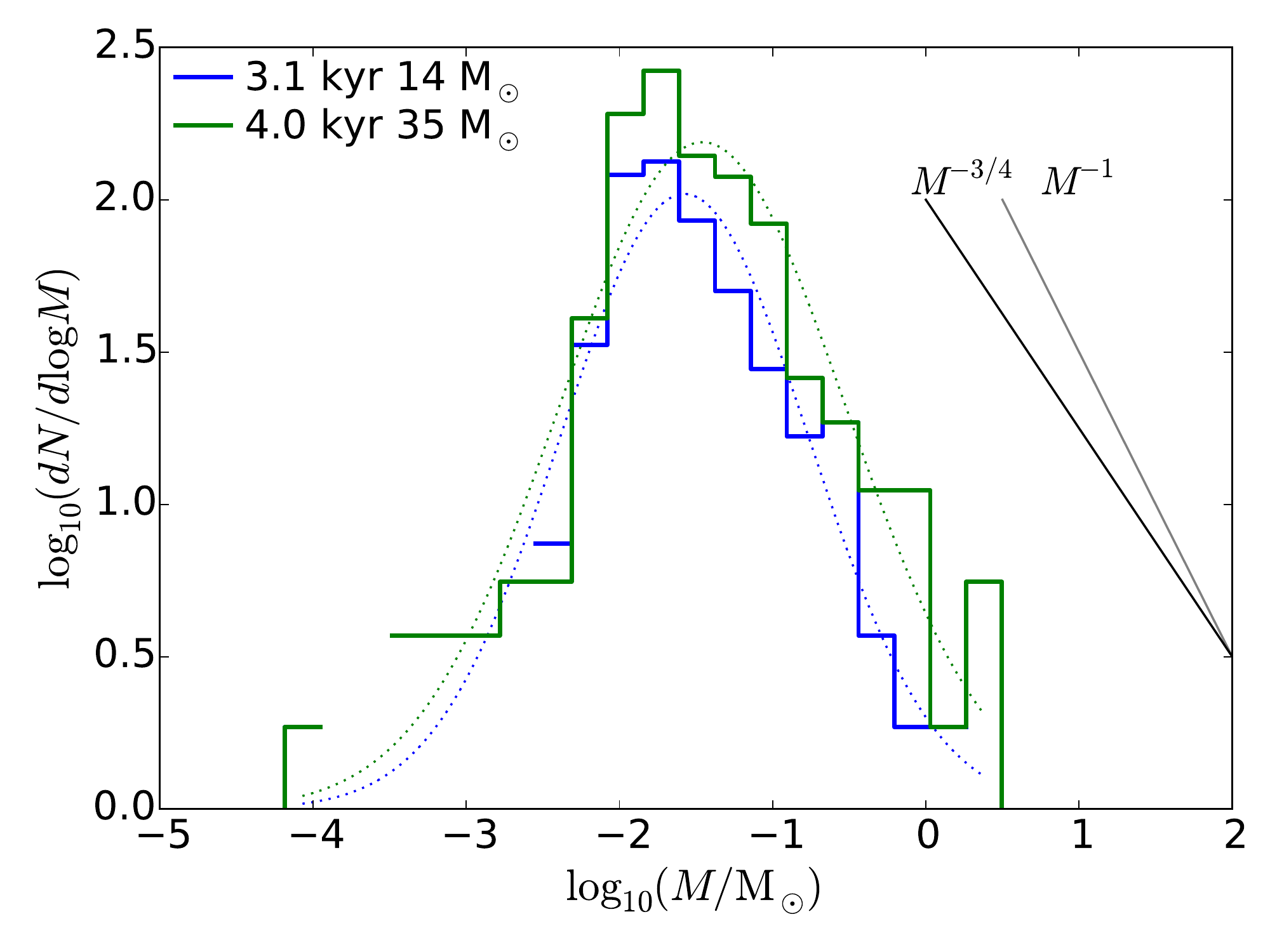}}  
\put(6.8,12.3){I--, 8 AU}
\put(0,0){\includegraphics[width=8.7cm]{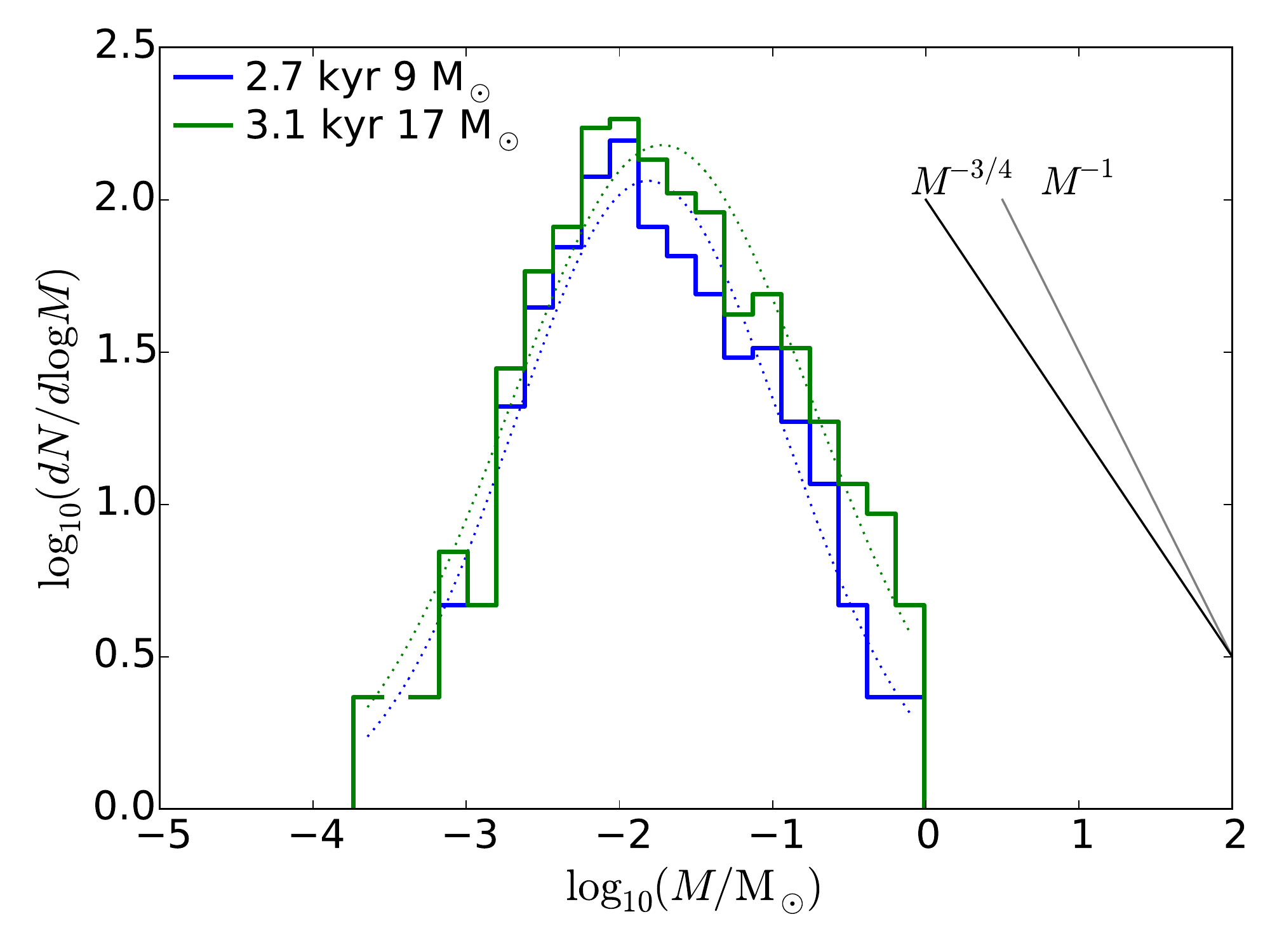}}  
\put(7.0,5.8){I, 4 AU}
\end{picture}
\caption{Stellar distribution of isothermal runs with varied resolutions of 17, 8, 4 AU in colored histograms, with time and total accreted mass shown in the legend. Lognormal fits are presented with dashed thin curves. Powerlaws of $M^{-3/4}$ and $M^{-1}$ are shown with black and gray lines, respectively, for comparison. The characteristic mass of the stellar distribution shifts to smaller values with increasing resolution.}
\label{fig_iso}
\end{figure}
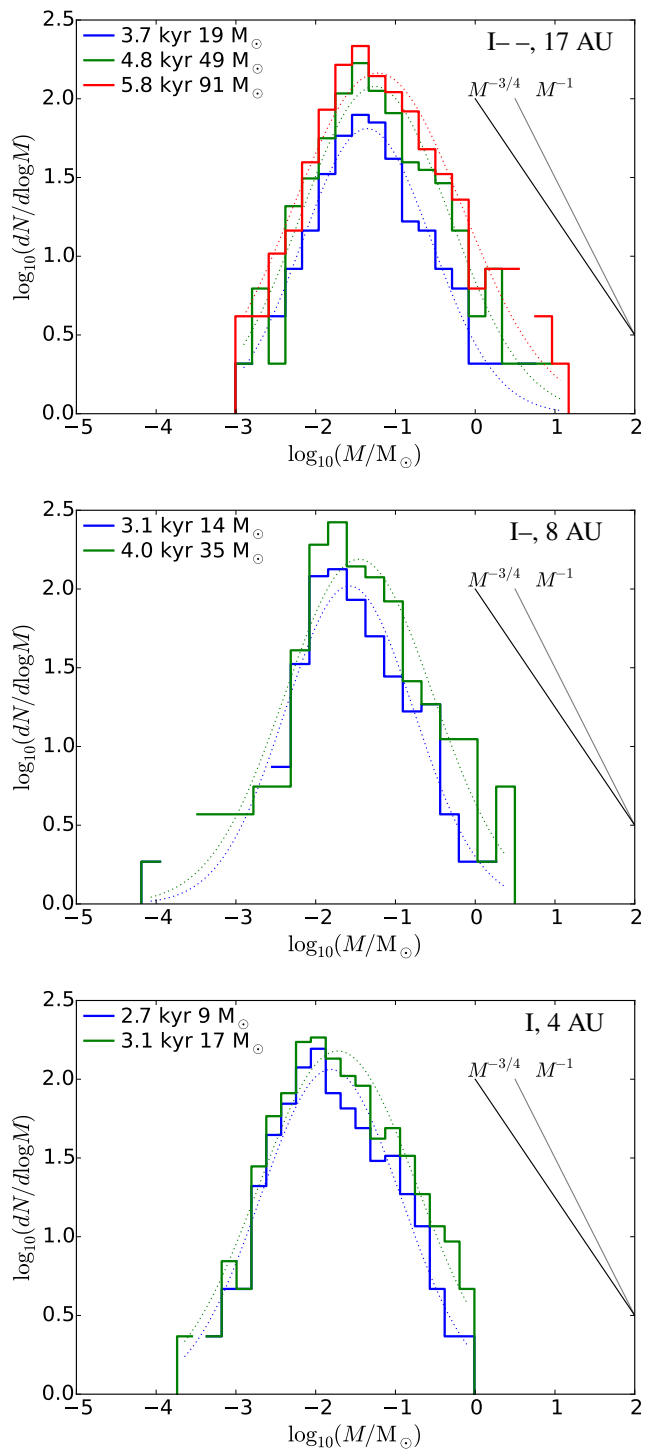

\section{Numerical setup and initial conditions}

\subsection{Numerical setup and runs performed}

\begin{table*}[t!]
\caption{Simulation parameters: the label, the density $n _\ad$ at which the eos switches from isothermal to 
adiabatic as defined in Eq.~(\ref{eq_eos}), the minimum density $n _\mathrm{sink}$ at which the sink is introduced, 
the polytropic index $\gamma$, the maximal refinement level $l_\mathrm{max}$, and the physical resolution. 
The temperature remains 10 K at density below $n _\ad$. When full eos is used, the second and third adiabatic index switches occur 
at $30~n _\ad$ and $30000~n _\ad$, corresponding roughly to 100 K and 1000 K, respectively. In the labels, `C/I' 
stands for adiabatic/isothermal eos, the number is the value of $n _\ad$, `h/s/f 'stands for hard/soft/full eos, 
`+/--' is for higher/lower resolution, `d' stands for higher sink density threshold, `*' represents a stricter clump 
merging scheme, and `o' stands for density threshold criteria only for sinks.}
\label{table_eos}
\centering
\begin{tabular}{l c c c c c c c}
\hline\hline
Label   & $n _\ad (10^9 \cc)$ & $n _\mathrm{sink} (10^9 \cc)$ &  $\gamma$  &  $l_\mathrm{max}$    & resolution(AU) \vspace{.5mm}\\
\hline
C10h   & $10$  & $10$ &  5/3  & 14 & 4   \\
C10h+   & $10$  & $10$ &  5/3  & 15 & 2   \\
C10h--   & $10$  & $10$ &  5/3  & 13  & 8  \\
C10h-- --   & $10$  & $10$ &  5/3  & 12 & 17   \\

C3h--     & $3$ & $1$ &  5/3  & 13  & 8    \\
C1h     & $1$ & $1$ &  5/3  & 14 & 4    \\
C1h--    & $1$ & $1$ &  5/3  & 13  & 8    \\
C1h--*    & $1$ & $1$ &  5/3  & 13  & 8  \\
C1h-- --     & $1$ & $1$ &  5/3  & 12 & 17     \\
C1hd--o     & $1$ & $100$ &  5/3  & 13   &8  \\
\hline
C10s--     & $10$  & $1$ &  4/3  & 13  & 8    \\
C1s--     & $1$  & $1$ &  4/3  & 13  & 8    \\
\hline
C10f     & $10$ & $10$ &  5/3, 7/5, 1.1 & 14 & 4    \\
C1f     & $1$  & $10$ &  5/3, 7/5, 1.1   & 14 & 4    \\
C10fd     & $10$ & $300$ &  5/3, 7/5, 1.1 & 14 & 4    \\
\hline
I     & $10^{4}$ & $10$ &  1  & 14 & 4    \\
I--     & $10^{4}$ & $1$ &  1  & 13  & 8    \\
I-- --     & $10^{4}$ & $1$ &  1  & 12 & 17     \\
\hline
\end{tabular}
\end{table*}

We use the adaptive mesh refinement (AMR) magnetohydrodynamics code RAMSES \citep{Teyssier02,Fromang06} to evolve the 
hydrodynamical equations. 
Simulations are initialized with Bonner-Ebert-like spherical molecular cloud with density profile 
$\rho(r) = \rho_0/\left[1+\left(r/r_0\right)^2\right]$, 
where $r_0$ is the size of central plateau and $r$ the distance to the cloud center, 
as described in paper I (see more details therein). 

The run C1 in paper I is used as the canonical run here and is labeled C10h as will be presented later. 
The cloud has $1000~\Ms$ and $0.084~\pc$ radius, 
with the simulation box twice the size of the cloud. 
The central density $n_0=\rho_0/ (\mu m_{\rm p}) \simeq 6 \times 10^7~\cc$, where $\mu=2.33$ is the mean molecular weight and $m_{\rm p}$ is the atomic hydrogen mass. 
Note that such density is very high and may not be realistic to describe most 
Milky-Way clouds. However, our goal here is to investigate
the physical processes rather than performing a detailed comparison with observations. 
A turbulent velocity field following Kolmogorov spectrum with random phases is seeded, 
and a relaxation phase without self-gravity is run during $\sim 30\%$ of the turbulent crossing time on the base grid ($2^8$) to prepare for coherent density and velocity fluctuations. Initially, the turbulence is a mix of compressive and solenoidal modes
with respective energy ratio 1:2. 
The AMR scheme requires that the local Jeans length is always resolved by ten cells. 
The canonical run has maximal refinement of 14 levels, corresponding to $4~\AU$ resolution, while higher and lower resolution 
runs are also performed.
The simulation parameters are listed in Table~\ref{table_eos}.

\setlength{\unitlength}{1cm}
\begin{figure*}
\begin{picture} (0,12.8)
\put(0,6.5){\includegraphics[width=8.7cm]{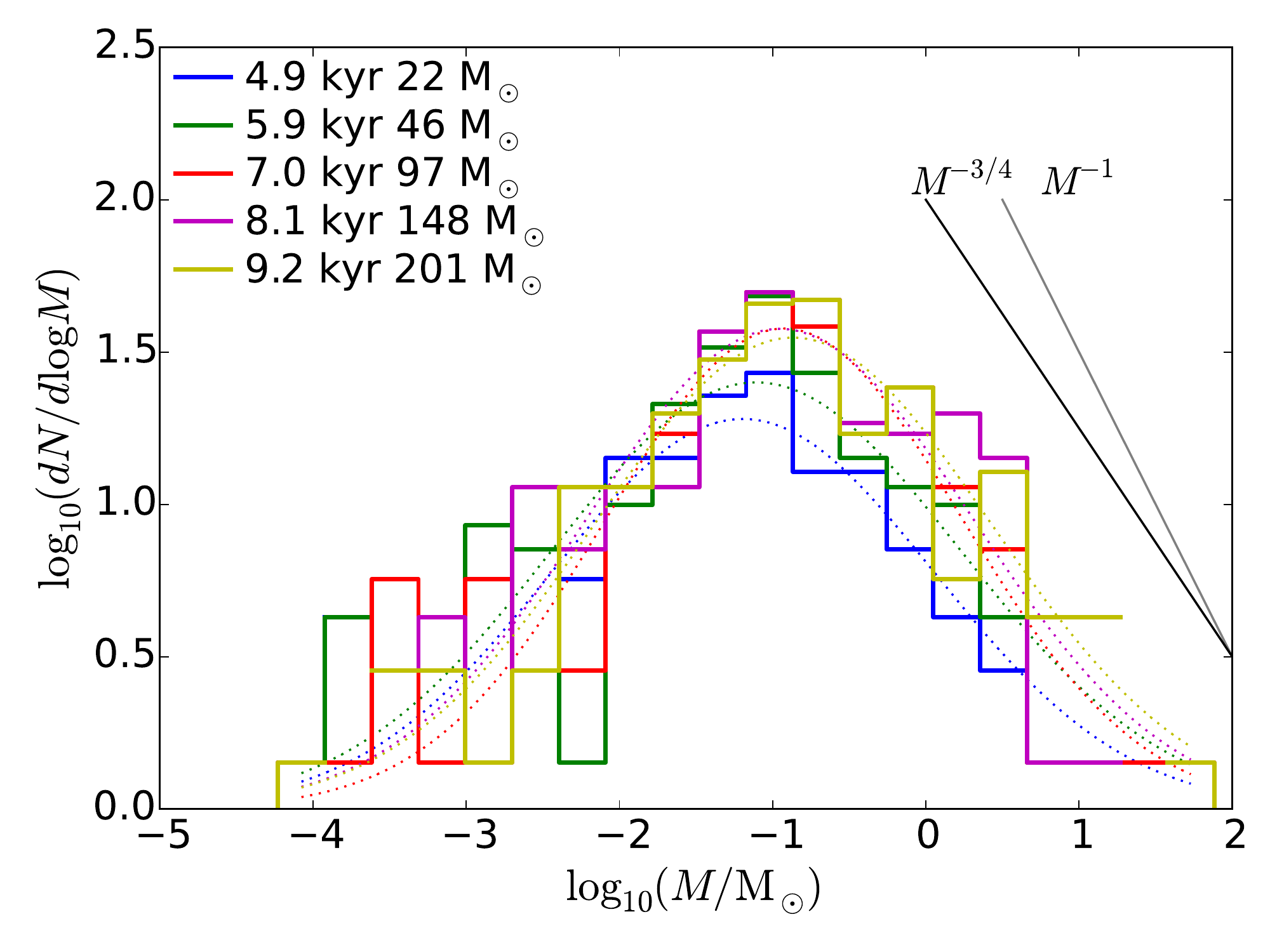}}  
\put(6,12.3){C10h-- --, 17 AU}
\put(9,6.5){\includegraphics[width=8.7cm]{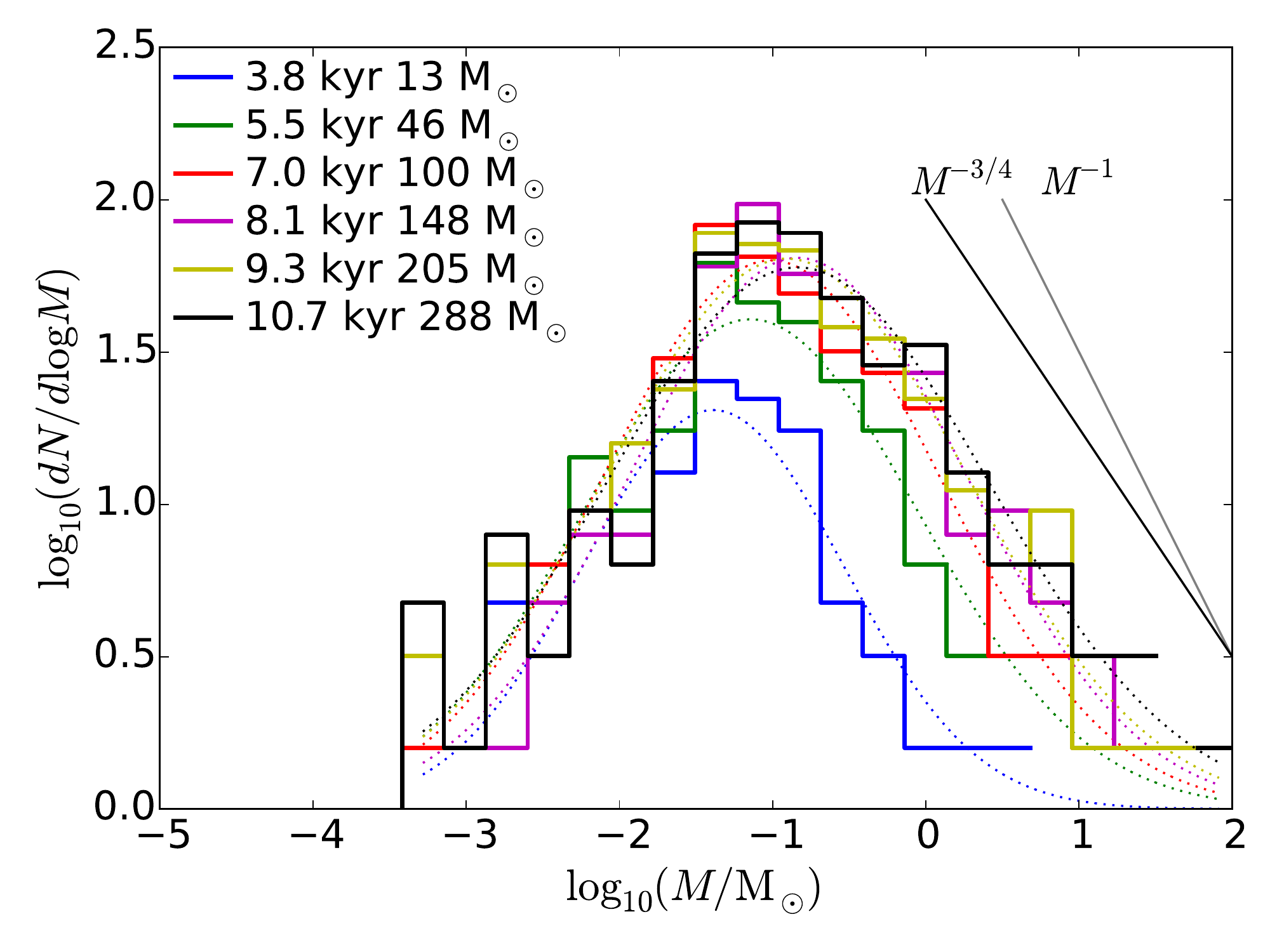}}  
\put(15.4,12.3){C10h--, 8 AU}
\put(0,0){\includegraphics[width=8.7cm]{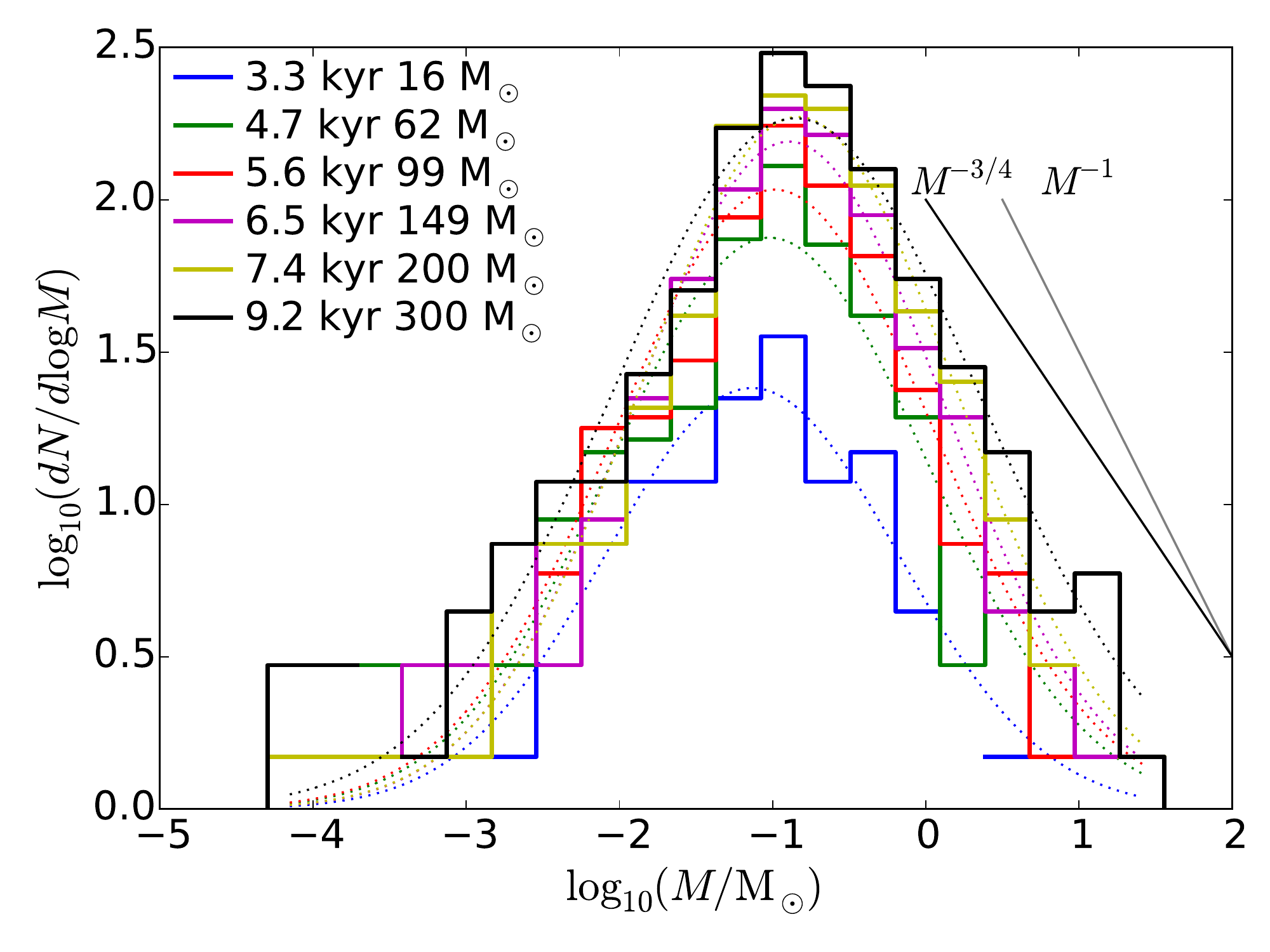}}  
\put(6.5,5.8){C10h, 4 AU}
\put(9,0){\includegraphics[width=8.7cm]{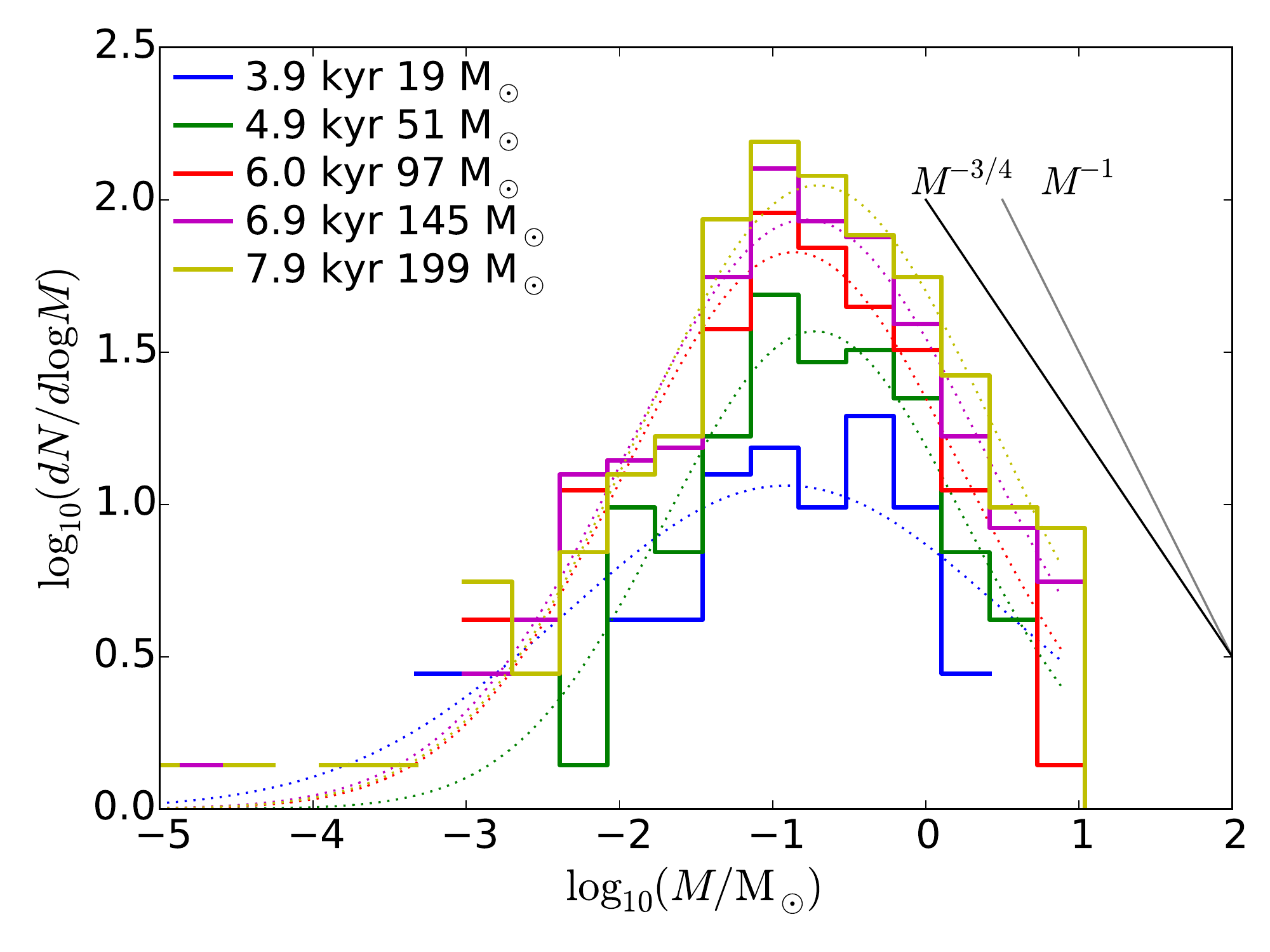}}  
\put(15.3,5.8){C10h+, 2 AU}
\end{picture}
\caption{Stellar distribution of runs with smoothed two-slope polytropic eos with $\gamma = 5/3$ and $n_\ad = 10^{10}~\cc$, with varying resolutions of 
17, 8, 4, 2 AU. The stellar distribution characteristic mass $\simeq 0.1~\Ms$ irrespective of numerical resolution. }
\label{fig_g53_10}
\end{figure*}
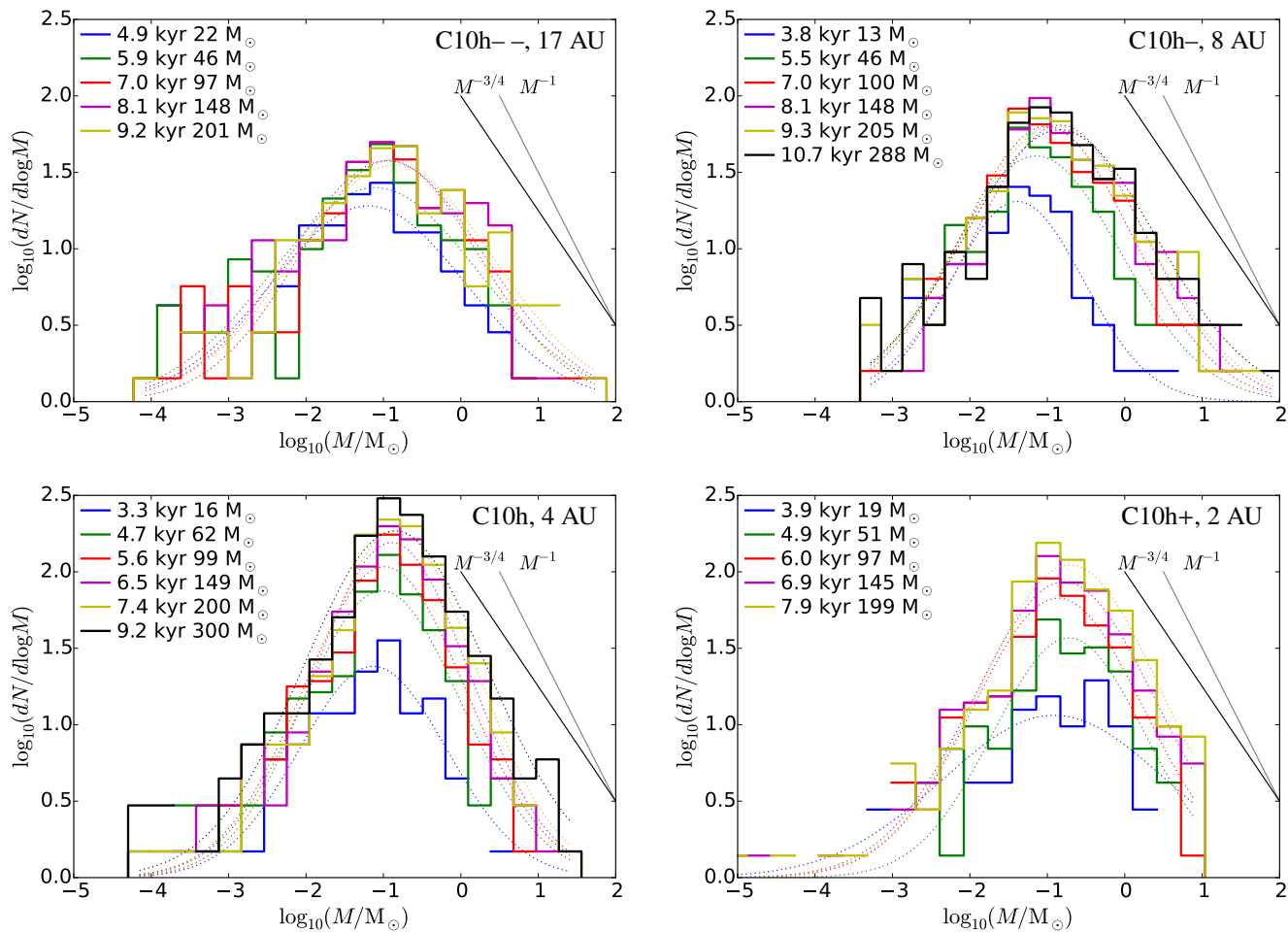

\subsection{Equation of state}
In  a dense core, the temperature is about 10 K except at high density 
where the gas is  optically opaque to its own radiation and is nearly adiabatic
\citep[e.g.][]{Masunaga98,vaytet2012,Vaytet17}. 
The temperature $T$, that depends on the gas density $\rho$, is described with the eos:
\begin{align}
T = T_0 \left[1 + (\rho / \rho_\ad)^{(\gamma-1)}\right], \label{eq_eos}
\end{align}
where the isothermal temperature $T_0$ at low density is set to 10 K. 
We study the effect of the eos by varying the critical turnover density $\rho_\ad$, 
that designates the change between isothermal behavior and adiabatic heating, 
and the polytropic index $\gamma$, that describes the heating rate. 
The polytropic index of molecular hydrogen is $\gamma_1=$5/3 and $\gamma_2=$7/5 before 
and after the excitation of rotation modes of hydrogen molecules at around 100 K. 
At $\sim 1000$ K, the H$_2$ molecules start to dissociate and $\gamma$ drops to $\gamma_3 \sim 1.1$.  
In this study, we aim to examine the impact of thermodynamics on the final mass that a star could reach. 
We also consider, therefore, a more complete description of full eos
\begin{align}
T = T_0 \left\{1 \!+\! { (\rho / \rho_{\ad})^{(\gamma_1-1)} \over \left[1+(\rho / \rho_{\ad,2})^{(\gamma_1-\gamma_2)n}\left(1+(\rho / \rho_{\ad,3})^{(\gamma_2-\gamma_3)n}\right)\right]^{1/n}}\right\}, 
\label{eq_full_eos}
\end{align}
where $n = 3$ is a smoothing parameter. 
This barotropic eos is employed for convenience. 
It signifies effectively $T\propto \rho^{\gamma-1}$ with $\gamma=1$ for $\rho<\rho_{\ad}$, $\gamma=\gamma_1$ for $\rho_{\ad}<\rho<\rho_{\ad,2}$, $\gamma=\gamma_2$ for $\rho_{\ad,2}<\rho<\rho_{\ad,3}$, and $\gamma=\gamma_3$ for $\rho>\rho_{\ad,3}$. 
In the following studies, we fix $\rho_{\ad,2} = 30 ~\rho_{\ad}$ and $\rho_{\ad,3} = 1000 ~\rho_{\ad,2}$.

As shown in Table~\ref{table_eos}, 
$n _\ad = \rho_\ad / (\mu m_p)$ is varied between $10^9~\cc$ and $10^{10}~\cc$, 
while using a hard eos $\gamma = 5/3$ or a soft one $\gamma=4/3$. 
We also study the effects of a full eos that describes the transition of $\gamma$ from 1 to 5/3, 7/5, and then 1.1. 
The two latter transitions correspond roughly to $T=100$ and 1000 K. 
We also set $n _\ad = 10^{13}~\cc$ in run I, 
that is practically isothermal since such high density is hardly reached at this resolution. 

\subsection{Sink particle algorithm}
Sink particle algorithm from \citet{Bleuler14} is used \citep[see also][]{krumholz04,federrath2010}, 
and we briefly describe it here. 
Sink particles are formed at the highest level of refinement. 
The algorithm first identifies local density concentrations, that are referred to as clumps, above a given density threshold (typically we use 1/10 of the density threshold for sink particles $n_\mathrm{sink}$). 
Connected clumps are merged if the density contrast between the peak and the saddle point is less then two. 
The final clumps are then passed on for sink formation check. 

The clump peak undergoes several criteria before a sink formation site is flagged. 
Firstly, a density threshold is imposed such that the peak must have density larger then $n_\mathrm{sink}$, 
that we vary between $10^{10}$ and $3 \times 10^{11}$ to investigate its influence. 
Secondly, extra criteria check whether the clump is virially bound and has converging flow. 
A sink particle is placed at the density peak when all criteria are satisfied. 
Simple density thresholding is used in runs C1hd--o to study the numerical effects of the algorithm. 
The sink then interacts gravitationally with the gas component as well as other sinks, 
and accretes from the surrounding with a threshold scheme. 
If the cells surrounding the sink particle exceeds $n_\mathrm{sink}$, 
75 percent (this numerical factor seems not to have a crucial impact on our results, see Appendix \ref{appen_acc}) of the excessive gas mass in the cells is accreted onto the sink particle. 

\subsection{Missing physics}
The physics is deliberately simplified because we are trying to conduct 
a systematic set of simulations to clarify the influence of the initial conditions, eos,
and resolution. Other processes not included here are nonetheless playing 
significant, possibly dominant roles. This is particularly the case for the accretion luminosity 
that emanates from the protostars \citep{krumholz07,Bate09b,commercon2011} and
heats the gas, in turn increasing the thermal support. Magnetic field is another  
important process \citep{hennebelle2011,peters2011,myers2013} which likely affects the fragmentation of clusters
through magnetic braking and magnetic support that we do not consider here.

\section{Results: mass spectra in numerical simulations}

\setlength{\unitlength}{1cm}
\begin{figure}
\begin{picture} (0,19.3)
\put(0,13){\includegraphics[width=8.7cm]{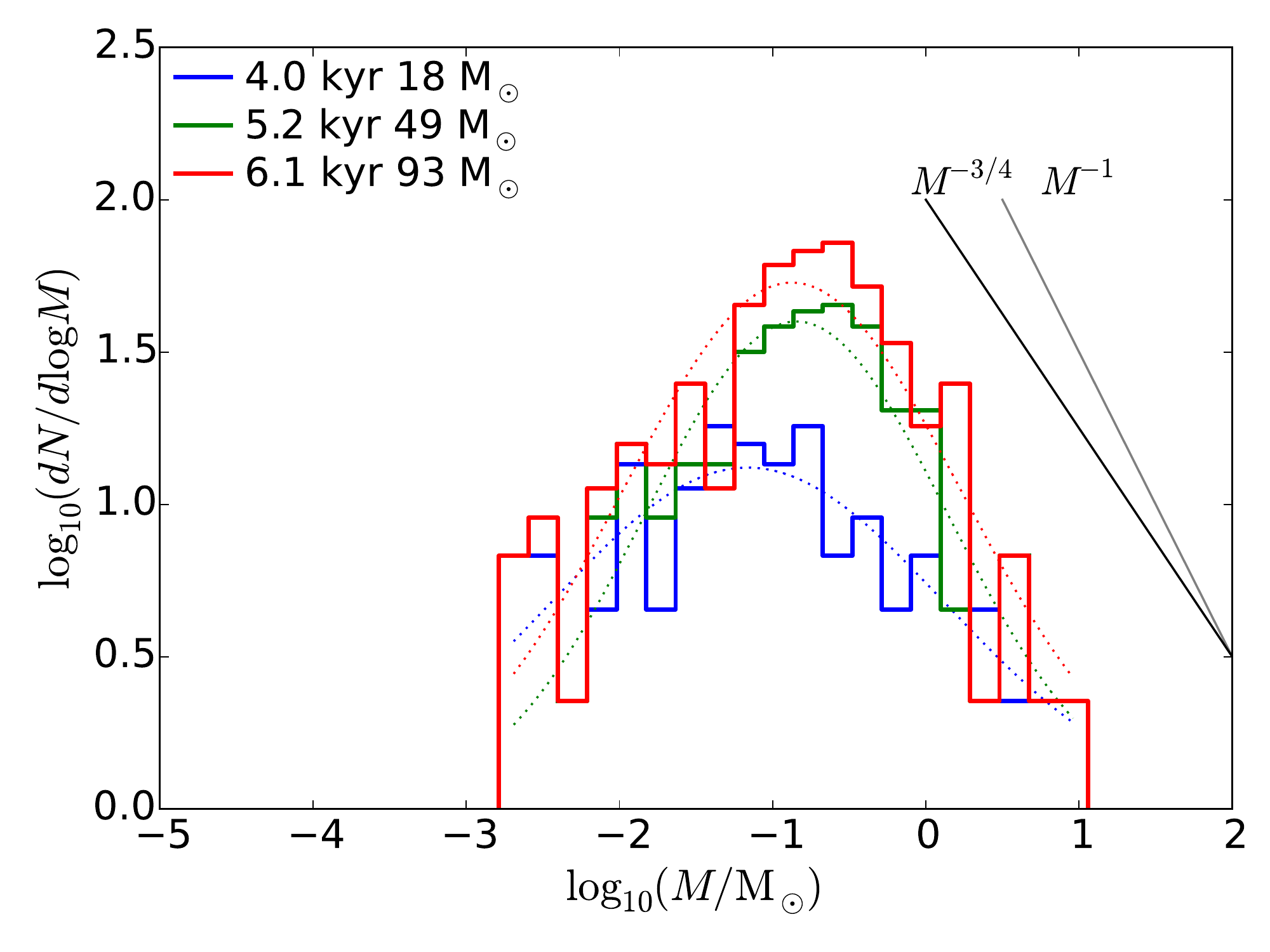}}  
\put(6.5,18.8){C3h--, 8 AU}
\put(0,6.5){\includegraphics[width=8.7cm]{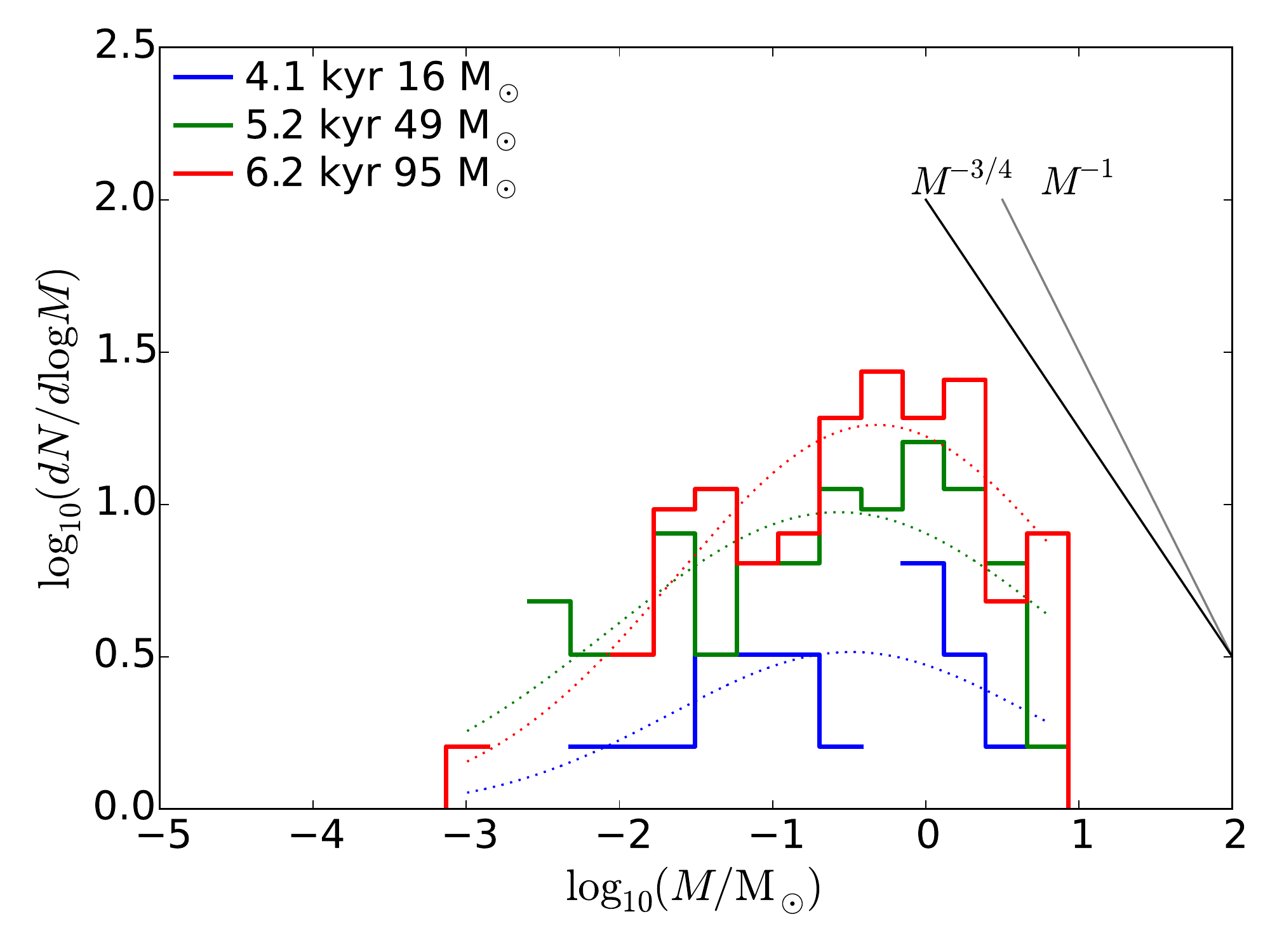}}  
\put(6.5,12.3){C1h--, 8 AU}
\put(0,0){\includegraphics[width=8.7cm]{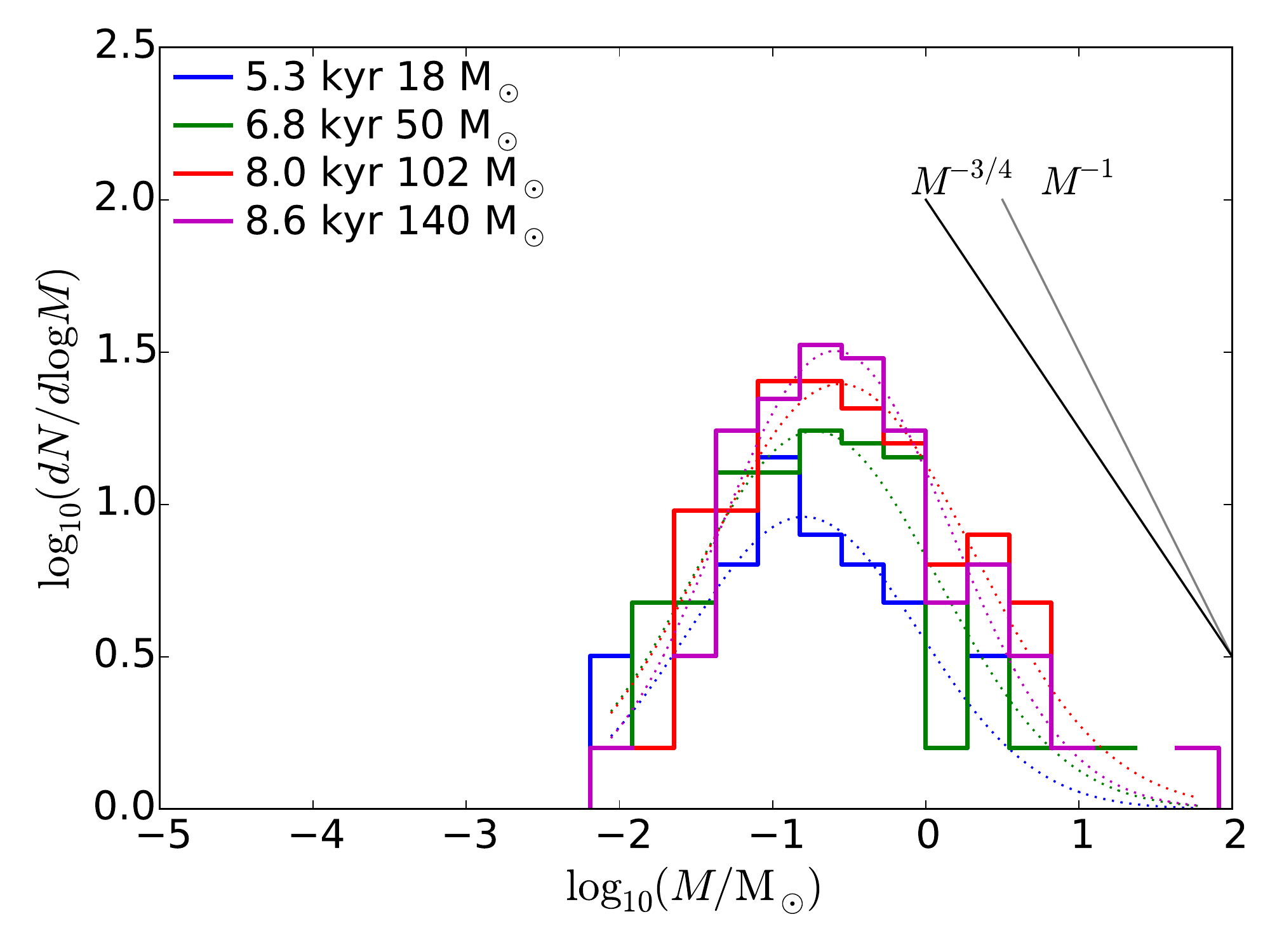}}  
\put(6.2,5.8){C1h-- --, 17 AU}
\end{picture}
\caption{Stellar distributions of runs with smoothed two-slope polytropic eos with $\gamma = 5/3$, lowered $n_\ad$ at $3 \times 10^9$ (C3h--) and $10^9~\cc$ (C1h--, C1h-- --). 
The stellar distribution characteristic mass increases with decreasing $n_\ad$. 
At $n_\ad=10^9~\cc$, the stellar distribution becomes top-heavy with improved resolution from 17 to 8 AU.}
\label{fig_g53}
\end{figure}
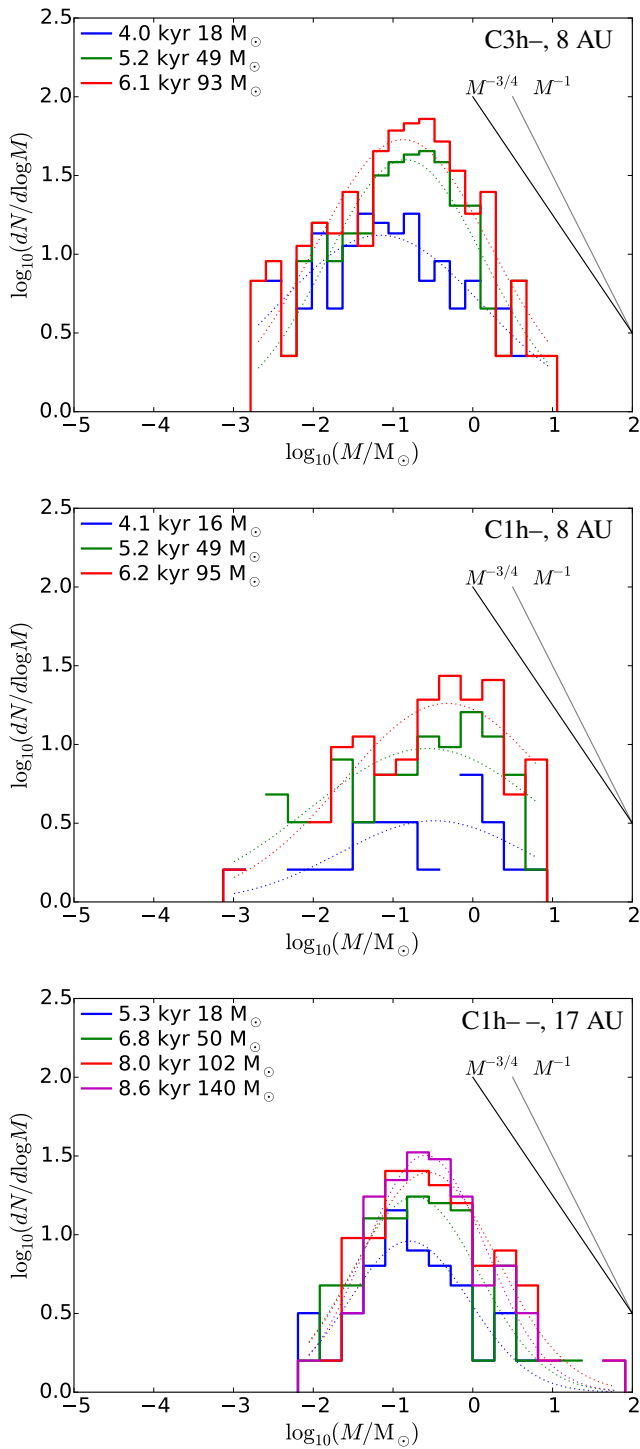

\setlength{\unitlength}{1cm}
\begin{figure}
\begin{picture} (0,12.8)
\put(0,6.5){\includegraphics[width=8.7cm]{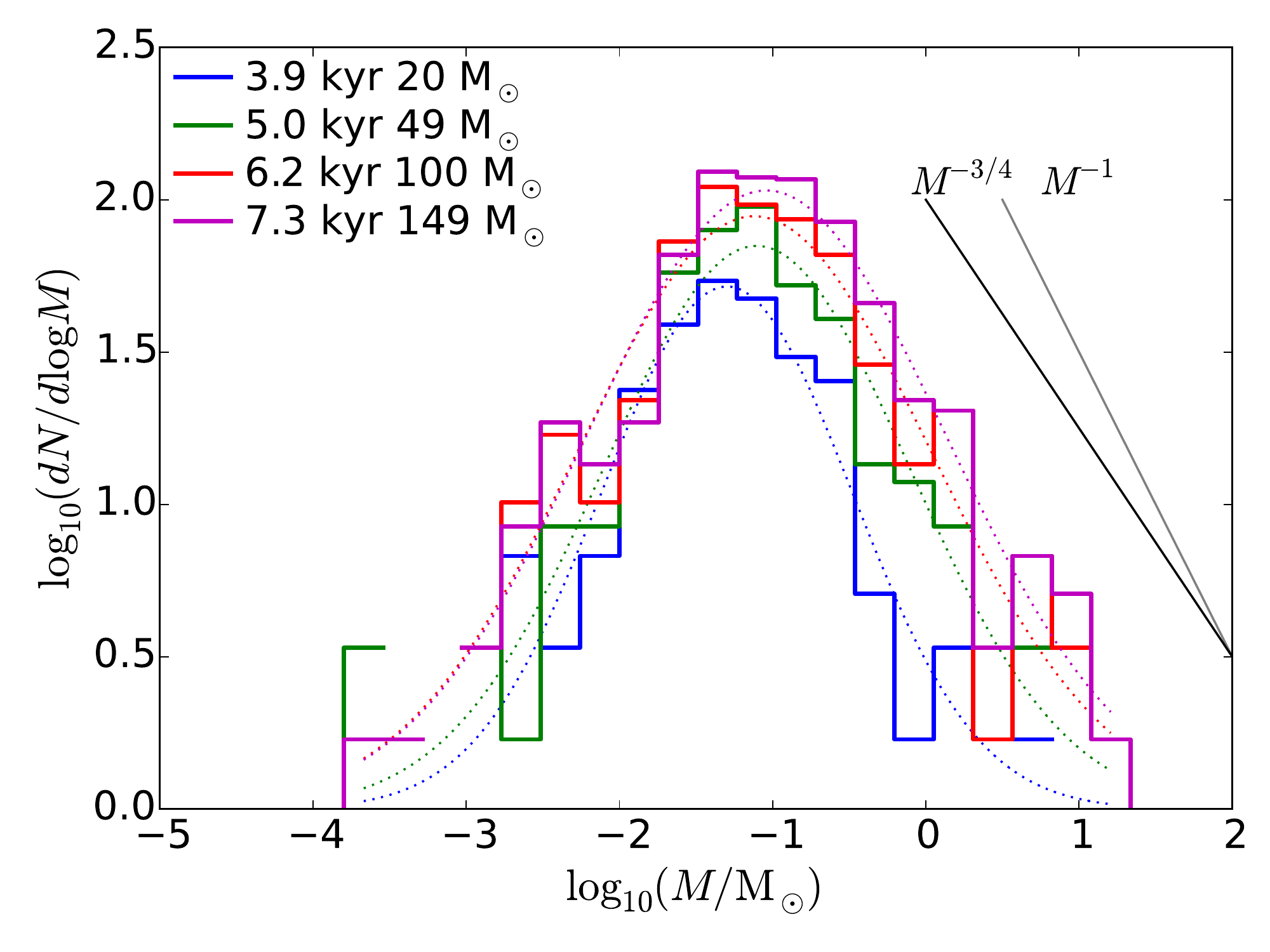}}  
\put(6.4,12.3){C10s--, 8 AU}
\put(0,0){\includegraphics[width=8.7cm]{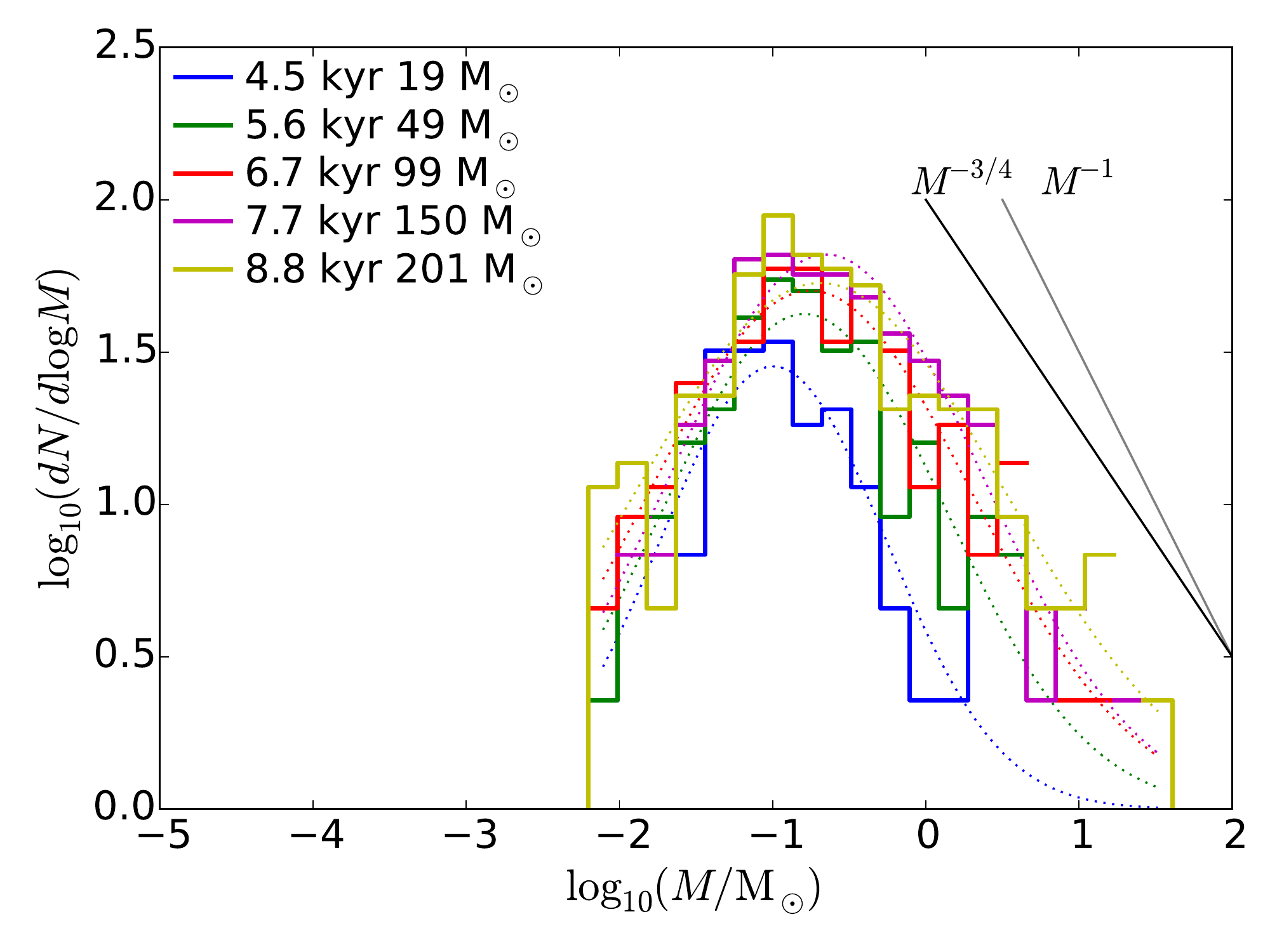}}  
\put(6.5,5.8){C1s--, 8 AU}
\end{picture}
\caption{Stellar distributions of runs with smoothed two-slope polytropic eos with $\gamma = 4/3$, $n_\ad = 10^{10}~\cc$ (C10s--) and $10^9~\cc$ (C1s--). 
The stellar distribution characteristic mass increases by a factor $\sim 3$ when $n_\ad$ is lowered by a factor ten.}
\label{fig_g43}
\end{figure}

\setlength{\unitlength}{1cm}
\begin{figure}
\begin{picture} (0,19.3)
\put(0,13){\includegraphics[width=8.7cm]{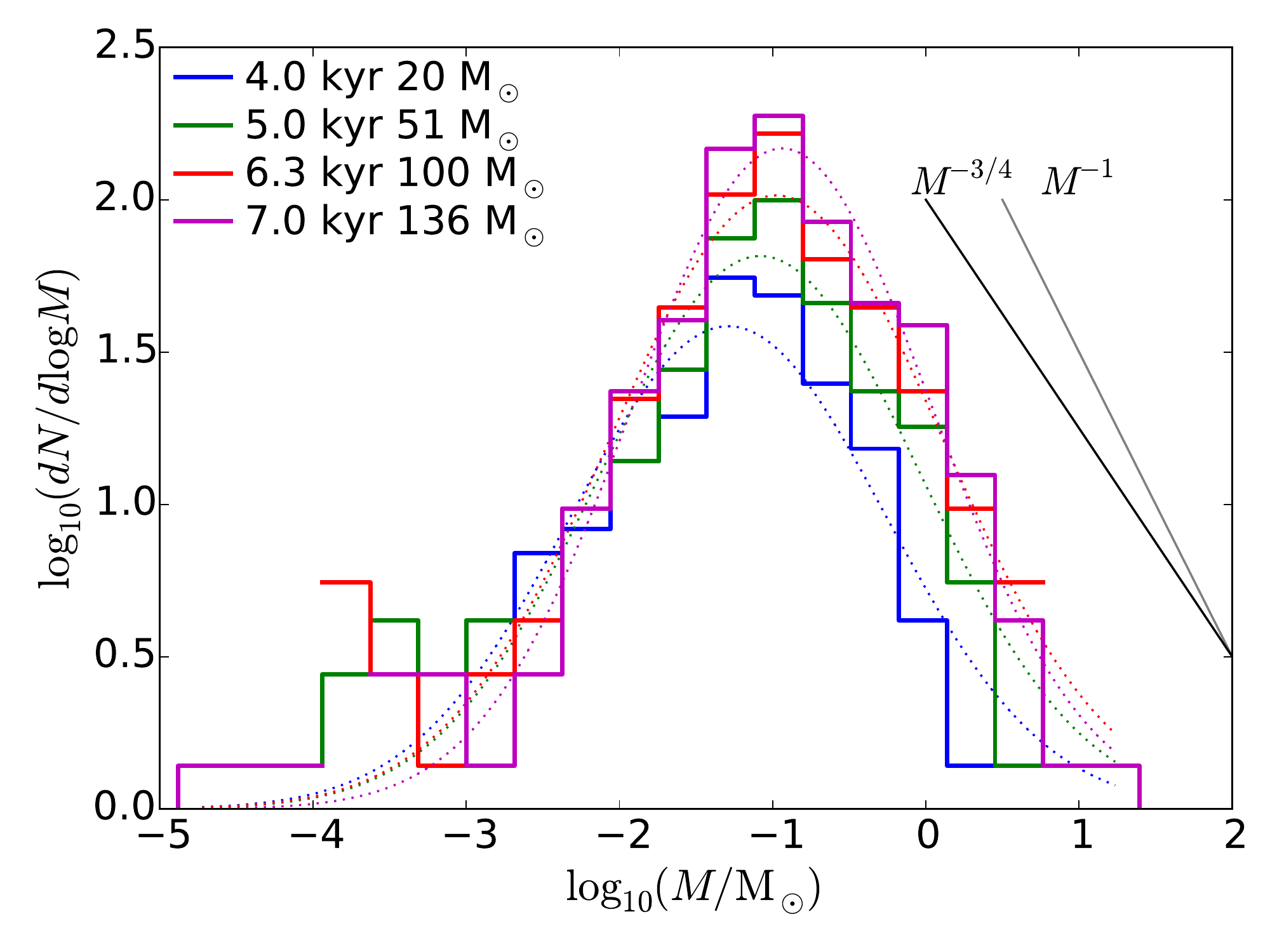}}  
\put(6.5,18.8){C10f, 4 AU}
\put(0,6.5){\includegraphics[width=8.7cm]{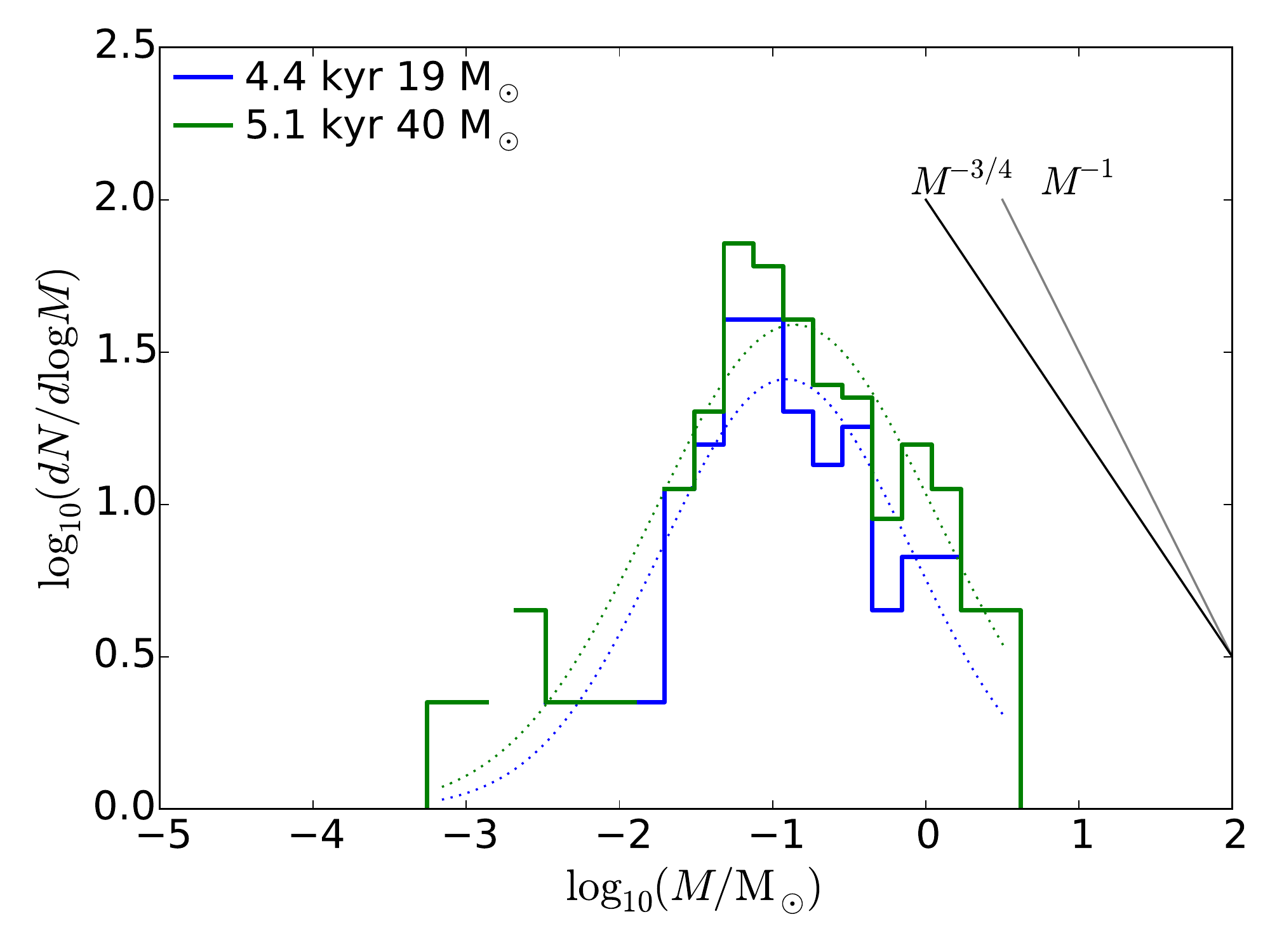}}  
\put(6.5,12.3){C10fd, 4 AU}
\put(0,0){\includegraphics[width=8.7cm]{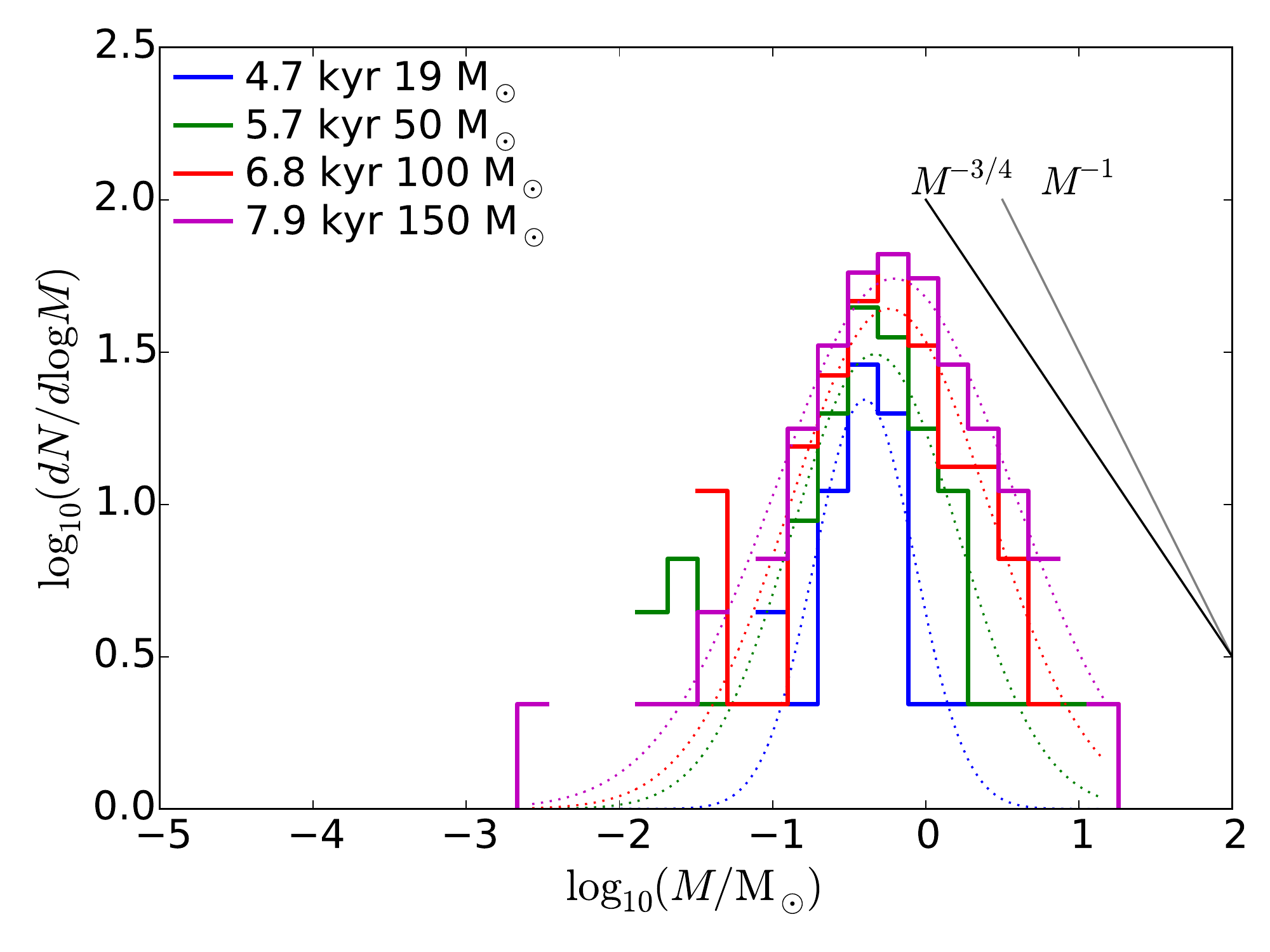}}  
\put(6.6,5.8){C1f, 4 AU}
\end{picture}
\caption{Stellar distributions of runs with full eos with $n_\ad = 10^{10}~\cc$ (C10f, C10fd) and $10^9~\cc$ (C1f). The transitions 
to $\gamma = 5/3, 7/5,$ and $1.1$ happen at $n_\ad$, $30~n_\ad$, and $30000~n_\ad$. 
Using higher sink density threshold (C10fd) results in fewer low mass stars, while not changing the peak of the stellar distribution. Lowering $n_\ad$ by a factor ten increases the peak mass by $\sim 10$.}
\label{fig_gfull}
\end{figure}

\subsection{Isothermal runs}
The sink distributions, displayed in Fig.~\ref{fig_iso}, present a well-defined peak and a powerlaw at large masses that is roughly $\propto M^{-1}$.
The isothermal runs show that the stellar distribution peak moves from $0.05$ to $0.01~\Ms$ when changing the resolution from 17 to 4 AU. 
As the collapse of isothermal gas is self-similar, 
there is no characteristic scale or mass, at high density,  in such simulations. 
Roughly speaking, when doubling the resolution, the highest density that can be described increases
by a factor of about $2^3=8$. One may, therefore, expect that the peak position shifts by $\simeq \sqrt{64}=8$
when two further refinement levels are introduced. This is not incompatible with the trend observed in Fig.~\ref{fig_iso}.

The mass of sink particles, therefore, depends on  the numerical resolution. 
This is likely what happens in simulations which recovered the stellar distribution in a collapsing 
massive and isothermal clouds  \citep[e.g.][]{Bonnell06, Girichidis11,BallesterosParedes15}. 
It is worth emphasizing that this is not a statement about the absence of convergence
of isothermal calculations in general. As shown in \citet{Girichidis11} and in paper I, initial conditions matter 
a lot. 
Typically, the dependence of the peak position is expected once the collapsing cloud is sufficiently cold. 
For example, \citet{gong2015} seem to get numerical convergence in their colliding flow calculations. 

\subsection{Runs with $\gamma=5/3$}
With $\gamma=5/3$, the fiducial run (C10h) in Fig.~\ref{fig_g53_10} exhibits very quickly a well-defined stellar distribution that peaks at $\simeq 0.1~ \Ms$. 
The runs with four resolutions, 2, 4, 8, and 17 AU, show that the  peak position is robust and 
does not change significantly with the resolution. 
This is a noticeable difference from the isothermal case. There is nevertheless some significant evolution 
of the mass spectra when the spatial resolution improves. First of all, the number of objects near the peak increases 
from the resolution improves from 17 to 4 AU (going from $\log_{10}(dN/d \log M) \simeq 1.6$ to $\simeq 2.1$ at the time when 200 $\Ms$ have been accreted), 
while this number does not seem to change significantly between the two highest resolutions (two bottom panels).
As in the isothermal simulations, the high-mass part is also described by a powerlaw $\propto M^{-1}$, 
though this is much clearer in the highest resolution runs. 

Figure~\ref{fig_g53} reveals that $n_\ad$ has an impact on the peak position. 
With $n_\ad = 10^9$, sinks are completely prevented from forming at the resolution of 4 AU (C1h, no mass spectra available). 
Decreasing the resolution allows sink particles to form. 
The run at 8 AU resolution (C1h--) produces top-heavy spectra with a loosely-defined peak at $\simeq 1~\Ms$, 
while a coarser resolution at 17 AU (C1h-- --) shifts the peak to $\simeq 0.3~\Ms$ and more sinks are formed. 
This effect results from the excessive heating at high densities where the approximated expression of one single $\gamma$ value stops being physical, 
and the high temperature gas is thus always thermally supported against collapse at the finest resolution scale and sink particles hardly form. 
This behavior will receive more quantitative interpretation in Sect.~\ref{first_larson}.
The run with $n_\ad = 3 \times 10^9$ at 8 AU resolution (C3h--) produces mass spectra that peak at smaller mass than that of run C1h--, at same resolution.

These results suggest that the eos has a strong impact on the stellar distribution characteristic mass. 
Moreover we find that the stellar distribution peak is significantly higher than the Jeans mass at which the gas 
becomes opaque to its radiation (typically on the order of $10^{-3}~\Ms$) or even the mass of the  
first Larson core (a few $10^{-2}~\Ms$). The link between the two will be addressed in 
Sect.~\ref{first_larson} and Sect.~\ref{st_point}.

\subsection{Runs with $\gamma=4/3$}
As revealed by Fig.~\ref{fig_g43} for runs using $\gamma = 4/3$, 
the peak shifts to slightly smaller masses and the stellar distribution is slightly wider than with $\gamma=5/3$. 
The run with $n_\ad = 10^9~\cc$ (C1s--) yields slightly higher peak mass than that with $n_\ad = 10^{10}~\cc$  
(C10s--) since the heating starts at lower density and thus the higher temperature at same density provides more support against gravity. 
With $\gamma = 4/3$, the gas heats up less quickly as density increases, compared to $\gamma = 5/3$, 
and the high-density region suffers less from over-heating (more sinks form in C1s-- than C1h--).

\subsection{Runs with a full eos}
The first panel of Fig.~\ref{fig_gfull} shows very similar behavior 
to the fiducial run (C10h), seemingly suggesting that it is the $\gamma=5/3$ part 
of the eos that matters the most.
To test the influence of the sink particle scheme, we have performed a run for which 
 $n_\mathrm{sink} = 3 \times 10^{11}~\cc$ (C10fd, second panel).  
Applying higher threshold for sinks does not affect the peak mass, 
while the low-mass end becomes less populated. This shows that the exact moment where the sink 
particles are being introduced does not control the peak position but has some influence on the 
distribution of small masses. This is compatible with the eos being the most important process 
that determines the peak position. 

Finally, we have also considered $n_\ad = 10^9~\cc$ using the full eos.
The shift of the peak is as obvious with a full eos (C10f and C1f) as that with $\gamma=5/3$. 
Note that in run C1f we overcome the over-heating problem at high resolution encountered in run C1h. 
Gas is allowed to collapse and form sink particles without difficulty. This confirms 
that the determining part of the eos is the one corresponding to $\gamma=5/3$.

\section{Mass of the first Larson core and peak of the sink distribution}
\label{first_larson}
As the previous section suggests that the thermodynamics is essential in determining the peak of the stellar distribution, 
we calculate the mass of the first Larson core and study its correlation with the peak mass obtained in the simulations. 

\subsection{The first Larson core}
As gravitational collapse proceeds, density increases and at some point the 
dust becomes opaque to its own radiation. At this point the gas becomes 
essentially adiabatic and a so-called first Larson core forms \citep{Larson69,Masunaga98}. 
While the gravitational energy can overcome the thermal support through contraction of an isothermal gas, 
gas with polytropic index $\gamma > 4/3$ can resist the collapse. 
The first Larson core grows in mass by accretion, 
leading to the increase of the central density and temperature. When the temperature 
reaches about 1500 K, molecular hydrogen starts dissociating, therefore
cooling efficiently the gas and the gravitational collapse restarts, leading 
eventually to the formation of the second Larson core, namely the protostar. 
We believe that to obtain a protostar, it is necessary to accumulate sufficient mass 
to trigger the second collapse and that this process is essential in setting the 
peak of the stellar distribution. First Larson cores with smaller masses 
cannot form protostars (or will do so through cooling but on a longer timescale). 
While this process is not explicitly resolved in our simulations, the introduction of
the sink particles somehow mimics this sequence since they are introduced only 
when enough mass has been accumulated for the gas to be gravitationally unstable. 

By integrating the hydrostatic equations of the gas, 
we deduce the density profile that a core develops before undergoing the secondary collapse.
From the density profile, the mass can be inferred, provided that some radius can be estimated or specified.
The equations of hydrostatic equilibrium are:
\begin{align}
{dM \over dr} &= 4\pi r^2 \rho, \label{eq_dMdr} \\
{1\over \rho}{dP \over dr} &= -{GM \over r^2}, \label{eq_dPdr}
\end{align}
where $M$, $\rho$, $P$, $r$, and $G$ are the mass contained within a given radius, density, pressure, radius, and the gravitational constant. 
The derivative of Eq.~(\ref{eq_dPdr}) with respect to $r$ yields
\begin{align} \label{eq_ddPdr} 
{d \over dr} \left({1\over \rho}{dP \over dr}\right) = {2GM \over r^3} - {G\over r^2}{dM\over dr}
= -{2\over \rho r}{dP \over dr} - 4\pi G \rho, 
\end{align}
where the second equality is obtained by substituting with Eqs. (\ref{eq_dMdr}) and (\ref{eq_dPdr}). 
The pressure is related to the density by the eos as that used in the simulations :
\begin{align}
P   = {k_\mathrm{B} \rho T(\rho)\over \mu m_\mathrm{p}},
\end{align}
where $k_\mathrm{B}$ is the Boltzmann constant and 
the temperature $T(\rho)$ is given by Eq.~(\ref{eq_eos}) or (\ref{eq_full_eos}).

This results in a second order differential equation of $\rho$ that we can solve numerically using standard Runge-Kutta methods:
\begin{align}\label{eq_hydro_T}
& \left[  {dT \over d\rho} \! +\!  {T \over \rho} \right] \! {d^2 \rho \over dr^2} \! =\! 
  \left[{T\over \rho^2} \! -\!  {1 \over \rho} \! {dT\over d\rho} \! -\!  {dT^2  \over d^2\rho} \right] \!   \left({d\rho\over dr}\right)^2 
\!  \! +\!  {2 G M \over r^3}   \! -\!  { 4 \pi G  m_\mathrm{p}\rho \over k_\mathrm{B}}.
\end{align}
Before the second collapse occurs, there is no singularity at the center and thus we use the boundary conditions
\begin{align}
\rho(r=0) = \rho_0 ~~\mbox{and}~~ {d\rho \over dr}(r=0) = 0, 
\end{align}
where $\rho_0$, the central density, remains a free parameter to be specified. 

\begin{figure}[]
\centering
\includegraphics[trim=0 20 0 10,clip,width=0.5\textwidth]{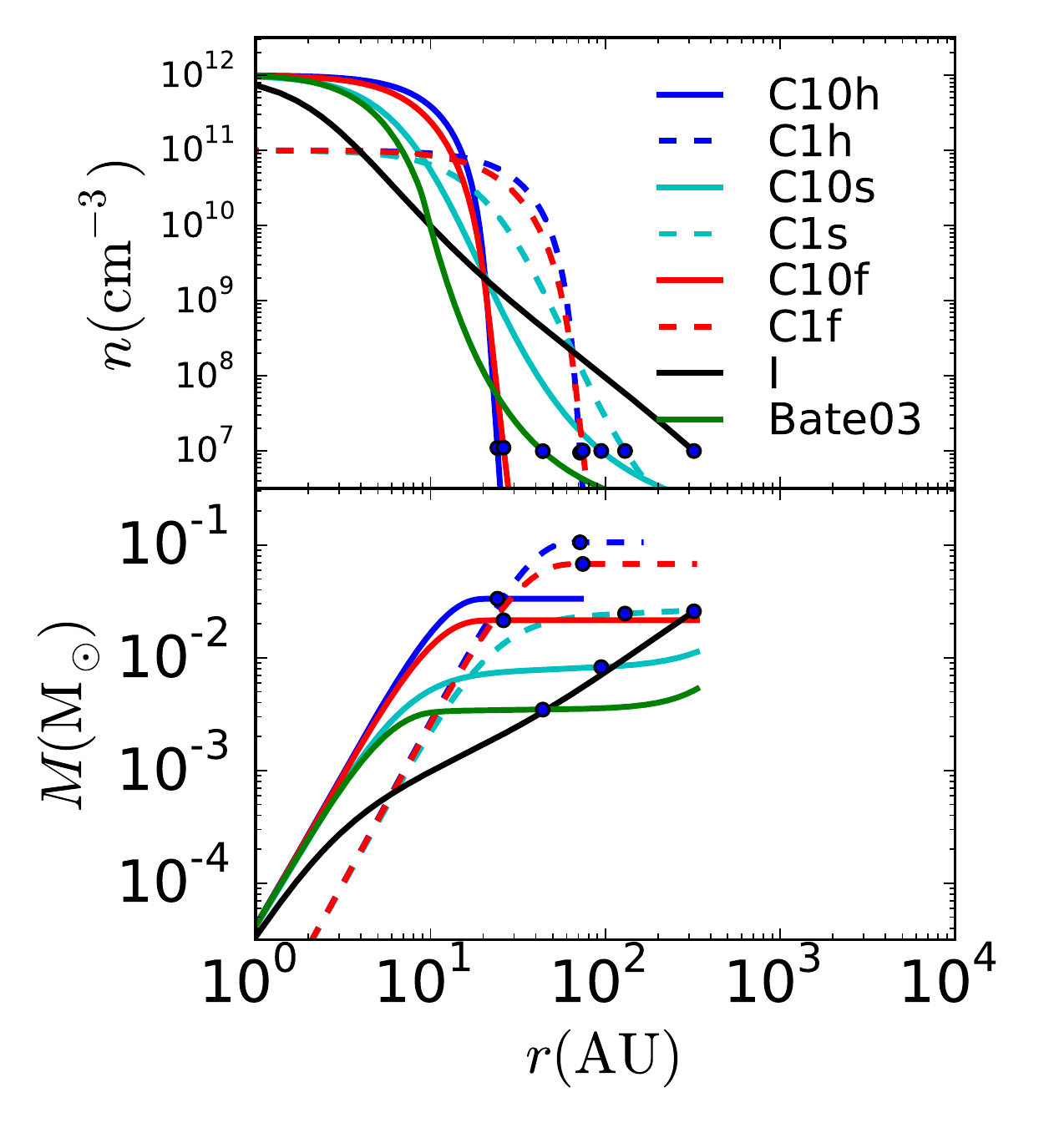}
\caption{First Larson cores for various equations of state. The hydrostatic equilibrium equations are integrated with $n_0 = 100 ~n_\ad$ as boundary condition. We present cases with $\gamma=5/3$ (blue), $\gamma=4/3$ (cyan), full eos (red), and isothermal (black). The legends correspond to the labels of the simulations, with solid lines for $n_\ad=10^{10}~\cc$ and dashed lines for $10^9~\cc$. We also show the case $\gamma=7/5$ with $n_\ad=2.45 \times 10^{10}~\cc$ by \citet[][green]{Bate03}.
The values of $n_0$ are chosen for illustrative purpose, and the actual values reached in simulations depend on the resolution (highest density resolved, see Fig. \ref{fig_rhoc_sim}). The circle indicates the radius at which the density reaches $10^7~\cc$, a reference ambient density that truncates the core. As long as the density profile is decreasing steeply enough, the choice of this truncation density has no strong impact on the resulting mass.}
\label{fig_examples}
\end{figure}

Figure~\ref{fig_examples} shows some examples of the density profile (upper panel) and integrated core mass (lower panel).  
The cases $\gamma=5/3$ (blue), $\gamma=4/3$ (cyan), and full eos (red) are plotted with $n_\ad = 10^{10}~\cc$ (solid) and $10^9~\cc$ (dashed), where the legends correspond to the labels of simulations. 
The black curves display the isothermal case. 
We also show the profile calculated with the eos used by \citet{Bate03} with $\gamma=7/5$ and $n_\ad = 2.45 \times 10^{10}~\cc$. 
The central density used as boundary condition for the integration is chosen to be $n_0 = 100~ n_\ad$, which is a reasonable value for illustrative purpose, while the actual value of $n_0$ that is reached in simulations depends on the resolution and sink formation algorithms. 

In general, the result of the hydrostatic equilibrium without singularity is a central density plateau with a decreasing envelope. 
Depending on the central density, the solution may decrease infinitely or reach negative values at large radius, 
while it should be connected to a confining ambient pressure in physical conditions. 
For sake of convenience, 
we define the boundary of this object where the density reaches the ambient density, 
namely $10^7~\cc$ in this study. 
Indeed, with a sharply decreasing density profile, 
this selection of density cutoff does not have a strong impact on the derived mass. 
As long as the outer density profile has radial dependence much steeper than $r^{-2}$, 
the object is clearly distinguishable from the isothermal envelope and its mass is well-defined.

\setlength{\unitlength}{1cm}
\begin{figure*}
\begin{picture} (0,18.7)
\put(1,9.5){\includegraphics[trim=0 20 0 10,clip,width=7cm]{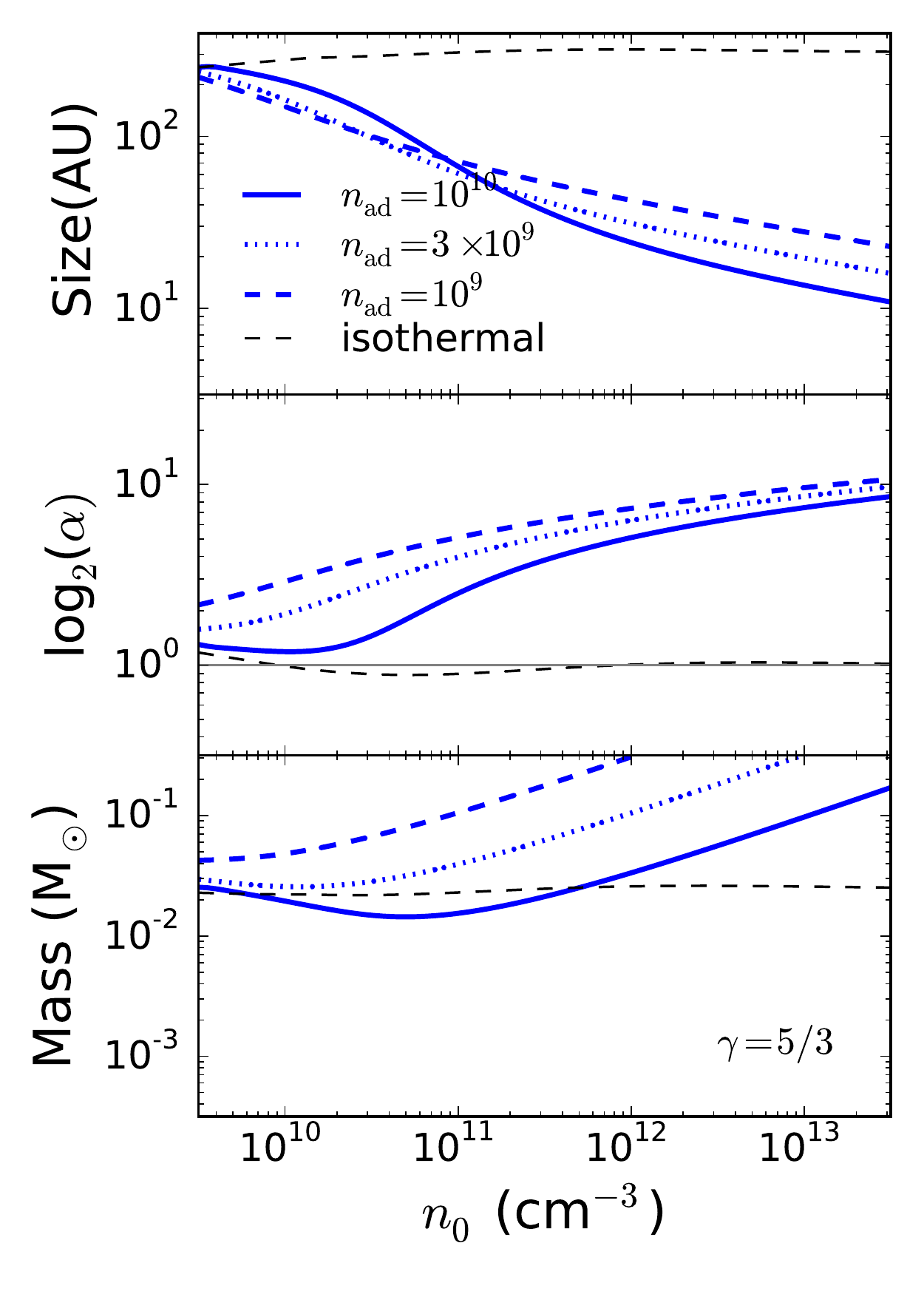}}  
\put(10,9.5){\includegraphics[trim=0 20 0 10,clip,width=7cm]{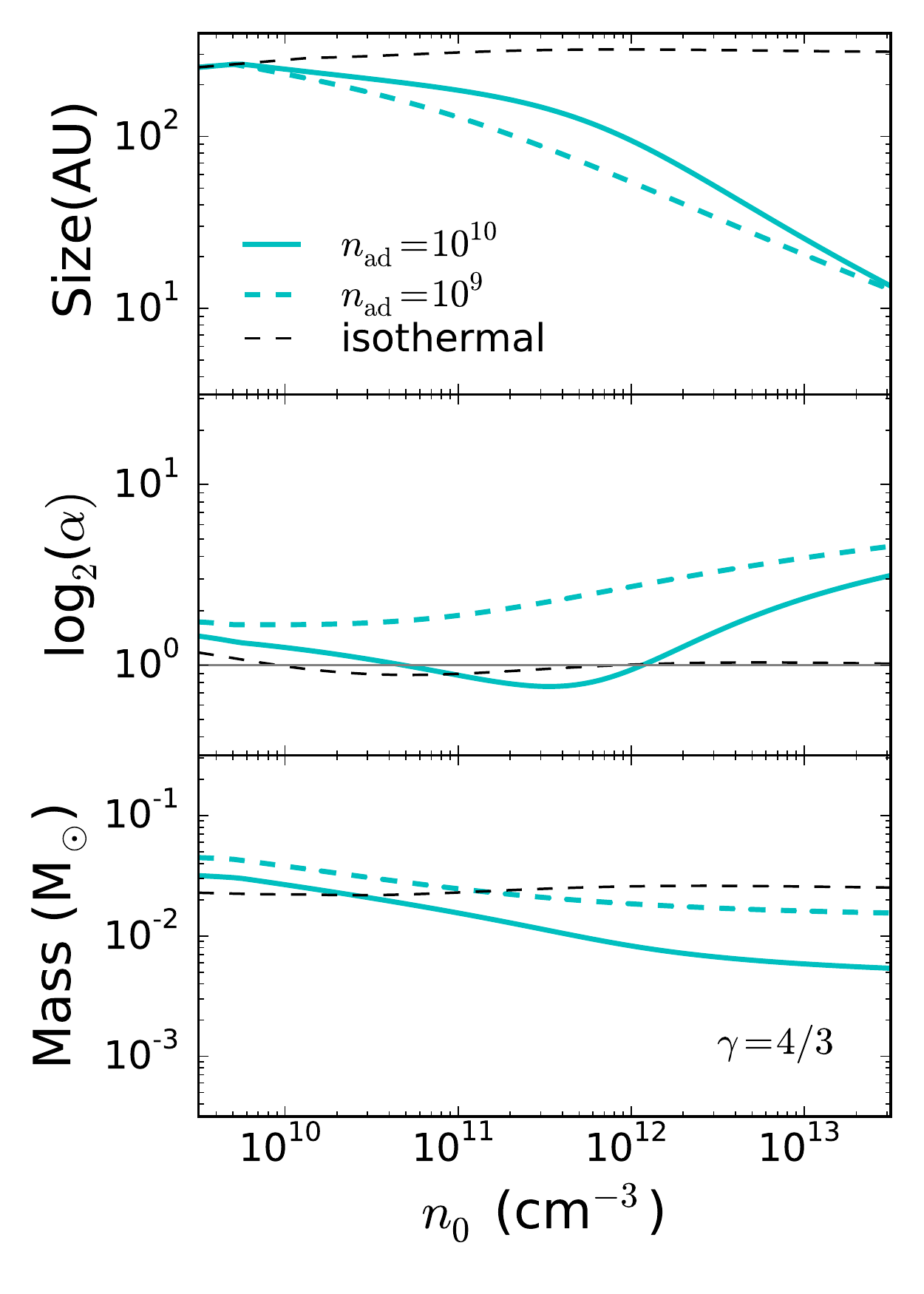}}  
\put(1,0){\includegraphics[trim=0 20 0 10,clip,width=7cm]{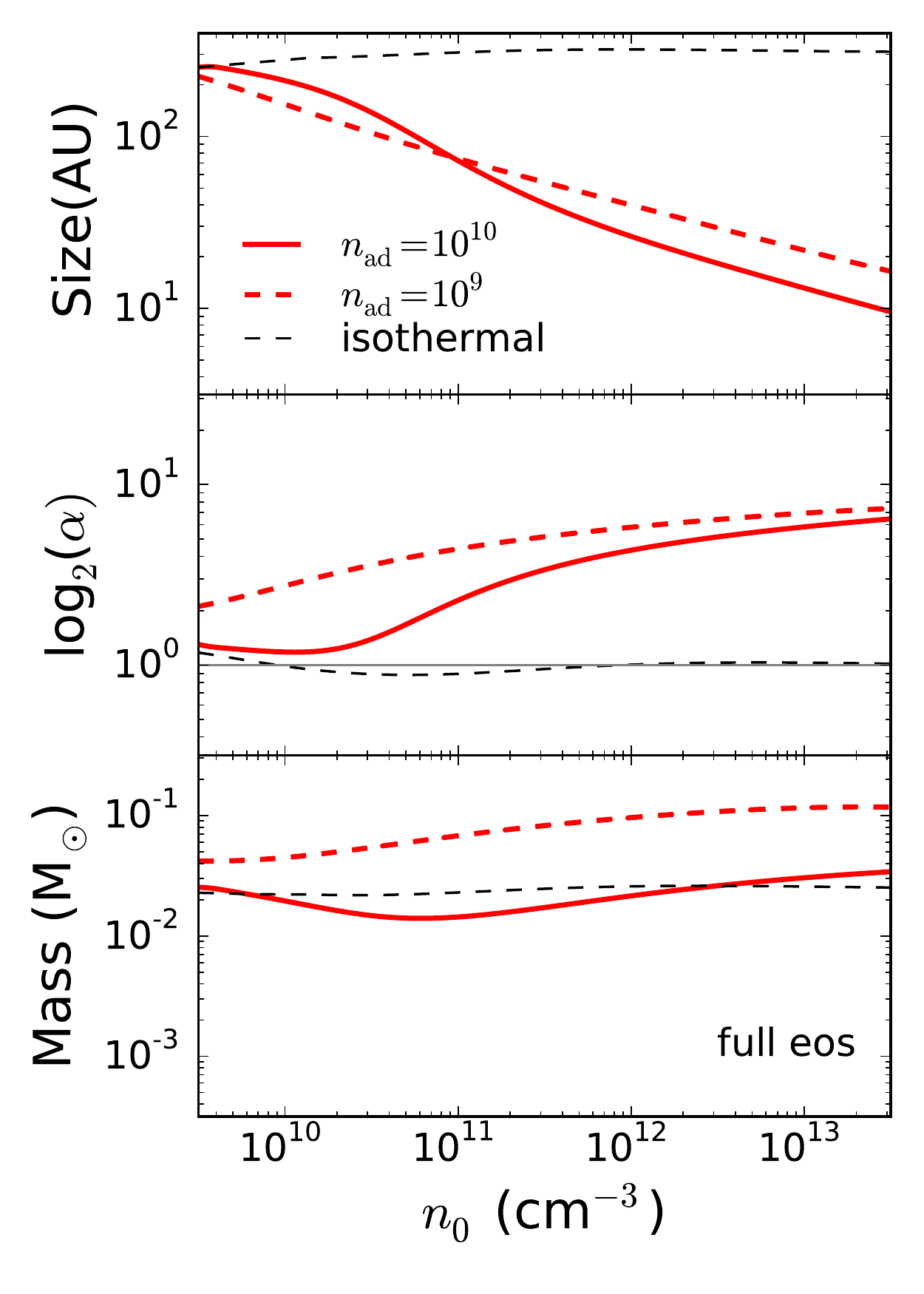}}  
\put(10,0){\includegraphics[trim=0 20 0 10,clip,width=7cm]{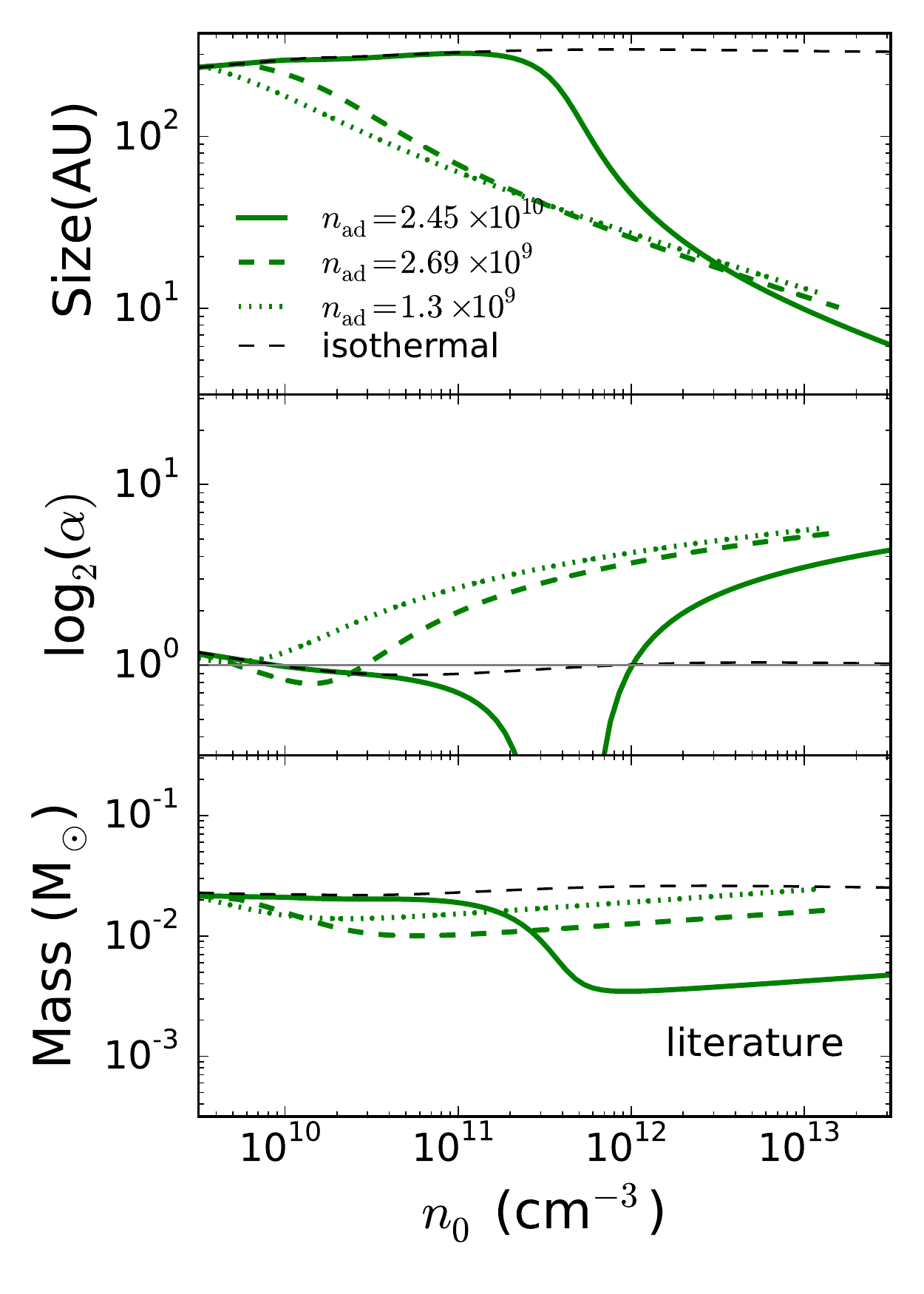}}  
\end{picture}
\caption{Properties of the first Larson core plotted against the central density $n _0$. The hydrostatic equilibrium 
equation (\ref{eq_hydro_T}) is solved with various $n _\ad$ and $\gamma$ values, and truncated at the ambient 
density $n = 10^7~\cc$. 
Three quantities are being shown:
the radius at which the density decreases to $n = 10^7~\cc$
the local powerlaw exponent $\alpha = -r dn/dr /n$, and the integrated mass.
\emph{Top-left}: $\gamma=5/3$, \emph{Top-right}: $\gamma=4/3$, \emph{Bottom-left}: full eos. 
\emph{Bottom-Right}:  eos taken 
from \citet[][solid]{Bate03}, \citet[][dashed]{Bate05b}, and \citet[][dotted]{Bonnell11} are shown.
Each time the isothermal case is plotted in black for comparison. 
 A core is well-defined only when 
$\alpha \gg 2$, and thus giving a characteristic mass of the first Larson core. Once a core is well-defined, its mass is almost 
constant irrespectively of $n _0$ when $\gamma=4/3$ or full eos is used. In the contrary, the core mass increases with 
increasing $n _0$ when $\gamma = 5/3$, explaining the peak mass shift between runs C1h-- and C1h-- --. } 
\label{fig_rhoc_sim}
\end{figure*}

Figure~\ref{fig_rhoc_sim} displays as function of the central density, $ n_0$, and
for various eos
the size of the core truncated at ambient density $10^7~\cc$ (top),  
the local powerlaw exponent of the density profile (middle) $\alpha = -r dn/dr /n$, 
such that locally $n \propto r^{-\alpha}$ with the gray line tracing $\alpha=2$, 
and the integrated mass (bottom) for several $n _\ad$ and $\gamma$ values corresponding to the simulations. 
Above certain value of $n _0$ (when entering the adiabatic regime), 
$\alpha$ starts to be significantly larger than 2 and a core is well defined, 
providing a clear definition of  the first Larson core mass, $M_\mathrm{L}$. 
Exceeding this mass, the gas becomes gravitationally bound, 
and this corresponds to the formation of sink particles in simulations. 
With small values of $n _0$, on the other hand, 
the integrated mass is sensitive to the ambient density cutoff and is almost constant irrespectively of $n _0$. 
At low densities. 
This corresponds to the isothermal regime of the eos and the density profile approaches that of a singular isothermal sphere \citep[SIS,][]{Shu77}.
The mass simply depends on the density cutoff and no obvious core could be identified. 
The size is, therefore, physically meaningful only at high $n _0$,
while at low $n _0$ it is merely a truncation of the SIS. 
In practice, we use $\alpha \gg 2$ to infer the radius of the Larson core. 
As can be seen, the size of the first Larson core is typically a few tens of AU, 
well resolved in our simulations. 
When a non-isothermal eos is used, the density is almost flat at the center and drops significantly at some radius. 
The larger the value of $n _0$, the smaller the size of the core, 
such that the integrated mass is not strongly increasing with increasing $n _0$, 
and even reaches a characteristic value, except for the cases with $\gamma=5/3$, 
where the mass increases with $n _0$ at rate of roughly $M_\mathrm{L} \propto n_0^{1/2}$. \\

An interesting question is whether rotation could modify these conclusions. In particular, since rotation 
adds another support, it could possibly lead to more massive first Larson cores. 
To investigate this possibility we have used the ESTER code \citep{espinosa2013,rieutord2013}, which 
computes bidimentional equilibria of rotating polytropes in solid-body rotation. We found that the 
maximum mass of a rotating polytrope increases only slightly (tens of percents). The reason is that the 
rotation support is significant only in the outer part while most of the mass lies in the central region.

\subsection{Interpreting the numerical simulation results}
Assuming, as will be shown later, that the peak of the stellar distribution is simply proportional 
to the mass of the first Larson core,  $M_\mathrm{L}$, 
these calculations can be used to interpret the results obtained from the numerical simulations.

For $\gamma=5/3$ and $n_\ad=10^{10}~\cc$, Fig.~\ref{fig_g53_10} shows that  
the peak of the stellar distribution does not change significantly with the resolution. 
Since increasing the resolution allows to probe higher densities, 
we can interpret this behavior as a consequence of the weak dependence of $M_\mathrm{L}$ on the core central density, $n _0$, 
that as can be seen from Fig.~\ref{fig_rhoc_sim} (blue solid lines) 
remains below $2 \times 10^{-2}~\Ms$ for  $n _0 < 10^{12}$, 
while noting that a self-gravitating core is well defined only when $n _0 \gtrsim 2 \times 10^{10}$ ($\alpha > 2$). 
With even higher resolution, 
we can expect that the stellar mass spectrum peak shifts to higher values as seen with the C1h runs at varied resolutions. 

When taking $n _\ad = 10^9~\cc$ and $\gamma = 5/3$ (Fig.~\ref{fig_g53}), 
the self-gravitating first Larson core is always well defined, 
with a mass that increases with $n_0$. 
This is reflected by the peak mass shift between runs C1h-- -- and C1h--. 
The eos with $n _\ad = 10^9~\cc$ gives a mass of the first Larson core larger than that with $n _\ad = 10^{10}~\cc$ at the same value of $n_0$, which is also in agreement with simulation results.

With $\gamma=4/3$, the effect of varyied $n _\ad$ on the peak of the stellar distribution has been found to be less significant (Fig.~\ref{fig_g43}). 
This again is in good agreement with Fig.~\ref{fig_rhoc_sim} (cyan lines), 
where at same value of $n_0$, the mass varies less with $n_\ad$ as compared to the $\gamma=5/3$ cases.
 
Finally, the runs with full eos (Fig.~\ref{fig_gfull}) present similar behavior to 
the runs with $\gamma=5/3$  (Fig.~\ref{fig_g53}) regarding the relation between peak mass and $n_\ad$. 
This is also expected as the mass dependence of $M_\mathrm{L}$ are relatively similar to the one 
of $\gamma=5/3$ for $n _0$ below $10^{12}~\cc$ (Fig.~\ref{fig_rhoc_sim}, red lines).

\subsection{Correlation between first Larson core and the stellar distribution peak}
\begin{figure}[]
\centering
\includegraphics[trim=12 15 12 10,clip,width=0.5\textwidth]{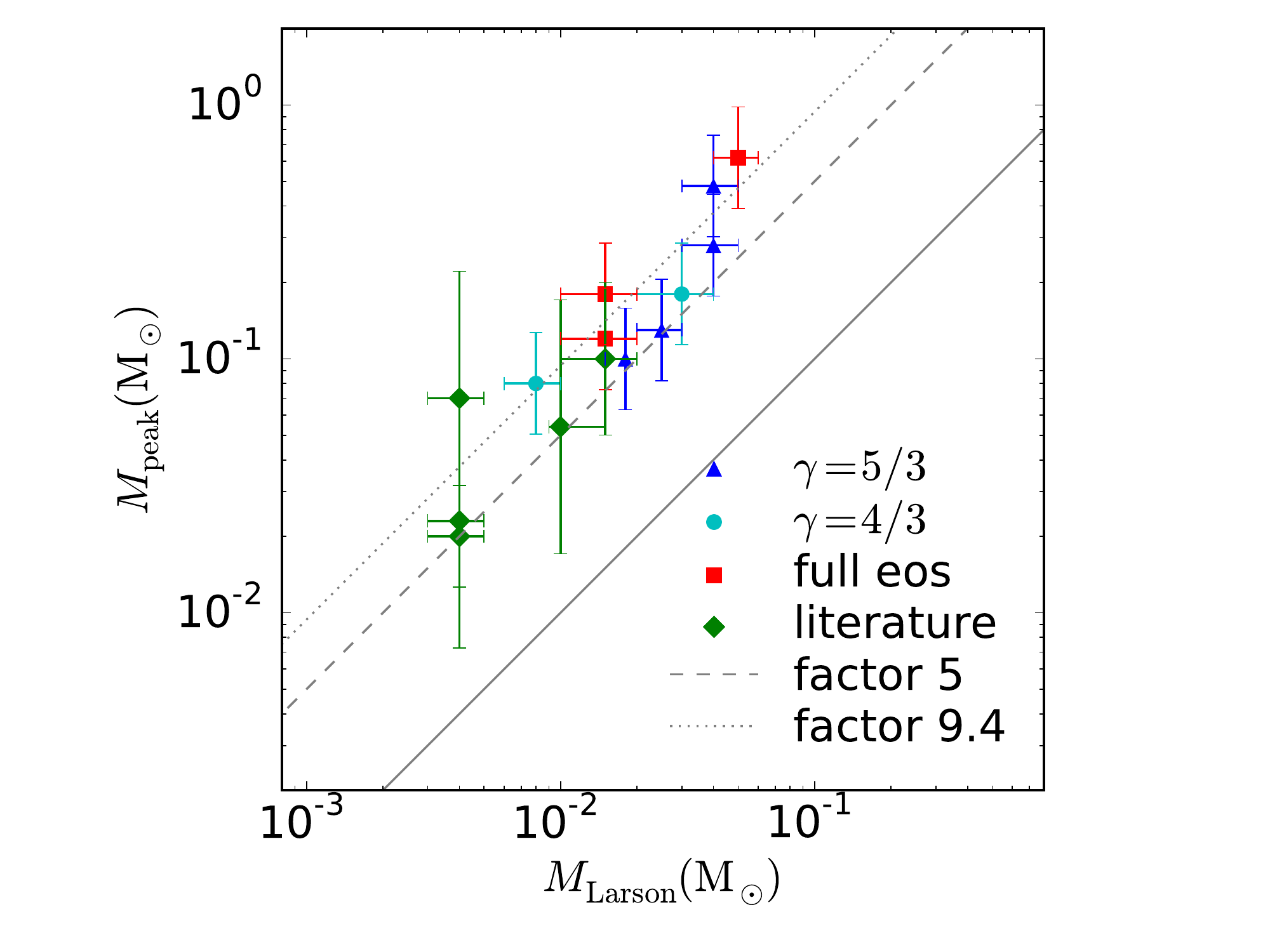}
\caption{Peak mass of sink mass spectrum plotted against first Larson core mass calculated with eos used in simulations. The peak mass is read from the 
stellar distribution, with the bin size as error. The error bars from literature results are larger as they have lower statistics and larger fluctuations in the spectrum. The first Larson core mass is read from Figs.~\ref{fig_rhoc_sim}, with error bars indicating the mass range with the probable $n_0$ range. Examining by eye, there is a correlation $M_\mathrm{peak} \sim 5 M_\mathrm{Larson}$. We also do a simple linear regression, without considering the errors, to find the scaling factor, and this leads to $\sum (M_\mathrm{peak}M_\mathrm{Larson})/ \sum M_\mathrm{Larson}^2=9.4$.} 
\label{fig_Mlr_Mpk}
\end{figure}

A link between the mass of the first Larson core, $M_\mathrm{L}$, 
and the mass at which the stellar distribution peaks is suggested in previous discussions. We now turn to a more systematic comparison. For that purpose, we have 
compiled our simulations with other results from the literature and 
we present in Fig.~\ref{fig_Mlr_Mpk} the correlation between the stellar distribution peak mass and the first Larson core mass deduced from the eos 
employed in the corresponding runs. 
The mass of the first Larson core is deduced from Figs.~\ref{fig_rhoc_sim}  around the point where $\alpha$ starts to be larger than two. 
Since the definition is not totally straightforward and also because  $M_\mathrm{L}$ depends on $n_0$, we estimated error bars
by considering that $n_0$ may vary by a factor 10. To estimate the peak position we took the mass corresponding to the  maximum values in the 
mass spectra. The error bars are given by the binning size.

We also included results from literature. \citet{Bate03} simulate $50~\Ms$ cloud with $\gamma = 7/5$ above $10^{-13}~\gcc$ and obtain 
median sink mass of $0.7~\Ms$. Increasing the density by decreasing the size \citep{Bate05a} or increasing the mass and velocity dispersion
 while keeping the same density \citep{Bate09a} give similar median masses 0.023 and 0.02 $\Ms$ that are significantly lower than the previous 
case. Alternatively, changing the critical density to $1.1 \times 10^{-14}~\gcc$ \citep{Bate05b} gives median mass $0.054~\Ms$. The median 
mass is used from these studies since there is not enough statistics and a peak is not always well-defined. The eos used in the simulation 
series by Bate et al. is a piece-wise temperature. To calculated the density profile of a mass concentration, the temperature as well as its 
derivative is required to be continuous, we thus use 
\begin{align}
T = T_0 \left[ 1+ \left(\rho/\rho_\ad\right)^{(\gamma-1)n} \right]^{1/n},
\end{align}
where $n=100 \gg 1$, to replace their eos when inferring the mass of the first Larson core.

Figure~\ref{fig_Mlr_Mpk} shows that the first Larson core mass is in good correlation with the stellar distribution peak
and we roughly get $M_\mathrm{peak} \simeq 10 M_\mathrm{L}$.
We try to explain for this extra factor of about ten in Section~\ref{st_point}. 

\section{Stabilizing effects of tidal forces around first Larson core: a model}
\label{st_point}

To link the mass of the first Larson core to the mass of the forming star, 
we present  a simple model discussing the effect of a point mass in a density field. 
The questions we want to address are: 
how the typical mass of a self-gravitating object embedded in a dense collapsing cloud
is limited in general to a small fraction of that of the latter? 
How does this fraction depend on the mass of the accreting object? 
Here we propose that there is a competition between  
the accretion onto the already formed self-gravitating object and the tendency to form new ones from gravo-turbulent fragmentation. 
What we want to estimate is thus the typical radius at which the probability to form another self-gravitating fragment equals to one. 
The mass enclosed within this radius is expected to be accreted by the central object, while the mass outside is expected to be 
distributed between forming fragments.

We discuss the conditions under which a density perturbation located near the point mass can become self-gravitating and form another core. 
The idea is that the first Larson core is going to ``screen'' itself by shearing out fluctuations through tidal forces.  
The modeling proceeds in essentially four steps:
\begin{itemize}
\item In Sect.~\ref{step1}, we describe a point mass, surrounded by a density profile, with a perturbation at distance $r\p$ and of size $\delta r$. 
 \item In Sect.~\ref{step2}, we develop the mass condition. The perturbation needs to have a mass of at least $M_\mathrm{L}$ such that its collapse can result in the formation of a stellar object. 
\item In Sect.~\ref{step3}, we infer the energy condition.
The gravitational field generated by the point mass, the density field, and the perturbation itself are used to calculate the gravitational 
energy of the perturbation. The idea is to compare the self-gravity of the perturbation to the destructive tidal field created by the 
point mass and the envelope. The gravitational energy of the perturbation is required to be negative such that it can contract to form another star.
\item In Sect.~\ref{step4}, we estimate the mass that should eventually be accreted onto the central object.
The mass and energy constraints provide a condition on the density perturbations as a function of distance. Assuming a lognormal 
distribution for the density fluctuations (on top of the local mean density which itself is $\propto r^{-2}$), 
we can infer at which radius the probability of finding a self-gravitating perturbation of 
mass at least equal to $M_\mathrm{L}$ is one. 
\end{itemize}

\subsection{Step 1: density profile around sink particles and perturbation}
\label{step1}
Extracting gas profiles around young sinks particles, that have not yet decoupled from their natal gas, 
we have shown in paper I that the cores forming stars have indeed some characteristic density profiles $\rho \propto r^{-2}$ as expected from theory \citep{Shu77}. 
Thus we take the density field $\rho\e(r\e)$ as a function of distance to the point mass  
\begin{align}
\rho\e &= {Ac_\mathrm{s}^2 \over 2\pi G} {1 \over r\e^2}, 
\label{rho_env}
\end{align}
where $A$ is the amplitude of the  singular isothermal sphere density profile \citep{Shu77}. 
In paper I, the density profiles around young sink particles are inferred from simulations and typical value of $A \sim 10$ is found
at early time when the sinks are still actively accreting (see left panels of Fig.~4 of paper I).
For the sake of simplicity, we consider a uniform density spherical perturbation $\rho\p$ at distance $r\p$:
\begin{align}
\rho\p &= \eta {Ac_\mathrm{s}^2 \over 2\pi G} {1 \over r\p^2} ~~\mbox{for}~~ |\mathbi{r}\e - \mathbi{r}\p| = |\delta\mathbi{r}|  
=  \delta r \leq  \delta r\p,
\end{align}
where $\mathbi{r}\e$ and $\mathbi{r}\p$ are the vectors pointing from the central object to a point in the envelope and to the center of perturbation
and $\eta$ is a constant for the local density contrast of the perturbation.

To determine whether the perturbation is prone to self-gravitating collapse, 
we use the virial theorem to probe its stability
\begin{align} \label{eq_int_vir} 
E_\mathrm{vir}&(r\p,\delta r\p,\eta) = E_\mathrm{g}(r\p,\delta r\p,\eta) + 2E_\mathrm{ther} \\
&= \int\limits_{V\p} \rho\mathbi{g} \cdot \delta\mathbi{r} ~dV + 3 M\p(r\p,\delta r\p,\eta) c_\sound^2 \nonumber\\
&= \int\limits_{V\p} (\rho\e+\rho\p)~(\mathbi{g}_\mathrm{L}+\mathbi{g}\e+\mathbi{g}\p) \cdot \delta\mathbi{r} dV+ 3 M\p(r\p,\delta r\p,\eta) c_\sound^2. \nonumber
\end{align} 

The virial energy $E_\mathrm{vir}$ is a combination of the gravitational energy $E_\mathrm{g}$ and the thermal energy $E_\mathrm{ther}$, with $M\p$ being the mass of the perturbation. 
The gravitational acceleration $\mathbi{g} = \mathbi{g}_\mathrm{L} + \mathbi{g}\e + \mathbi{g}\p$ has three components, 
coming from the central first Larson core that is described with a point mass, 
the powerlaw density envelope, and the perturbation itself. 
Note that to correctly calculate the virial energy, only the tidal tensor part of the gravity should be considered and the mean value exerted on the center of the object should be removed. Thus in the following we always use the relative gravitational acceleration with respect to the center of the perturbation. 
The collapse criteria are, therefore,
\begin{align} \label{eq_cond_core}
M\p(r\p,\delta r\p,\eta) &\ge M_\mathrm{L} ~~~\mbox{and}~~~ \\
E_\mathrm{vir}(r\p,\delta r\p,\eta) &\le 0.
\end{align}

\subsection{Step 2: the  mass condition}
\label{step2}

The mass contained inside the perturbation radius $\delta r\p$ around the center of perturbation is 
\begin{align}
\label{eq_M_u}
M\p(r\p,\delta r\p,\eta) = \int\limits_{V\p} \rho(\mathbi{r}\e) ~dV = \int\limits_{V\p}\left( \rho\e+\rho\p\right) ~dV. 
\end{align} 
We designate $\theta$ the angle between $\delta\mathbi{r}$ and $-\mathbi{r}\p$.
With the trigonometric relation
\begin{align}
r\e^2 &= r\p^2 + \delta r^2 - 2r\p\delta r \cos \theta, 
\end{align}
we can integrate
\begin{align} \label{eq_M_u}
M\p(r\p,\delta r\p,\eta)  = {Ac_\sound^2 \over G} r\p \mathfrak{m}\p(u\p,\eta),
\end{align} 
where the normalized quantity $u\p=\delta r\p/r\p$. 
The normalized mass is expressed as $\mathfrak{m}\p(u\p,\eta)$ and is given in Appendix \ref{calc_mass}.

As explained above, the mass of the perturbation must be typically equal to $M_\mathrm{L}$ since no star could form below this value. 
By requiring $M\p(r\p,\delta r\p,\eta)=M_\mathrm{L}$, we obtain a relation between the size $u\p$ and amplitude $\eta$ of a perturbation at distance $r\p$.

\subsection{Step 3: the energy condition}
\label{step3}
Now turning to the calculation of the virial parameter, 
we first show the importance of tidal forces in the vicinity
of the sink particles, 
and then, to get physical intuition, we perform a simplified 1D estimate 
before carrying out a full 3D calculation. 

\subsubsection{Tidal forces}
Tidal forces have been advocated in various context to either limit or quench the star formation
\citep{bonnell2008,ballesteros2009}, or on the contrary to trigger it \citep{jog2013,renaud2014}.
\citet{ntormousi2015} discuss the influence of the density distribution and show that 
in a collapsing cloud, the tidal forces tend to be initially compressive, promoting fragmentation, and then become
stabilizing.
The key quantity to study the influence of tidal forces is the gravitational stress tensor, $\partial _i g_j$, 
that is characterized by the three eigenvalues, $\lambda_i$.
From Poisson equation, we know that $\sum \lambda_i = - 4 \pi G \rho$. Thus there is at least 
one negative eigenvalue, say $\lambda_1$, and in most cases two of them are negative. 
It is well known that certain gravitational fields present a value of $\lambda_3$ that is positive
and that such fields render the formation of self-gravitating perturbations less likely (see Appendix~\ref{tidal} for tidal forces in the simulation).  
Below, the tidal effect is taken into account by including the corresponding term in the virial theorem.

\begin{figure}[t]
\centering
\includegraphics[trim=0 20 0 12,clip,width=0.5\textwidth]{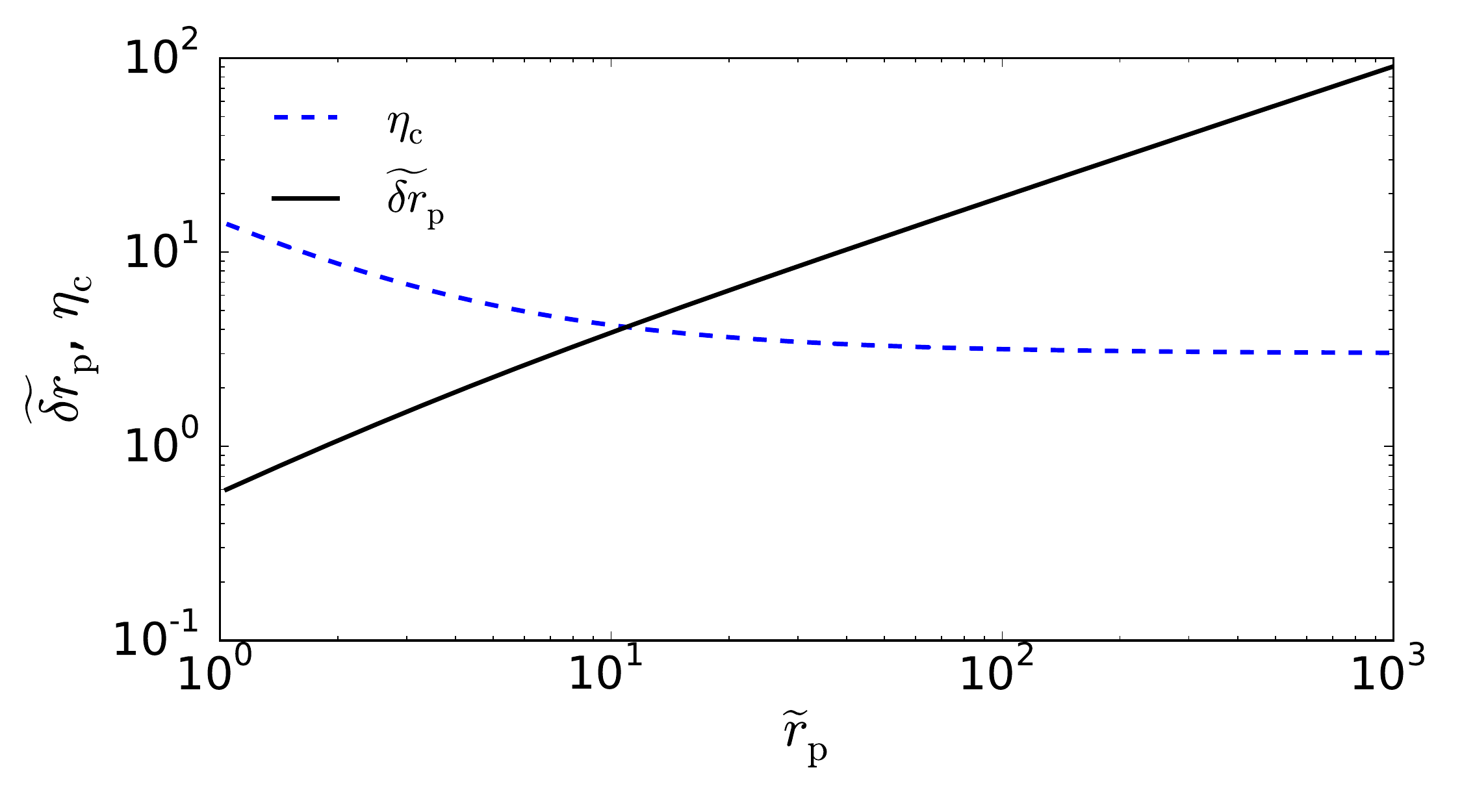}
\caption{Critical value of $\eta_\mathrm{c}$ ({\it blue dashed}) and corresponding $\widetilde{\delta r\p}$ ({\it black solid}) as functions of the normalized distance $\widetilde{r}\p$. Gravitational potential energy is calculated in 1D.} 
\label{fig_upeta_rp_1D}
\end{figure}

\subsubsection{1D calculation of gravitational energy}
The direction along $\mathbi{r}\p$ is the relevant dimension as the gravity of the central mass and the envelope generates diverging tidal forces and prevents the over-dense perturbation from self-gravitating collapse. 
Provided the reasonable illustrative results and the simplicity, 
we first perform a virial integration along this direction on the axis of the perturbation, ignoring thermal energy, 
and try to obtain some physical insight from the simplified calculations. 
We define the position $\delta r$ with respect to the center of perturbation at $r\p$, 
with the positive direction pointing away from the central mass.
The gravitational acceleration, relative to the center of the perturbation, along the radial direction is 
\begin{align}
g(\delta r) = -{GM_\mathrm{L} \over (r\p\! +\! \delta r)^2}+{GM_\mathrm{L} \over r\p^2} - {2A c_\sound^2 \over r\p\! +\! \delta r} +{2A c_\sound^2 \over r\p} - \eta{2A c_\sound^2 \over 3r\p^2}\delta r.
\end{align}
By removing the acceleration in the moving coordinate of the perturbation, the tidal force generated by the gravity is adequately considered. 
The first two terms on the right-hand side represent the gravitational acceleration generated by the central star, third and fourth terms acceleration by the envelope, and last term the gravity of the perturbation itself. 
The density is
\begin{align}
\rho(\delta r) = {Ac_\mathrm{s}^2 \over 2\pi G} \left({1 \over (r\p+\delta r)^2} + { \eta  \over r\p^2} \right).
\end{align}
The virial integration is, therefore, 
\begin{align}
\label{eq_vir_1D}
\int\limits_{-\delta r\p}^{\delta r\p}   \rho(\delta r) ~ g(\delta r) ~\delta r ~d\delta r 
={(Ac_\sound^2)^2 \over 2\pi Gr\p} \mathcal{E}_\mathrm{1D,g}(r\p, u\p,\eta), 
\end{align}
where $\mathcal{E}_\mathrm{1D,g}$ is the non-dimensionalized energy (detailed derivation is given in Appendix~\ref{appen_1D}). 
This is a line integration, and the quantity has the dimension of energy per unit surface. 
We define a normalizing radius
\begin{align}\label{r_lars}
r_\mathrm{L} = {GM_\mathrm{L} \over  2A c_\sound^2} = 1.1 \times 10^2 ~\AU \left({1\over A}\right)\left({M_\mathrm{L} \over 0.01~\Ms}\right), 
\end{align}
such that 
\begin{align}\label{M_lars}
\widetilde{r}\p = r\p / r_\mathrm{L} = M\e(r\p) / M_\mathrm{L},
\end{align} 
where $M\e(r\p) = 2Ac_\sound^2r\p/G$ is the mass of the envelope within $r\p$. 
The energy  $\mathcal{E}_\mathrm{1D,g}$ can be expressed as
\begin{align}
\mathcal{E}_\mathrm{1D,g}(\widetilde{r}\p,\! u\p,\! \eta) = {2\over \widetilde{r}\p}\mathcal{E}_\mathrm{1D,L}(u\p,\! \eta)+\mathcal{E}_\mathrm{1D,e}(u\p,\! \eta)+\mathcal{E}_\mathrm{1D,p}(u\p,\! \eta),
\end{align}
with contributions from the three components. 

The solutions are found by solving simultaneously for all $\widetilde{r}\p$
\begin{align}
\mathfrak{m}\p(u\p,\eta)={2\over \widetilde{r}\p} ~~~\mbox{and}~~~ 
\mathcal{E}_\mathrm{1D,g}(\widetilde{r}\p,u\p,\eta)=0.
\end{align}
The first condition is simply the condition stated by Eq.~(\ref{eq_cond_core}), while the normalized expression stated 
by Eq.~(\ref{eq_M_u}) is used. 

We plot the critical density contrast $\eta_\crit$ and the corresponding size of the perturbation, $\widetilde{\delta r\p} = \delta r\p/r_\mathrm{L}$,
 against $\widetilde{r}\p$ in Fig.~\ref{fig_upeta_rp_1D}, 
showing that, as we get away from the central star, lower density contrast (smaller $\eta$) and lower level of mass concentration (larger $\delta r\p$) are needed for the perturbation to be self-gravitating. 
As $\eta$ increases at fixed mass, $u\p$  and $\mathcal{E}_\mathrm{1D,g}(\widetilde{r}\p,u\p,\eta)$ decrease. 
Therefore, at any given distance to a first Larson core, 
we can derive a critical local density contrast $\eta_\crit$ above which the perturbation is self-gravitating and is prone to collapse. 
Within this distance, all density fluctuations smaller than $\eta_\crit$ are prevented from collapsing by the tidal field of 
the central mass and its envelope. 
The value of $\eta_\crit$ is typically larger than a factor of a few, while the radius of the perturbation must be significantly smaller than the distance from the central object. 
These numbers indicate that the conditions for a perturbation to become unstable are not 
easy to satisfy and therefore most perturbations, which in the absence of tidal forces would be unstable, are rendered stable. 
This process favors further accretion onto the central object other than distributing the mass between new fragments.

\begin{figure}[t]
\centering
\includegraphics[trim=0 20 0 12,clip,width=0.5\textwidth]{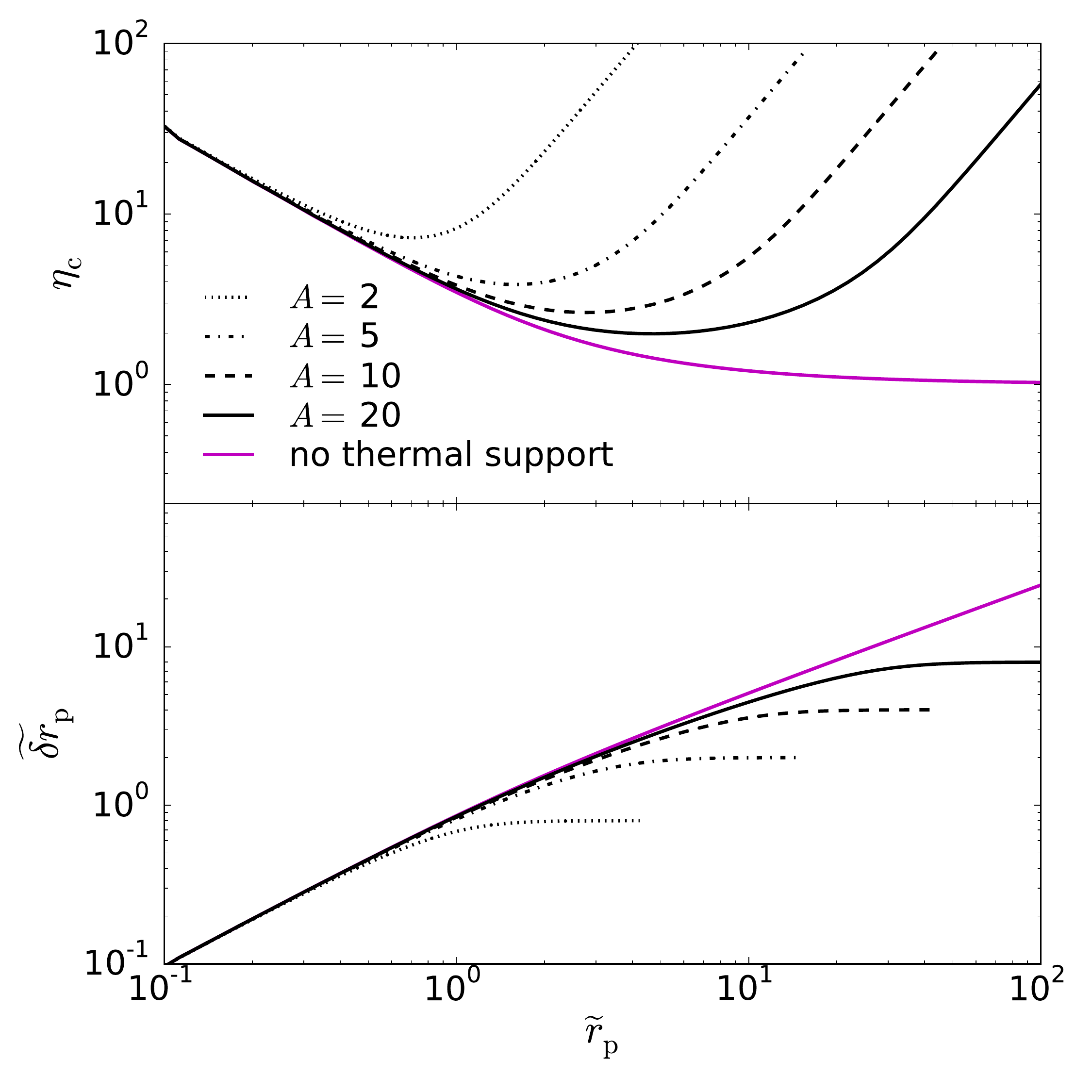}
\caption{Critical value of the perturbation amplitude, $\eta_\mathrm{c}$ ({\it upper panel}) and corresponding 
normalized size of the perturbation $\widetilde{\delta r\p}$ ({\it lower panel}) as function of the normalized distance $\widetilde{r}\p$. 
The 3D gravitational energy is used and  thermal support  is included for various density profiles. 
Fluctuations have to be at least several times the local density to be unstable. The low amplitude
perturbations are stabilized by tidal forces or thermal support.
} 
\label{fig_upeta_rp}
\end{figure}

\subsubsection{Full 3D calculation}
The SIS density profile generates positive tidal force in the radial direction, 
while the field is compressing in the other two directions. 
A full 3D calculation can be, therefore, different from the 1D case and it is necessary to clarify this effect.  
For conciseness, the full integration of Eq.~(\ref{eq_int_vir}) in 3D space is presented in Appendix \ref{appen_3D}, 
and here we simply discuss the results: 
\begin{align}\label{eq_3D_norm}
E_\grav&(r\p,\delta r\p,\eta) = {(Ac_\sound^2)^2 \over G}r\p ~\mathcal{E}_\mathrm{3D,g}(\widetilde{r}\p,u\p,\eta)\\
&=  {(Ac_\sound^2)^2 \over G}r\p\left[ {2\over \widetilde{r}\p}\mathcal{E}_\mathrm{3D,L}(u\p,\eta)+\mathcal{E}_\mathrm{3D,e}(u\p,\eta)+\mathcal{E}_\mathrm{3D,p}(u\p,\eta)\right], \nonumber
\end{align}
where $\mathcal{E}_\mathrm{3D,g}$ is the normalized gravitational energy. 
The results are qualitatively similar to the 1D calculations. 
The values of $\eta_\crit$ and corresponding $\widetilde{\delta r\p}$ are plotted in Fig. \ref{fig_upeta_rp} as function of the normalized distance $\widetilde{r}\p$ ({\it magenta}). 
At short distance they are very close to the 1D approximation and $\eta_c$ must be larger than several but at large distance $\eta_c$
drops to $\sim 1$, implying that the perturbation is easily unstable. 

However one should also consider the thermal support $3M\p c_\sound^2$ against self-gravity. 
We can infer the condition $\mathcal{E}_\mathrm{3D,g}+3\mathfrak{m}\p/A \le 0$ from Eqs.~(\ref{eq_int_vir}) and~(\ref{eq_M_u}).
Including the thermal support does not change much the conclusions at small $r\p$, 
while at larger distance, where the envelope density is low, 
thermal support becomes dominant and tidal forces unimportant. 
In this context, $\eta_\crit$ becomes an increasing function of $r\p$, thus imposing a lower limit on $\eta_\crit$ ({\it black} lines in Fig. \ref{fig_upeta_rp}).

The result sensitively depends on $A$, the amplitude of the density field. When $A$ is equal to a few, the combination 
of thermal support and tidal forces stabilize most perturbations. Only the one denser than about 10 times 
the background density can eventually collapse. This is in good agreement with simulations such as the ones 
performed by \citet{Girichidis11}, which show that clouds with $r^{-2}$ density profiles are not prone to fragmentation. 

\subsection{Step 4: final mass from integrated probability of nearby first Larson core formation}
\label{step4}
In the previous paragraph we have obtained the critical density contrast, $\eta_\crit$, of forming a second self-gravitating core near a point 
mass $M_\mathrm{L}$ as function of the distance. The question is now under which circumstances would one encounter such fluctuations, 
or more precisely what is the probability for a given perturbation amplitude to occur?  
To answer this question, one would need to know the fluctuation spectrum within a collapsing clump around an accreting point mass. Since there is no available statistics, this last step 
is certainly the least quantitative. 
To get some estimate, we now use the typical density fluctuations produced by supersonic 
turbulence to determine the characteristic distance at which 
a perturbation with sufficient amplitude could be found and a second core 
could form. The mass within this radius is expected to be protected from fragmentation and, therefore, should be accreted.

We use the simple assumption of a lognormal PDF \citep[e.g.][]{Vazquez94,Federrath08,HF12}, 
that is, the probability of finding perturbations of $\eta$ at distance $\widetilde{r}\p$ is
\begin{align}
P(\widetilde{r}\p, \delta) = {1\over \sqrt{2\pi} \sigma(\widetilde{r}\p)} \exp \left[{-\left(\delta+\sigma^2(\widetilde{r}\p)/2\right)^2 \over 2\sigma^2(\widetilde{r}\p)}\right],
\end{align}
where $\delta = \log (1+\eta)$ and $\sigma$ is the width of the density PDF.
The density fluctuation is related to the local turbulent Mach number such that 
$\sigma^2(\widetilde{r}\p) = \log \left(1+b^2\mathcal{M}(\widetilde{r}\p)^2\right)$, with $b=0.5$.
To estimate the local Mach number, 
we assume that the turbulent Mach number is proportional to the that obtained from the infall velocity. 
This implies that the collapse is able to amplify existing perturbations until local energy equipartition is reached. 
Given the mass of the central core and the envelope, the turbulent Mach number is, therefore, 
\begin{align}\label{eq_mach_2}
\mathcal{M}(\widetilde{r}\p) 
&= {1 \over c_\sound} \sqrt{{G [M_{\rm L} + M\e(r\p)] \over r\p} \epsilon}
= {1 \over c_\sound} \sqrt{{G M_{\rm L} (1+\widetilde{r}\p) \over r\p}\epsilon} \\
&= \sqrt{2 A (1+{1 \over \widetilde{r}\p}) \epsilon}, \nonumber
\end{align} 
where the second equivalence is obtained using Eq.~(\ref{M_lars}) and the third using Eq.~(\ref{r_lars}). 
The coefficient, $\epsilon \le 1$, represents the amount of released gravitational energy that goes into the turbulent fluctuations, rather than the infalling component. 
As estimated in paper I, $A \simeq 10$, leading  to $\mathcal{M} \simeq 4 \sqrt{\epsilon} $.
We refer to this assumption as model I.
We integrate the mass that exceeds the critical density at all $r\p$, 
using the density of the fluctuation $ \rho(\widetilde{r}\p^\prime)\exp(\delta)$, the local background multiplied by the  relative density fluctuations. 
Dividing by $M_\mathrm{L}$, we obtain the typical number of self-gravitating cores with a mass at least equal to $M_\mathrm{L}$ contained within $r\p$
\begin{align}\label{eq_N_r}
\mathcal{N}(\widetilde{r}\p) &=  {1 \over M_\mathrm{L}}\int\limits_0^{r\p} \int\limits_{\delta_\crit}^\infty P(\widetilde{r}\p^\prime, \delta) \rho(\widetilde{r}\p^\prime)\exp(\delta)4\pi {r\p^\prime}^2 d\delta  dr\p^\prime \\
&= \int\limits_0^{r\p} {1\over 2}\left[1+\mathrm{erf}\left({\sigma^2(\widetilde{r}\p^\prime)-2\delta_\crit \over 2\sqrt{2}\sigma(\widetilde{r}\p^\prime)}\right)\right] {Ac_\sound^2 \over 2\pi G {r\p^\prime}^2} {4\pi {r\p^\prime}^2 \over M_\mathrm{L}} dr\p^\prime \nonumber\\
&= \int\limits_0^{\widetilde{r}\p} {1\over 2}\left[1+\mathrm{erf}\left({\sigma^2(\widetilde{r}\p^\prime)-2\delta_\crit \over 2\sqrt{2}\sigma(\widetilde{r}\p^\prime)}\right)\right] d\widetilde{r}\p^\prime, \nonumber
\end{align}
where we have used Eq.~(\ref{r_lars}) to normalize the expression. With the definition of ~$\widetilde{r}\p$, the mass 
contained within a sphere of radius $a \widetilde{r}\p$ is equal to $a M _\mathrm{L}$. 
When $\mathcal{N}(\widetilde{r}\p^\ast)=1$, the chance of getting another fragment with mass $M_ \mathrm{L}$ is equal to 1. 
The mass enclosed within this radius is accreted by the central object since this mass 
is protected from fragmentation. Thus the mass of the final object is certainly 
larger than  $M_\mathrm{F} = (1+\widetilde{r}\p^\ast)M_\mathrm{L}$.

\begin{figure}[t]
\begin{picture} (0,4.65)
\put(0,0){\includegraphics[trim=0 18 0 12,clip,width=0.5\textwidth]{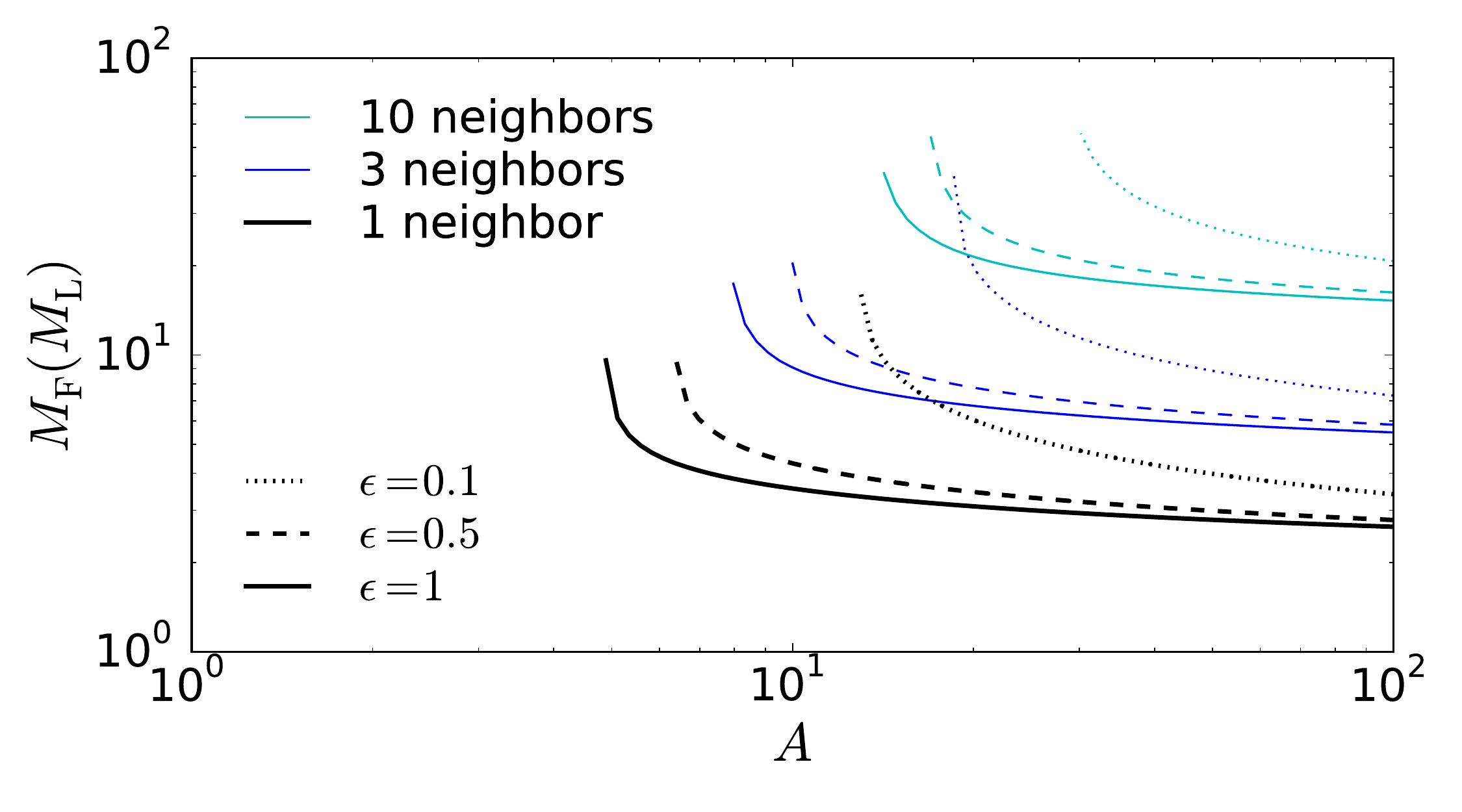}}
\put(7,1){Model I}
\end{picture}
\caption{The final mass $M_\mathrm{F}$, in units of $M_\mathrm{L}$, 
plotted against $A$, the amplitude of the density envelope (Eq.~(\ref{rho_env})), and for various $\epsilon$, the efficiency at which infall triggers turbulent fluctuations. 
The critical radius $\widetilde{r}\p^\ast$ is equivalent to the amount of protected envelope mass in units of $M_\mathrm{L}$, 
and thus $M_\mathrm{F} = (\widetilde{r}\p^\ast+1) M_\mathrm{L}$.  
Black lines show the expected mass if the envelope is truncated at $\mathcal{N}(\widetilde{r}\p)=1$, and $M_\mathrm{F}$ is expected to be larger than 4 for typical values of $A$. Blue thin lines show the same result, while truncated at $\mathcal{N}(\widetilde{r}\p) =3$, and the expected $M_\mathrm{F}$ is increased  to $\sim 8-9$. Cyan lines show results for $\mathcal{N}(\widetilde{r}\p) =10$.} 
\label{fig_MF_M0}
\end{figure}

Figure~\ref{fig_MF_M0} shows $M_\mathrm{F}$   as a function of $A$. 
The typical values of $A \sim 10-20$ and $\epsilon \sim 0.5$  give a final stellar mass of $\sim 4 ~M_\mathrm{L}$
if we integrate up to $\mathcal N =1$ ({\it black lines}), $\sim 8-9$ or $20-30 ~M_\mathrm{L}$ is obtained if we consider $\mathcal N = 3$ ({\it blue lines}) or $\mathcal N = 10$ ({\it cyan lines}).
Up to which radius one should integrate, that is to say how far the protostar is able to accrete, is not an easy question. 
Typically, one can estimate that about 6 fragments (one for each direction) should prevent any further accretion, but 
on the other hand these fragments will attract mass to them that should not be attributed to the central object. Therefore, 3 fragments sound like a reasonable number. 
In appendix~\ref{modelII} another model for the Mach number dependence is explored and leads to similar conclusions.  

Remember that the mass is integrated without considering if they are spatially connect. 
The value of $\mathcal{N}$ is thus a lower limit, and so is the derived $\widetilde{r}\p^\ast$.
Although the model we presented is not very accurate, it nevertheless suggests that the 
non-linear factor of $\simeq 10$ inferred from the simulations is entirely reasonable. Further work is certainly 
required here. 

\subsection{Tidal protection in simulations}
\begin{figure}[]
\begin{picture} (0,6.6)
\put(0,0){\includegraphics[trim=0 10 0 10,clip,width=0.5\textwidth]{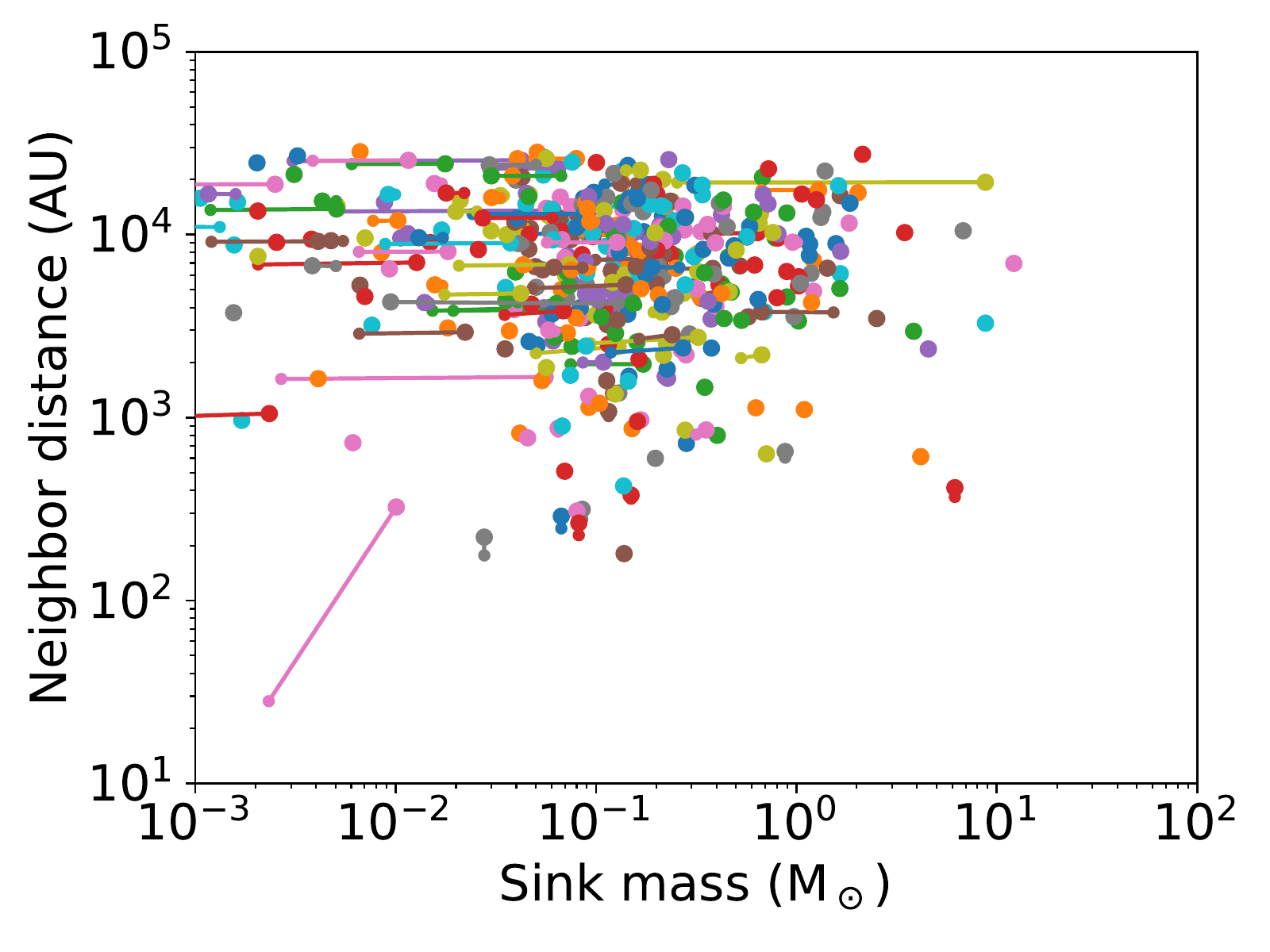}}
\end{picture}
\caption{Distance of nearby sink formation plotted against sink mass. The distance is calculated at the moment of the neighboring sink formation. The distances from the each sink to its first and tenth (small and large dots interlinked by a line) closest neighbors are plotted against its mass at the moment of the formation of the neighbors. In the presence of one sink, any neighbor can hardly form within the distance of a few hundreds of AU. } 
\label{fig_neighbor}
\end{figure}

To verify our model of tidal protection that inhibits nearby core formation, we examine the distance between the sinks in our simulations. For each sink, the distance of formation position is calculated for all younger sinks. In Fig. \ref{fig_neighbor}, the distances of the first and tenth closest neighbors are plotted against the instantaneous sink mass at the moment of neighbor formation for run C10h+. The distances of the first few closest sinks are similar and almost identical in most cases (which suggests that our model needs further refinement). The formation of a nearby sink is generally inhibited within the distance of several hundreds of AU for established sinks. 

\section{Discussions}

\subsection{Necessary developments}
In our simulations, the physics of the first Larson core is 
setting the peak of the stellar distribution and we infer $M_\mathrm{peak} \simeq 10 ~M_\mathrm{Larson}$. 
Since the mass of the first Larson core at the point where it undergoes 
second collapse is typically on the order of a few 0.01 $\Ms$ \citep{Masunaga98,vaytet2012,Vaytet17}, 
which is very close to the observed peak of the IMF \citep{bastian2010}.
The most recent radiative simulations of core collapse by \citet{Vaytet17} show that the mass of the first 
Larson core is $\sim 0.02~\Ms$. One of the difficulty encountered by our calculations is that it is not possible 
 to properly describe the second collapse and the disappearance of the first Larson core as it is 
being accreted. Therefore, there are  some uncertainties in the value of the first Larson core that should be used. 
Smoothed two-slope polytropic eos representing the results of radiative transfer balancing in dense gas, 
detailed simulations with complete radiation calculations would give more accurate results, 
though numerical resolution is a major obstacle. On the other hand, varying the 
density at which the sink particles are introduced (runs C10f and C10fd in Fig.~\ref{fig_gfull}) 
does not change the position of the peak despite reducing the number of objects. This may 
indicate that the exact way the first core is accreted on to the protostars may be not 
too critical in determining the peak position.

Finally, we reiterate that the physics used in the present paper and its companion, may be 
over simplified in other aspects since it does not include the magnetic field and heating by accretion luminosity. Both 
are known to reduce fragmentation \citep{krumholz2007,Bate09b,commercon2011,hennebelle2011,myers2013}
and may modify drastically the fragmentation we obtain.

\subsection{What sets the peak of the IMF: is there a cosmic conspiracy ?}
Unlike predicted by simple gravo-turbulent fragmentation models \citep[e.g.][]{HC08,HC13,hopkins2012} that the IMF peak 
depends on the turbulent Jeans mass, the IMF characteristic mass in high density regime has a lower limit 
imposed by the local thermal balancing and the formation of the first Larson core. 
The traditional view that the typical fragmentation mass is linked to the Jeans mass is no longer valid 
in these conditions because self-gravity triggers the development of a powerlaw density PDF. 
As discussed in paper I, the existence of a peak may however depend on the environment. 
In somewhat diffuse clouds, or more precisely when the local thermal support dominates over the 
turbulent one, the stellar distribution is flat ($dN/d\log M \propto M^0$) and the mass of the first Larson core, $M_\mathrm{Larson}$, 
may truncate the distribution rather than leading to a peak. 

One important issue, however, is the initial conditions and the exact way the matter is assembled. 
Although we verified in paper I (see the appendix) that the initial fluctuations are not crucial, 
setting up a dense reservoir without pre-existing cores may be a significant over simplification. 
In other words, the initial conditions that should be used to describe the ISM 
are not well understood yet and one must caution that the ones used in this paper may introduce some biases. 
In particular, as stressed in paper I, the high mass part of the mass spectra tends to present a powerlaw that is too
shallow compared to the canonical Salpeter distribution. 
If dense clusters turn out to be assembled from material where cores are already 
developed, the pre-existing cores may play a stronger role and  the peak of the IMF 
could be inherited from the peak of the core mass function (CMF) 
as proposed in \citet{padoan1997}, \citet{HC09}, \citet{Hennebelle12}, \citet{hopkins2012}, \citet{lee2016a} and \citet{lhc2017}. 
Such picture is suggested by the observations of the CMF which 
resembles the IMF \citep{motte98,alves2007,andre2010,konyves2015} and could be the dominant modes in some environments. 
It is remarkable that the peak of the CMF (particularly if an efficiency of $\simeq 30-50 \%$
is taken into account) and the peak of the stellar distribution imposed by the physics 
of the first Larson core, as discussed in this paper, lead to values that are similar within a factor of 
2-3. 
This coincidence may be for a large part responsible of the apparent universality of the IMF.

\section{Conclusions}

We have performed a series of numerical simulations  describing the collapse of a 1000 $\Ms$ cloud. 
These numerical experiments explicitly ignore magnetic field and radiative transfer at this stage. 
The transition from the isothermal to the adiabatic phase is of great importance and several 
effective equations of state have been explored. This includes varying $\gamma$, the effective adiabatic 
exponent, and $\rho_\ad$, the density at which the transition occurs.
We paid great attention to the numerical spatial resolution and its influences on the resulting stellar mass spectrum, 
in particular its peak position. 
The influence of the initial conditions are studied in paper I. 

We found that the isothermal simulations show no sign of convergence as the peak position is shifting toward lower masses 
when the spatial resolution improves. 
On the other hand, when a {polytrope-like} eos is used (typically with $\gamma > 4/3$), the peak position becomes independent of the numerical resolution if high enough. Using simple 1D hydrostatic models, we
infer the expected mass of the first Larson core that we compare to the measured peak position of the stellar distribution, 
compiling our simulations and various published results. 
We found a clear correlation between the two, with the peak position typically about ten times higher than 
the mass of the first Larson core.

We presented an analytical model that may account for this factor of roughly ten. It is based 
on the idea that in the vicinity of a point mass, such as a first Larson core, tidal 
forces are important and tend to quench further gravitational fragmentation by 
shearing out the density fluctuations. Therefore the mass located in the neighborhood of 
such point mass tends to be accreted, increasing the mass of the central objects in consequence. 
As discussed in paper I, this effect is true only when the gas is initially dense and turbulent, generating many fluctuations that eventually form stars close one to the other. 

Since the physics of the first Larson core is not expected to significantly vary from place to place, this
mechanism may provide a robust explanation for the apparent universality of the IMF peak. 
It should however be emphasized that an IMF-like distribution is produced only if the cloud is sufficiently massive and cold.

\begin{acknowledgements}
We thank the anonymous referee for comments which have improved the manuscript.
We thank Gilles Chabrier for a critical reading of the manuscript.
We thank Michel Rieutord for providing and helping to run the ESTER code to 
explore the influence of the rotation on the mass of a  polytrope.
This work was granted access to HPC
   resources of CINES under the allocation x2014047023 made by GENCI (Grand
   Equipement National de Calcul Intensif). 
   This research has received funding from the European Research Council under
   the European Community's Seventh Framework Programme (FP7/2007-2013 Grant  Agreement no. 306483). Y.-N. Lee acknowledges the financial support of the UnivEarthS Labex program at Sorbonne Paris Cit\'e (ANR-10-LABX-0023 and ANR-11-IDEX-0005-02)
\end{acknowledgements}

\appendix

\section{Sink accretion scheme} \label{appen_acc}
The thresholding scheme for sink accretion by \citet{Bleuler14} is used in this study. 
Within the sink accretion radius, four times the smallest cell size, 
a fraction $c_{\rm acc}$ of the mass exceeding the density threshold $n_{\rm sink}$ is accreted on to the sink, 
and by default $c_{\rm acc} = 0.75$. 
In the case where the mass from a cell is accreted by multiple sinks, 
the algorithm treats the sinks in decreasing order in terms of mass 
to ensure that more massive sinks accrete more mass. 
We perform a simulation with all parameters identical to C10h except that $c_{\rm acc} = 0.1$. 
Shown in Fig. \ref{fig_c01}, the stellar mass distribution does not differ significantly to that of run C10h (Fig \ref{fig_g53_10}), 
ensuring that the accretion scheme does not have a strong consequence on the final mass of the stars. 

\setlength{\unitlength}{1cm}
\begin{figure}
\begin{picture} (0,6.5)
\put(0,0){\includegraphics[width=8.7cm]{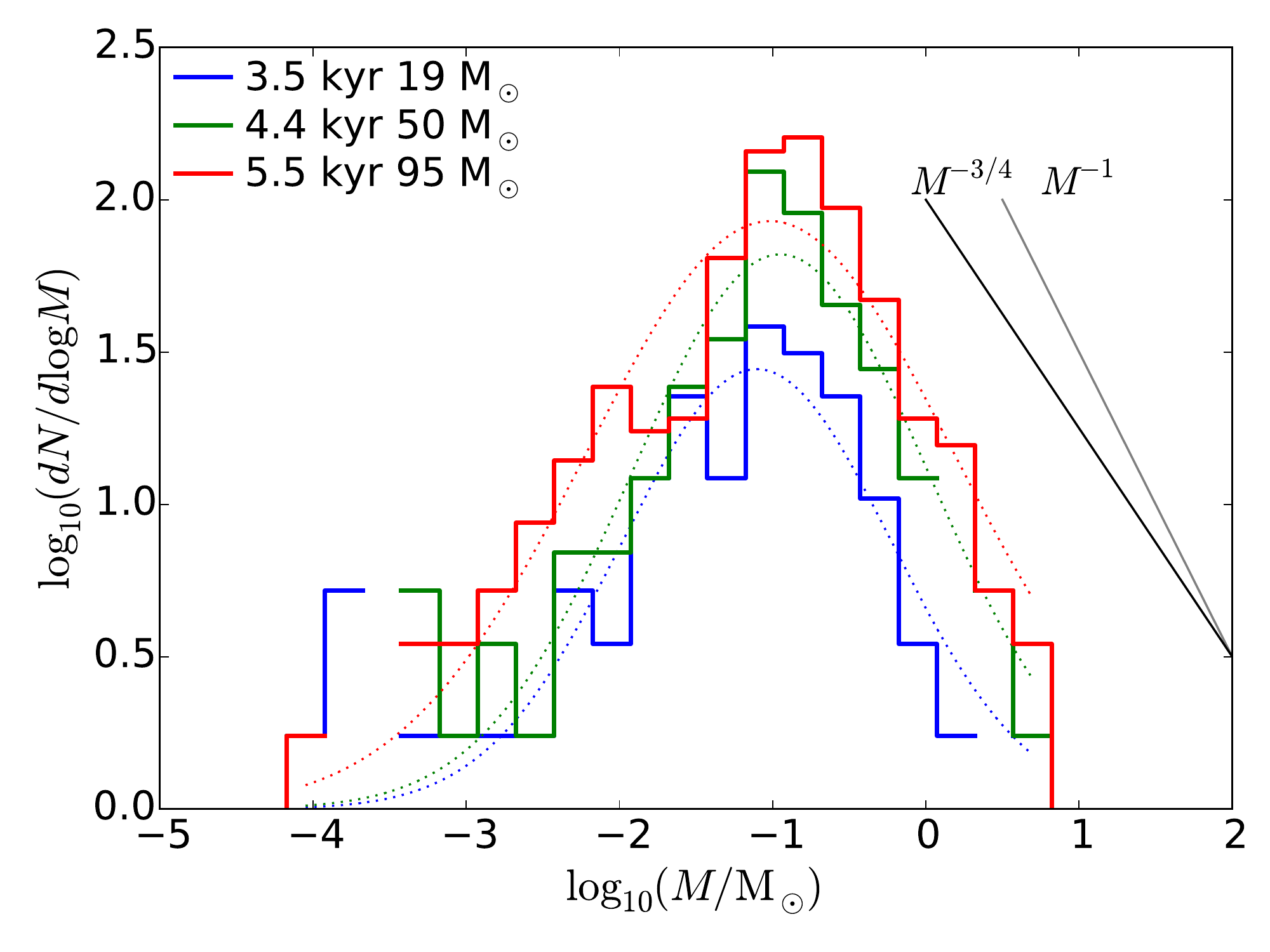}}  
\end{picture}
\caption{Stellar distributions of runs with smoothed two-slope polytropic eos with $\gamma = 5/3$, $n_\ad 10^{10}~\cc$, and $c_{\rm acc} = 0.1$ instead of 0.75 as in all the other runs. 
The peak of the distribution is unvaried compared to run C10h. }
\label{fig_c01}
\end{figure}

\section{Perturbation mass} \label{calc_mass}
The mass of the perturbation is equal to

\begin{align}
M\p(r\p,\delta r\p,\eta) &={Ac_\sound^2 \over G} \int\limits_{\delta r=0}^{\delta r\p}\int\limits_{\theta=-0}^{\pi } \left({1\over r\e^2} + {\eta\over r\p^2}\right)\sin\theta d\theta ~\delta r^2 ~d\delta r \\
& = {Ac_\sound^2 \over G}  \int\limits_{\delta r = 0}^{\delta r\p} \int\limits_{\cos\theta=-1}^1 \left({1\over r\e^2} + {\eta\over r\p^2}\right)d\cos\theta~\delta r^2 ~d\delta r \nonumber\\
& = {Ac_\sound^2 \over G}  \int\limits_{\delta r = 0}^{\delta r\p} \left.\left[-{\log r\e\over r\p\delta r}+{\eta \cos\theta\over r\p^2} \right]\right\vert_{\cos\theta=-1}^1\delta r^2 d\delta r \nonumber\\
& = {Ac_\sound^2 \over G} r\p \int\limits_{u=0}^{u\p} \left[\log\left({1+u\over 1-u}\right)u + 2\eta u^2 \right]du \nonumber\\
& = {Ac_\sound^2 \over G} r\p \left[u\p-{1-u\p^2 \over 2}\log\left({1+u\p\over 1-u\p}\right) + {2\over 3} \eta u\p^3 \right]\nonumber\\
& = {Ac_\sound^2 \over G} r\p \mathfrak{m}\p(u\p,\eta), \nonumber
\end{align} 
where the normalized quantities $u=\delta r/r\p$ and $u\p=\delta r\p/r\p$.

\section{Tidal forces in the simulations}
\label{tidal}
In the vicinity of an accreting point mass, two contributions
 have to be considered. First the $r^{-2}$ gravitational force induced by the 
central object and second the $r^{-1}$ force associated to the $r^{-2}$ density profile envelope. Both 
have a positive eigenvalue, $\lambda_3$, and two negative ones, $\lambda_1= \lambda_2$. 
It can be shown that for the $r^{-2}$ gravitational field, $\lambda_3 / \lambda_1 = -2$ while for  the second one  $\lambda_3 / \lambda_1 = -1$.

\begin{figure}[t]
\centering
\includegraphics[trim=0 15 0 10,clip,width=0.5\textwidth]{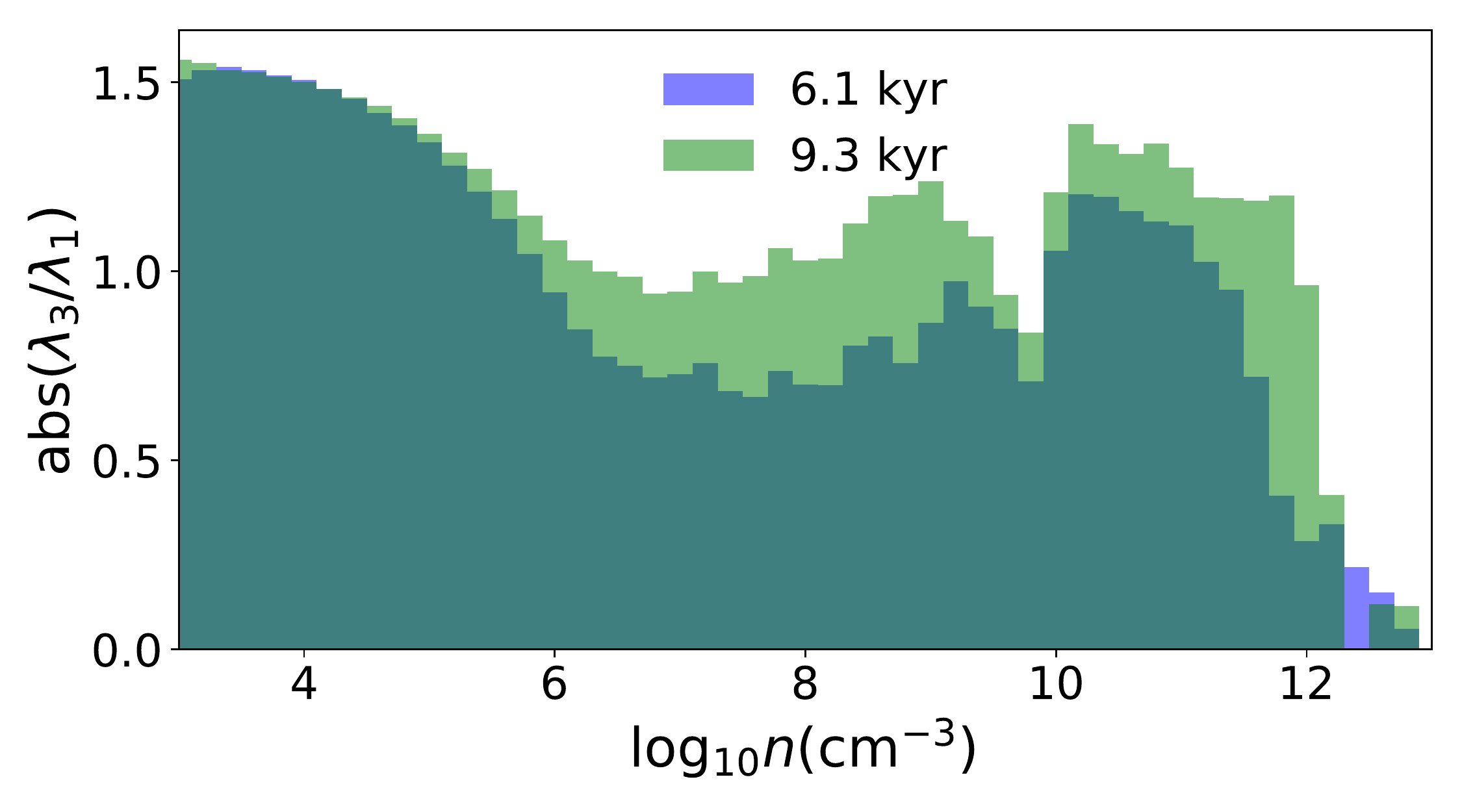}
\caption{Tidal forces in run C10h. The absolute value of the ratio between largest and smallest eigenvalues of the gravitational stress tensor $\lambda_3/\lambda_1$ is plotted against local density at 6.1 and 9.3 kyr. At very low density, corresponding to the outer region of the cloud, the gravitational field is dominated by the cloud global density distribution and the central cluster, and thus $-2<\lambda_3/\lambda_1$<-1. The tidal forces become less destructive when reaching slightly higher densities. At extremely high density ($> n_\mathrm{sink}=10^{10}~\cc$), corresponding to regions at the vicinity of existing sinks, the tidal forces becomes very destructive locally, preventing nearby sink formation.} 
\label{fig_hist_tidal}
\end{figure}

In order to assess the importance of tidal forces in our simulations, we have calculated the gravitational stress tensor. 
We first average $\lambda_3$ and $\lambda_1$ in density bins, and then plot the absolute value of their ratio. 
Figure \ref{fig_hist_tidal} shows result from run C10h at 6.1 and 9.3 kyr. 
Low densities correspond to outer region of the cloud that is dominated by the tidal field of cloud global density distribution and the central cluster. 
At high densities ($> n_\mathrm{sink}=10^{10}~\cc$), corresponding to regions in the vicinity of sink particles, the destructive tidal force is strong such that no core formation is allowed nearby. This illustrates the dominating effect of the gravitational field generated by the sink particle.

\section{1D calculation of gravitational energy}\label{appen_1D}
The expression of the 1D gravitational energy is
\begin{align}
\label{eq_vir_1D_comp}
&\int\limits_{-\delta r\p}^{\delta r\p}   \rho(\delta r) ~ g(\delta r) ~\delta r ~d\delta r = \\
&{(Ac_\sound^2)^2 \over 2\pi Gr\p} \left\{ -{GM_\mathrm{L} \over Ac_\sound^2r\p}\left[ {-8u\p^3\over 3(1-u\p^2)^3} - {2(\eta-1) u\p \over 1-u\p^2} + (\eta-1)\log\left({1+u\p\over 1-u\p}\right) \right]  \right. \nonumber \\
&+ {4u\p^3\over (1-u\p^2)^2} - {16\over 3}\eta u\p - {4u\p \over 1-u\p^2}- {4\eta \over 3}{u\p\over 1-u\p^2} - {4\over 9}\eta^2u\p^3 \nonumber\\
&\left. + 2\log \left({1+u\p\over 1-u\p}\right)  + {10\over 3}\eta \log\left({1+u\p\over 1-u\p}\right) \right\} \nonumber\\
&={(Ac_\sound^2)^2 \over 2\pi Gr\p} \mathcal{E}_\mathrm{1D,g}(r\p,u\p,\eta), \nonumber
\end{align}

\section{3D calculation of gravitational energy}\label{appen_3D}

We define $\xi$ the angle between $\mathbi{r}\e$ and $\delta\mathbi{r}$, and $\theta$ the angle between $\delta\mathbi{r}$ and $-\mathbi{r}\p$.
Projected onto the direction of $\delta\mathbi{r}$, 
the gravitational fields have the values: 
\begin{align}
g_\mathrm{L}(\mathbi{r}\e) &= g_\mathrm{L,a}(\mathbi{r}\e) -g_\mathrm{L,c}(\mathbi{r}\e)=-{GM_\mathrm{L} \over r\e^2} \cos \xi -{GM_\mathrm{L} \over r\p^2} \cos \theta, \\
g\e(\mathbi{r}\e) &=g_\mathrm{e,a}(\mathbi{r}\e) -g_\mathrm{e,c}(\mathbi{r}\e) = -{2Ac_\sound^2 \over r\e} \cos \xi -{2Ac_\sound^2 \over r\p} \cos \theta, \\
g\p(\mathbi{r}\e) &= -\eta{2Ac_\sound^2 \over 3 r\p^2} \delta r,
\end{align}
where the subscripts a and c signify the value at the point considered and the one at the center of the perturbation. 
The local density is
\begin{align}
\rho(\mathbi{r}\e) = {Ac_\mathrm{s}^2 \over 2\pi G} \left({1 \over r\e^2} + {\eta \over r\p^2}\right)
\end{align}
We have the trigonometric relations 
\begin{align}
r\e^2 &= r\p^2 + \delta r^2 - 2r\p\delta r \cos \theta ~~\mbox{and} \\
r\p^2 &= r\e^2 + \delta r^2 - 2r\e\delta r \cos \xi, 
\end{align}
and we deduce $\cos \xi /r\e= [1+ (\delta r^2 - r\p^2)/r\e^2]/(2\delta r)$. 
The contribution to the gravity at the center of the perturbation form the central star and from the envelop behave the same way, and the normalized integration gives
\begin{align}
&\mathcal{E}_\mathrm{3D,L,c} = \mathcal{E}_\mathrm{3D,e,c}  \\
&=\int\limits_{\delta r = 0}^{\delta r\p} \int\limits_{\theta= 0}^{\pi}- {\cos \theta \over r\p^2} \left({1\over r\e^2} +{\eta \over r\p^2} \right) \delta r^3 \sin \theta d\theta d\delta r  \nonumber\\
&= {u\p(3-u\p^2)\over 4} + {u\p^4+2u\p^2-3 \over 8}\log \left({1+u\p\over 1-u\p}\right) \nonumber
\end{align}
Considering the integration of Eq. (\ref{eq_int_vir}) separately for each gravitation term without subtracting the mean gravity field, we obtain:
\begin{align}
& \mathcal{E}_\mathrm{3D,L,a} = {1\over M_\mathrm{L}Ac_\sound^2} \int\limits_{V\p} \rho \mathbi{g}_\mathrm{L,a} \cdot \delta\mathbi{r} dV \\ 
&= -\int\limits_{\delta r = 0}^{\delta r\p} \int\limits_{\theta= 0}^{\pi} {\cos \xi \over r\e^2} \left({1\over r\e^2}+{\eta \over r\p^2}\right)\sin \theta d\theta \delta r^3 d \delta r \nonumber\\
&=  -\int\limits_{\delta r = 0}^{\delta r\p} \int\limits_{\cos\theta=-1}^1{\delta r^2\over 2} \left({1\over r\e^3}+{\eta \over r\e r\p^2}\right)\left(1+{\delta r^2-r\p^2 \over r\e^2}\right)d\cos\theta d\delta r \nonumber \\
&= - \int\limits_{\delta r = 0}^{\delta r\p} {\delta r\over 2r\p} \left.\left[{1\over r\e}-{\eta r\e\over r\p^2} - \left(r\p^2-\delta r^2\right)\left({1\over 3r\e^2}+{\eta \over r\e r\p^2}\right)\right]\right\vert_{\cos\theta=-1}^1  d\delta r \nonumber \\
& = - \int\limits_{u=0}^{u\p} \left({u\over 1-u^2} - {(3+u^2)u\over 3(1-u^2)^2}\right)udu \nonumber \\
& = - \left[-{4\over 3}u\p + \log\left({1+u\p\over 1-u\p}\right) -{2\over 3}{u\p \over 1-u\p^2} \right], ~\text{and thus}\nonumber 
\end{align}
\begin{align}
&\mathcal{E}_\mathrm{3D,L} = \mathcal{E}_\mathrm{3D,L,a} + \mathcal{E}_\mathrm{3D,L,c},\label{eq_gL_u} 
\end{align}
\begin{align}
&\mathcal{E}_\mathrm{3D,e,a} = {G \over (Ac_\sound^2)^2r\p} \int\limits_{V\p} \rho \mathbi{g}_\mathrm{e,a} \cdot \delta\mathbi{r} dV \\ 
&= -{1\over r\p} \int\limits_{\delta r = 0}^{\delta r\p} \int\limits_{\theta= 0}^{\pi } {2\cos \xi \over r\e} 
\left({1\over r\e^2}+{\eta \over r\p^2}\right)\sin \theta d\theta \delta r^3 d \delta r \nonumber\\
&= -{1\over r\p} \int\limits_{\delta r = 0}^{\delta r\p} \int\limits_{\cos\theta=-1}^1  \left({1\over r\e^2}+{\eta \over r\p^2}\right) 
\left(1+{\delta r^2-r\p^2 \over r\e^2}\right)d\cos\theta\delta r^2 d\delta r \nonumber \\
&= \int\limits_{\delta r = 0}^{\delta r\p} \!{\delta r^2\over r\p} \! \left.\left[{\log r\e\over r\p\delta r}\!-\!{\eta \cos\theta\over r\p^2} \!+\! {r\p^2\!-\!\delta r^2\over r\p \delta r}\left({1\over 2r\e^2}\!-\!{\eta \log r\e\over r\p^2}\right)\right]\right\vert_{\cos\theta=-1}^1 \!\!\! d\delta r \nonumber \\
& = \int\limits_{u=0}^{u\p}\! \left\{\!-u\log\left({1\!+\!u\over 1\!-\!u}\right)\!+\!{2u^2\over 1\!-\!u^2} \!+\! \eta\left[u\left(1\!-\!u^2\right)\log\left({1\!+\!u\over 1\!-\!u}\right) \!-\! 2u^2\right] \right\}du \nonumber \\
& = \left\{\!{3{-}u\p^2\over 2}\log\left({1\!+\!u\p\over 1\!-\!u\p}\right) \!-\!3u\p \!+\!\eta\left[ {u\p\over 2} {-} {5u\p^3 \over 6} {-} \left({1 \!-\! u\p^2\over 2}\right)^2\!\log\left({1\!+\!u\p\over 1\!-\!u\p}\right) \right] \right\}, \nonumber  
\end{align}
leading to
\begin{align}
&\mathcal{E}_\mathrm{3D,e} = \mathcal{E}_\mathrm{3D,e,a} + \mathcal{E}_\mathrm{3D,e,c},\label{eq_ge_u} 
\end{align}
\begin{align}
\label{eq_gp_u}
&\mathcal{E}_\mathrm{3D,p} = {G \over (Ac_\sound^2)^2r\p} \int\limits_{V\p} \rho \mathbi{g}\p \cdot \delta\mathbi{r} dV \\ 
& = -\eta {2\over 3}{1\over r\p^3} \int\limits_{\delta r = 0}^{\delta r\p} \int\limits_{\theta= 0 }^{\pi  }\left({1\over r\e^2}+{\eta \over r\p^2}\right)\sin \theta d\theta \delta r^4 d \delta r \nonumber\\
&= -\eta {2\over 3}{1 \over r\p^3} \int\limits_{\delta r = 0}^{\delta r\p} \int\limits_{\cos\theta=-1}^1\left({1\over r\e^2}+{\eta \over r\p^2}\right)d\cos\theta\delta r^4 d\delta r \nonumber \\
&= -\eta {2\over 3}{1\over r\p^3} \int\limits_{\delta r = 0}^{\delta r\p}  \left.\left[-{\log r\e\over r\p\delta r}+{\eta \cos\theta\over r\p^2} \right]\right\vert_{\cos\theta=-1}^1 \delta r^4 d\delta r \nonumber \\
& = -\eta {2\over 3} \int\limits_{u=0}^{u\p} \left[u^3\log\left({1+u\over 1-u}\right)+2\eta u^4\right]du \nonumber \\
& = -\eta {2\over 3} \left[{u\p\over 2} + {u\p^3\over  6} + \eta{2u\p^5 \over 5}-{1-u\p^4\over 4}\log\left({1+u\p\over 1-u\p}\right)\right],  \nonumber
\end{align}
where $u = \delta r/r\p$, and $u\p=\delta r\p/r\p$.
The condition for a second core forming near the initial central core is that $E_\grav({r\p, \delta r\p}) \leq 0$ and that $M(r\p,\delta r\p,\eta) = M_\mathrm{L}$. Substituting Eqs. (\ref{eq_gL_u},  \ref{eq_ge_u}, \ref{eq_gp_u}) and (\ref{eq_M_u}) into Eq. (\ref{eq_int_vir}) reduces the latter to Eq. (\ref{eq_3D_norm})

\section{Density fluctuations: model II} \label{modelII}

\begin{figure}[]
\begin{picture} (0,9.3)
\put(0,4.65){\includegraphics[trim=0 20 0 12,clip,width=0.5\textwidth]{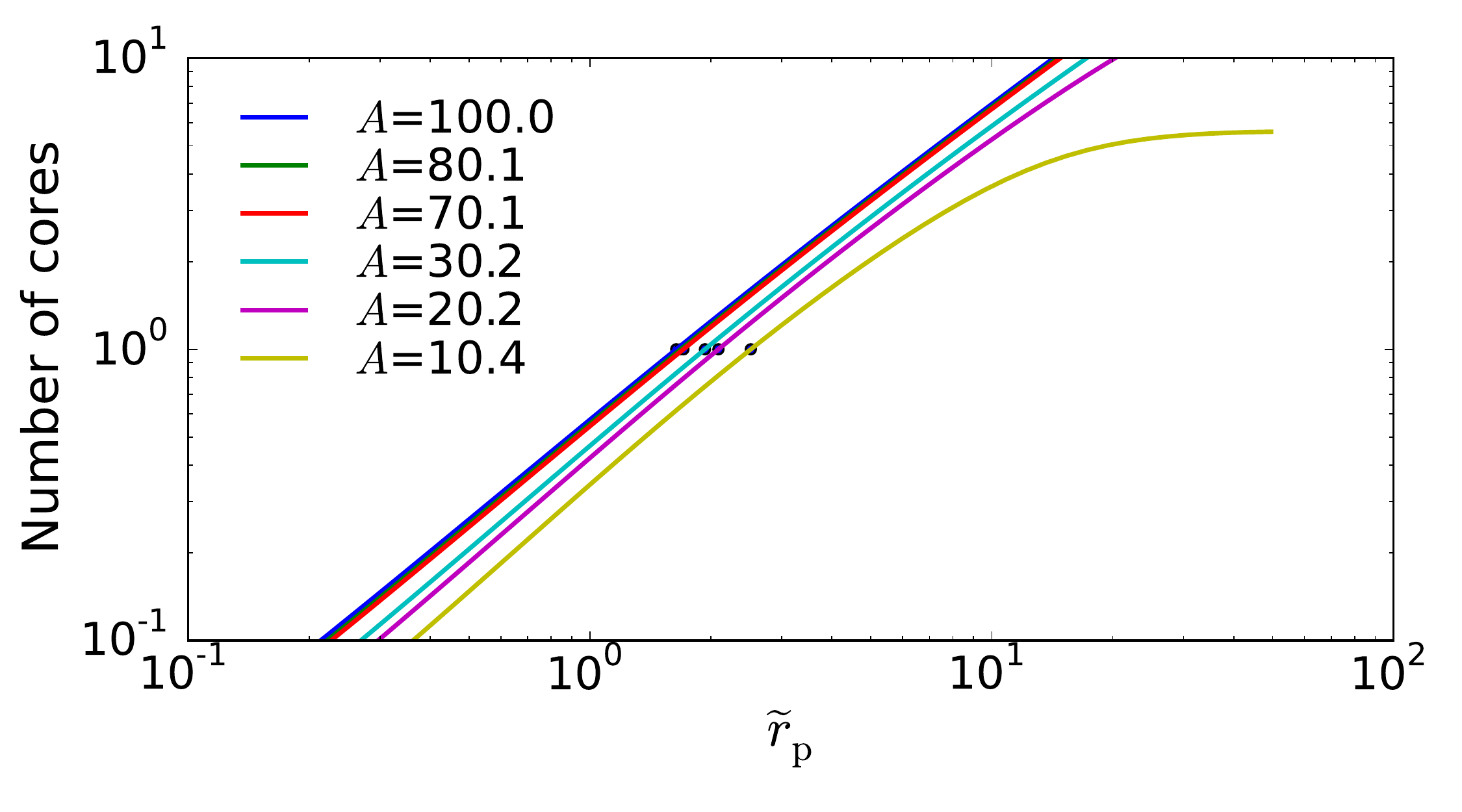}}
\put(7,5.65){Model I}
\put(0,0){\includegraphics[trim=0 20 0 12,clip,width=0.5\textwidth]{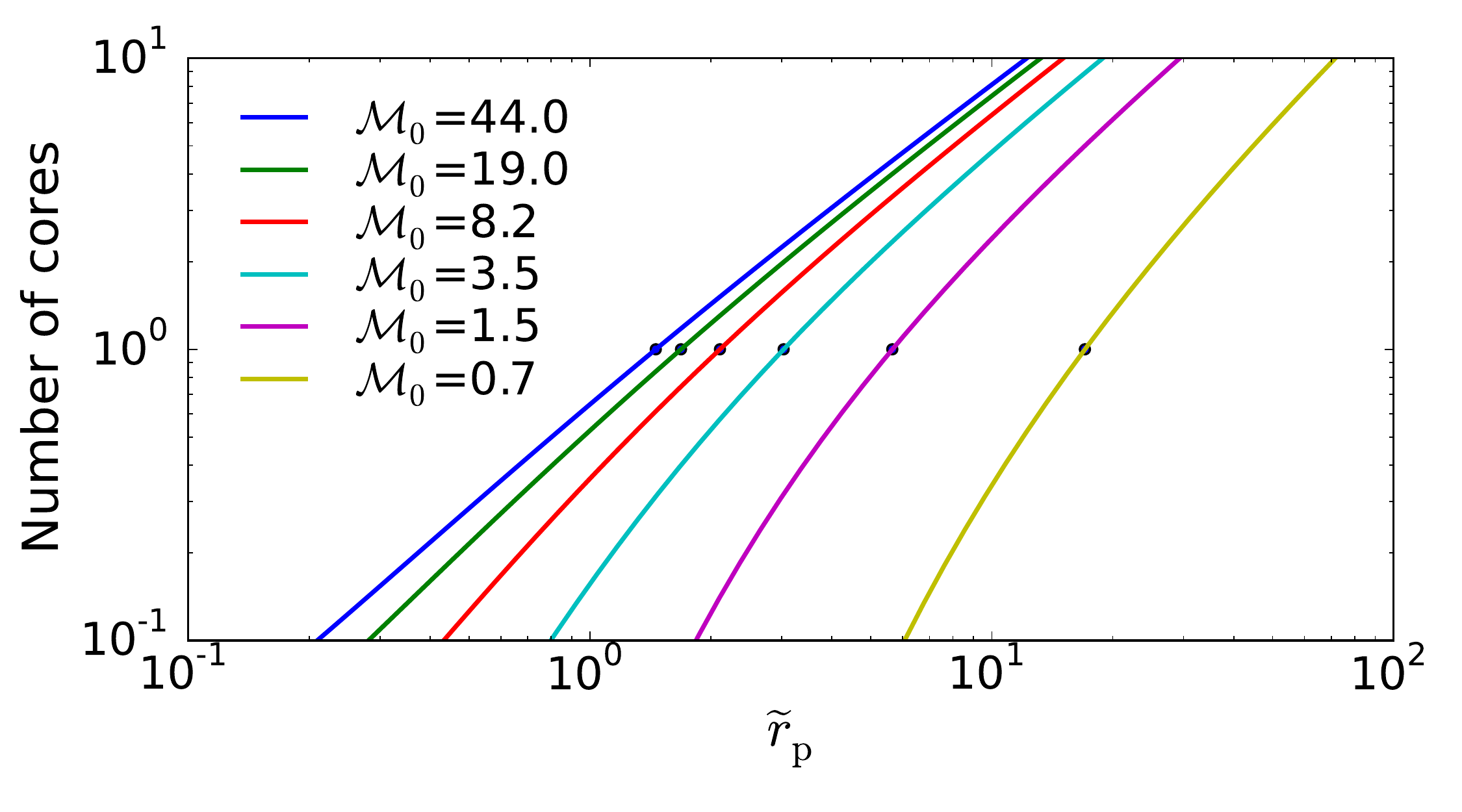}}
\put(7,1.){Model II}
\end{picture}
\caption{Probable number of self-gravitating cores found within radius $\widetilde{r}\p$. {\it Upper panel:} Model I plotted with $\epsilon=1$ and for several values of the density envelope amplitude, $A$. {\it Lower panel:} Model II without thermal support for several values of global cloud Mach number, $\mathcal{M}_0$. The circles mark the values where one self-gravitating core is found. In the range of our interest, that is, $\mathcal{N}(\widetilde{r}\p)$ equals a few, there is an almost linear dependance between $\mathcal{N}(\widetilde{r}\p)$ and~ $\widetilde{r}\p$.}
\label{fig_N_rp}
\end{figure}

\begin{figure}[]
\begin{picture} (0,4.6)
\put(0,0){\includegraphics[trim=0 18 0 12,clip,width=0.5\textwidth]{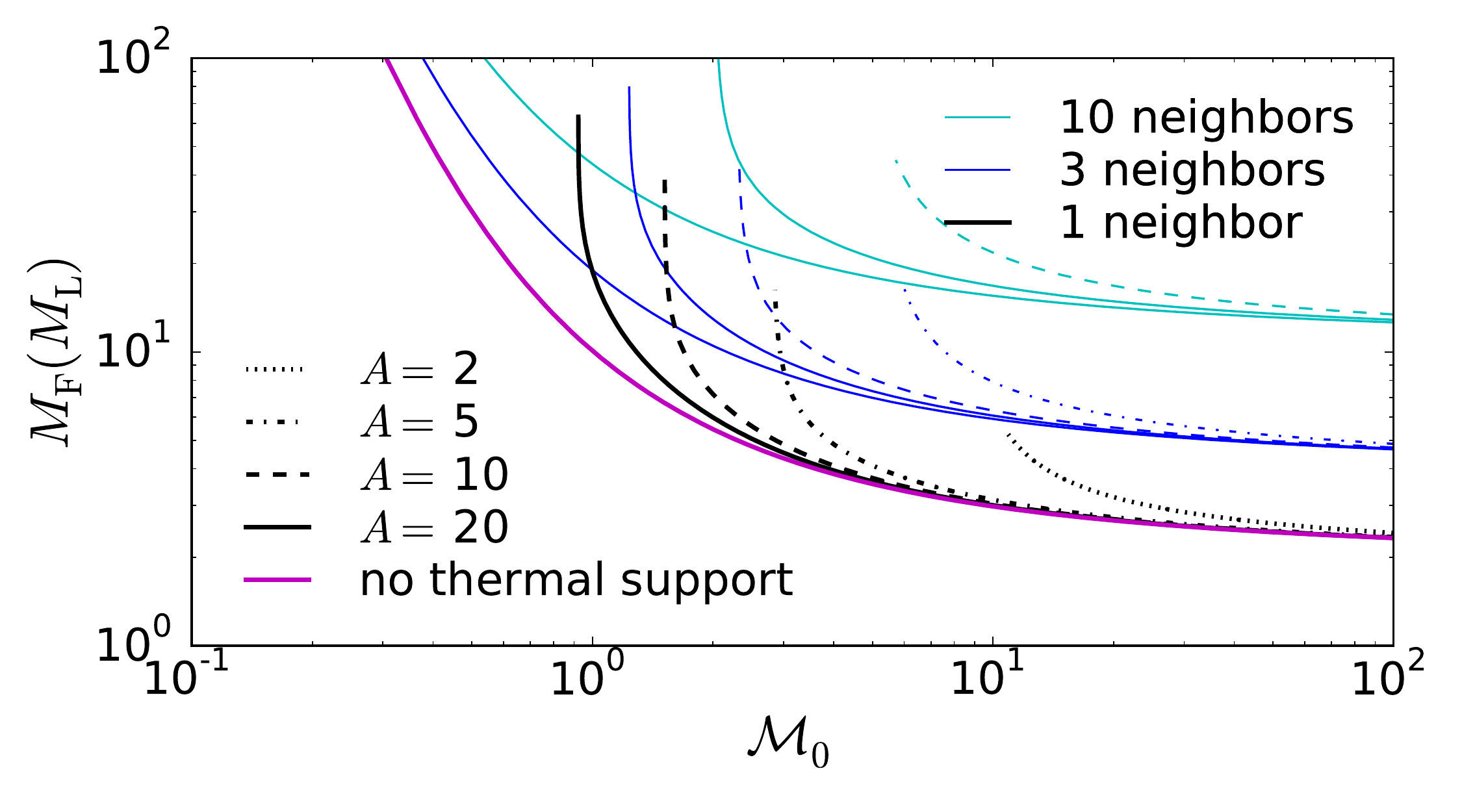}}
\put(7,1){Model II}
\end{picture}
\caption{The final mass $M_\mathrm{F}$ in units of $M_\mathrm{L}$ for model II.
 $M_\mathrm{F}$ is 
plotted against the level of turbulence for cases with $A=2, 5, 10, 20$, and without thermal support ($A \rightarrow\infty$, magenta). 
Model II also shows that $M_\mathrm{F}$ is expected to be larger than 3 
when the envelope is truncated at $\mathcal{N}(\widetilde{r}\p)=1$, and $> 5$ when $\mathcal{N}(\widetilde{r}\p)=3$ (yellow thin lines). } 
\label{fig_MF_M0_modelII}
\end{figure}

Since the density fluctuations are important and they control the fragmentation around the accreting objects, 
we explore a second model (model II) to explore the robustness of our conclusions. 
We assume that the turbulence follows a Larson relation inside the cloud, 
and is conserved during collapse (this assumption is possibly not accurate as the collapse likely generates 
turbulent fluctuations), we have 
\begin{align}
\mathcal{M}(\widetilde{r}\p) = {v_\turb(\widetilde{r}\p) \over c_\sound} = {v_{\turb, \mathrm{cloud}} \over c_\sound} \left({r_0(r\p) \over r_\mathrm{cloud}}\right)^{1\over 2} = \mathcal{M}_\mathrm{cloud} \left({M(r\p) \over M_\mathrm{cloud}}\right)^{1\over 6}, 
\end{align}
where $v_\turb$, $r_0$, $v_{\turb,\mathrm{cloud}}$, $r_\mathrm{cloud}$, 
and $\mathcal{M}_\mathrm{cloud}$ are the turbulent velocity at $r\p$, the radius that corresponds to the mass enclosed 
inside radius $r\p$ before the local mass concentration (at cloud average density), the turbulent velocity at cloud scale, 
the radius of the cloud, and the Mach number of the cloud. 
The mass $M(r\p)$ contained inside $r\p$ takes into account the envelope and the central mass, and can be express with the normalized form $M(r\p) =  M_\mathrm{L}(\widetilde{r}\p+1)$. This leads to
\begin{align}\label{eq_mach_1}
\mathcal{M}(\widetilde{r}\p) = \mathcal{M}_\mathrm{cloud} \left({M_\mathrm{L} \over M_\mathrm{cloud}}\right)^{1\over 6} (1+\widetilde{r}\p)^{1\over 6} = \mathcal{M}_0  (1+\widetilde{r}\p)^{1\over 6}.
\end{align}
The local lognormal dispersion is therefore
\begin{align}
\sigma^2(\widetilde{r}\p) = \log \left(1+b^2\mathcal{M}_0^2(1+\widetilde{r}\p)^{1\over 3}\right).
\end{align}
The initial Mach number of our canonical run is $\sim 20$, and $M_\mathrm{L} / M_\mathrm{cloud} \sim 10^{-2}/10^3 = 10^{-5}$. 
Above values yield typically $\mathcal{M}_0 = \mathcal{M}_\mathrm{cloud} (M_\mathrm{L} / M_\mathrm{cloud})^{1/6} \sim 2-3$. 
An important difference between the two assumptions made in models I and II is that the second estimate depends on the initial conditions 
while the first is entirely local.

We perform numerical integrations for several values of $\mathcal{M}_0$.  
Figure~\ref{fig_N_rp} shows the resulting $\mathcal{N}(\widetilde{r}\p)$, without considering the thermal support. 
We define the critical radius $\widetilde{r}\p^\ast$ where there is probability of finding one self-gravitating mass $M_\mathrm{L}$, as marked with circles in the figure (model I is also shown for comparison). 

The resulting  $M_\mathrm{F} = (1+\widetilde{r}\p^\ast)M_\mathrm{L}$
 is displayed in Fig.~\ref{fig_MF_M0_modelII} and should be compared to Fig.~\ref{fig_MF_M0}.  
When further considering the thermal energy of the perturbation, 
the protected radius is increased since more support is present. 
In  Fig. \ref{fig_MF_M0_modelII}, we also show results with thermal support using $A=2, 5, 10,$ and $20$. 
The effect of thermal support is more prominent when $A$ is small, imposing a lower limit on $\mathcal{M}_0$, 
below which a star prevents its surrounding gas form fragmenting and will grow to become massive. 
The physical interpretation is that when the envelope density is low, 
stronger fluctuations are needed to create multiple stars, otherwise one star will dominate and prevent any further formation once it is formed. 
This is in good agreement with the result of \citet{Girichidis11} that clumps with an $r^{-2}$ profile and weak turbulence do not fragment. 
For $A \sim 10$, as measured in paper I (left panels of Fig.~4), the condition $\mathcal{N}=1$ gives at minimum $M_\mathrm{F} \sim 10 M_{\rm L}$ for any $\mathcal{M}_0$ lower than $\sim 2$. 
As seen from Fig.~\ref{fig_Mlr_Mpk}, this is in  good agreement with the value inferred from the simulations in the sense that 
the model predicts that the accreting envelope, protected from  gravitational fragmentation, is several times the 
mass of the first Larson core.

\bibliographystyle{aa}
\bibliography{lars}

\end{document}